\documentclass[conference]{IEEEtran}
\IEEEoverridecommandlockouts

\usepackage{amsmath,amssymb,amsfonts}
\usepackage{algorithmic}
\usepackage{graphicx}
\usepackage{textcomp}
\usepackage{xcolor}
\usepackage{bm}
\usepackage{epstopdf}
\usepackage{lineno,hyperref,float,amsfonts,amsmath,epsfig,graphics,booktabs,multirow,cite,stfloats}
\usepackage{comment}

\usepackage{longtable}
\usepackage{caption}
\usepackage{array}
\newcommand{\tabincell}[2]{\begin{tabular}{@{}#1@{}}#2\end{tabular}}

\usepackage{tablefootnote}

\PassOptionsToPackage{hyphens}{url}
\usepackage{pifont}
\newcommand{\cmark}{\ding{51}}%

\usepackage{color,xspace}

\newcommand{\eg}{\emph{e.g.,}\xspace}

\newcommand{\ie}{\emph{i.e.,}\xspace}

\newcommand{\remove}[1]{}

\def\BibTeX{{\rm B\kern-.05em{\sc i\kern-.025em b}\kern-.08em
    T\kern-.1667em\lower.7ex\hbox{E}\kern-.125emX}}
\begin{document}

\title{Data Sharing, Privacy and Security Considerations in the Energy Sector: A Review  from Technical Landscape to Regulatory Specifications}

\author{Shiliang Zhang, Sabita Maharjan, Lee Andrew Bygrave, Shui Yu
	\thanks{Shiliang Zhang, Sabita Maharjan, and Lee Andrew Bygrave are with University of Oslo, Norway (e-mail: \{shilianz, sabita, l.a.bygrave\}@uio.no).}
	\thanks{Shui Yu is with University of Technology Sydney, Australia (e-mail: Shui.Yu@uts.edu.au).}
}

\maketitle

\begin{abstract}
Decarbonization, decentralization and digitalization are the three key elements driving the twin energy transition. The energy system is evolving to a more data driven, intelligent and advanced ecosystem, leading to the need of communication and storage of large amount of data of different resolution from the prosumers and other stakeholders in the energy ecosystem. While the energy system is certainly advancing, this paradigm shift is bringing in new privacy and security issues related to collection,  processing and storage of data - not only from the technical dimension, but also from the regulatory perspective. Although data privacy and security are well-investigated areas and the energy system is not an exception in this regard, understanding data privacy and security in the evolving energy system, regarding regulatory compliance, is an immature field of research. Contextualized knowledge of how related issues are regulated is still in its infancy, and the practical and technical basis for the regulatory framework for data privacy and security is not clear. In this study, we attempt to fill this gap by taking a holistic approach towards a comprehensive review of the data-related issues - including privacy and security - for the energy system, by integrating both technical and regulatory dimensions. We start by reviewing open-access data, data communication and data-processing techniques for the energy system, and use it as the basis to connect the analysis of data-related issues from the integrated perspective. We classify the issues into three categories: (i) data-sharing among energy end users and stakeholders (ii) privacy of end users, and (iii) cyber security, and then explore these issues from a regulatory perspective. We analyze the evolution of the regulatory framework for energy systems, and introduce the relevant regulatory initiatives for the categorized issues in terms of regulatory definitions, concepts, principles, rights and obligations in the context of energy systems. Finally, we provide reflections on the gaps that still exist, and guidelines for regulatory frameworks for a truly participatory energy system. To this end, this study provides (i) technical basis and insight into the gap between the current regulatory framework and the technical challenges related to data-sharing, privacy, and cybersecurity in the energy system (ii) systematic regulatory knowledge and contextualized guidelines for energy end-users and other stakeholders regarding categorized data-related issues, and (iii) open research challenges for specific potential applications where the gap between the technical and the regulatory dimensions is still wide, and also the guidelines towards facilitating a truly participatory approach to the  decentralized, distributed and secure energy system.
\end{abstract}


\section{Introduction}
Ambitious energy goals have been set aiming at sustainability, efficiency and security of energy supply in the future, \eg the 2020\footnote{\url{https://energy.ec.europa.eu/topics/renewable-energy/renewable-energy-directive-targets-and-rules/renewable-energy-targets_en\#the-2020-targets}}, 2030\footnote{\url{https://climate.ec.europa.eu/eu-action/climate-strategies-targets/2030-climate-energy-framework_en}}, and 2050\footnote{\url{https://climate.ec.europa.eu/eu-action/climate-strategies-targets/2050-long-term-strategy_en}} EU energy goals. 
Smart grid and transactive energy systems incorporate more renewables from distributed generations, and also create scope for prosumer-oriented energy market, with reduced energy cost and overall improved energy efficiency with the help of bilateral communications between the stakeholders at different levels of the energy system~\cite{TOOKI2024100596}. Distributed and participatory energy paradigms like microgrid, smart neighborhoods, energy community, vehicle-to-grid (V2G), and peer-to-peer energy trading~\cite{leal2023electricity,RODRIGUES2023112999} can reduce energy cost related to long-distance transmission. Novel energy paradigms also yield significant cost savings in terms of the required investment in infrastructure upscaling of the grid. However, the decentralized architecture and the distributed energy service provisioning induce information exchange and data storage at a much higher granularity than the conventional exchange. The availability, security, sharing, and processing of data that are critical to operation and maintenance, marketing and pricing, and various other services in the energy system~\cite{KHALID2024110253,10374424}. The emerging AI approaches play a crucial rote in facilitating the twin transition~\cite{daehlen2023twin} 
through data exchange and interactions, creating novel service and business models for the energy sector. Nevertheless, the digitalization and decentralization of the energy system, along with the availability of advanced AI techniques, induce significant concerns about data security and privacy from both technical and regulatory aspects~\cite{rao2023security,Papakonstantinou2015}.

Privacy is one of the well-recognized barriers to the acceptance of emerging prosumer-centric distributed energy system architectures.~\cite{BUGDEN2019137}. Smart meters can generate and transfer fine-grained user data for operation and maintenance purposes~\cite{DBLP:journals/tochi/JakobiPRSW19,DBLP:journals/tsg/EiblE15}. Such transfer can disclose sensitive information~\cite{DBLP:conf/wpes/TudorAP13}, \eg social class and employment status~\cite{RAZAVI2018312}, that may limit consumer's trust towards the energy system~\cite{BALTAOZKAN201465,DBLP:journals/taasm/FellSHE14,WILSON201772}. Consumers/prosumers with privacy concerns are conservative towards participatory approaches~\cite{HMIELOWSKI2019189,HORNE201564,NAUS2015125} due to the risk of data leakage~\cite{HANSEN2017112,KRISHNAMURTI2012790}. This situation is deteriorated further by the the risk of unauthenticated/insecure sharing of consumer data through the increasing connections and inter-dependencies amongst IoTs of utility companies and third parties~\cite{uppuluri2023secure}. Beyond privacy, issues with data security in the energy system, \textit{e.g.}, data jamming, spoofing, tampering, injection, eavesdropping, can lead to failure in critical information communication, illegal access to the energy system and consumer information, and even damage to the operation of the energy system that can lead to severe physical consequences such as power blackout in large scale~\cite{GHIASI2023108975}.

Data-related issues in the energy system can induce operational obscures to the energy stakeholders and risk of regulatory compliance, \textit{e.g.}, violations against General Data Protection Regulation (GDPR)~\cite{10.1145/3655693.3656546}. Addressing those challenges requires a thorough understanding of technical aspects of the data-related issues~\cite{DBLP:journals/tse/IwayaBR23}, including the availability of data, communication, and existing data processing techniques in the energy system. Moreover, it is critical to provide a clear articulation of data-related regulations~\cite{ogunniye2023survey}, which is anticipated to provide concrete guidelines that the stakeholders, \textit{e.g.}, prosumers, aggregators, DSOs, and third parties, need to follow to avoid compliance confusions in their data acquisition, transferring, processing, and storage. When considering, \textit{e.g.}, the scenario where stakeholders are legislatively involved in personal data processing via different communication and processing techniques, the technical and the regulatory dimensions are essentially closely connected~\cite{MACEDO2023138102}. We emphasize that an integrated approach with technical and regulatory understanding of the data-related issues is essential to develop comprehensive knowledge that the stakeholders need and can actually adopt.

Though there are efforts to check and review data-related issues in the energy sector~\cite{10.1093/ijlit/eau001,harvey2013smart,LESZCZYNA201862,ANANDAKUMAR2014126}, such efforts are subject to specific and separated aspects like privacy or cyber security. The aspects related to, \eg privacy and security, are intertwined~\cite{10.1145/3645091,DBLP:journals/misq/SmithDX11}. However, in this rapidly changing data sharing landscape, it is essential also to add the regulatory and legislative privacy and data security aspects, and to analyze from the integrated perspective for a comprehensive understanding. While the existing studies focus on the techniques and applications involved in data processing for the energy system~\cite{rao2023security}, aspects like data source and data availability, and the role and responsibilities of the entities involved in data exchange \textit{etc.}, are left largely unexplored. Such aspects play a crucial role in connecting between the technical and the regulatory dimensions, and thus in understanding and analyzing regulatory compliance for data-sharing amongst the stakeholders of the emerging distributed and decentralized energy ecosystem.

Legislation provides a concrete basis and establishes formal requirements for data-related issues. However, compared to the volume of technical literature, regulatory studies on data-related issues in the energy sector are sparse. To the best of our knowledge, there is no systematical review and analysis of data sharing, privacy and cyber security from a regulatory perspective, thus making it challenging to formulate a comprehensive understanding of the current status and development of privacy or security regulations in the distributed energy system. This is partially because (i) the related regulations along with the energy sector are still evolving~\cite{XIA2023104771,VEZZONI2023103134}, leading to a gap between the legacy and updated regulation articulation, and (ii) there is insufficient effort to map regulatory provisions to the energy sector~\cite{HAJIBASHI2023113055,FRILINGOU2023102934,MCCLEAN2023113378}, though regulations applicable to energy data related issues are available to a limited extent. There is a lack of clarification and contextualization for how data is regulated in the energy sector, \textit{i.e.}, general privacy regulations like GDPR are available and fully applicable to the energy sector, however, it is nontrivial to map the regulatory requirements in the general regulations into energy systems for specific scenarios and applications~\cite{LEE2021101188}. \textit{E.g.}, what data is personal data in the energy system, and what obligations are imposed to whom when processing personal data for energy trading and other applications. Such gaps hinder the development of regulation-complied algorithms and techniques for data sharing, and introduce uncertainty to stakeholders in concretely defining their roles and responsibilities and their decision-making~\cite{HEUNINCKX2023103040}. Furthermore, the lack of concrete provisioning of the regulatory compliance criteria and compliance guidelines at large might obscure the stakeholders particularly towards their acceptance and adoption of new energy paradigms like transactive energy systems and peer-to-peer energy trading, thus slowing down the twin transition~\cite{CARLANDER2023100097}.

In this work, we systematically categorize the available data, data exchange, and techniques of data-processing for the energy system, which is the technical basis for contextualizing data security and privacy from the regulatory perspective in the energy sector. Based on this analysis, we categorize the data-related issues into (i) data-sharing (ii) privacy, and (iii) cyber security. We then review each category from a regulatory perspective, and provide insights into the structure, provision, development of the regulatory framework relating to the three categorized issues. Particularly, we (i) clarify the concepts, definitions, roles, rights and obligations, principles, and organizational structure related to legislative compliance on data-sharing, privacy, and cyber security in the energy sector, (ii) look into the current landscape in terms of regulatory development on data-sharing, privacy, and cyber security, and (iii) extract the learned lessons from the regulatory analysis and provide guidelines towards a conducive regulatory framework for the twin energy transition. 

The rest of the paper is organized as follows. In Section~\ref{Smart grid information issues}, we review open-access data, data communication and data processing techniques for the energy system, and we check the issues with data sharing, privacy, and cyber security and the existing regulatory analysis on those issues. In Section~\ref{Results of the survey}, we review the regulatory provisions for data-sharing, privacy, and cyber security, and articulate those provisions in the context of energy system. In Section~\ref{sec:reflection}, we reflect on the current regulations and provide insights into the challenges associated with the regulatory framework, and we analyze the open research questions yet to be addressed towards the distributed and digitalized energy sector. We conclude this work in Section~\ref{Conclusions}.

\section{Technical basis for regulatory compliance in data processing in the energy system}\label{Smart grid information issues}

\begin{figure*}[tbh]
	\centering
	\includegraphics[width=0.95\textwidth]{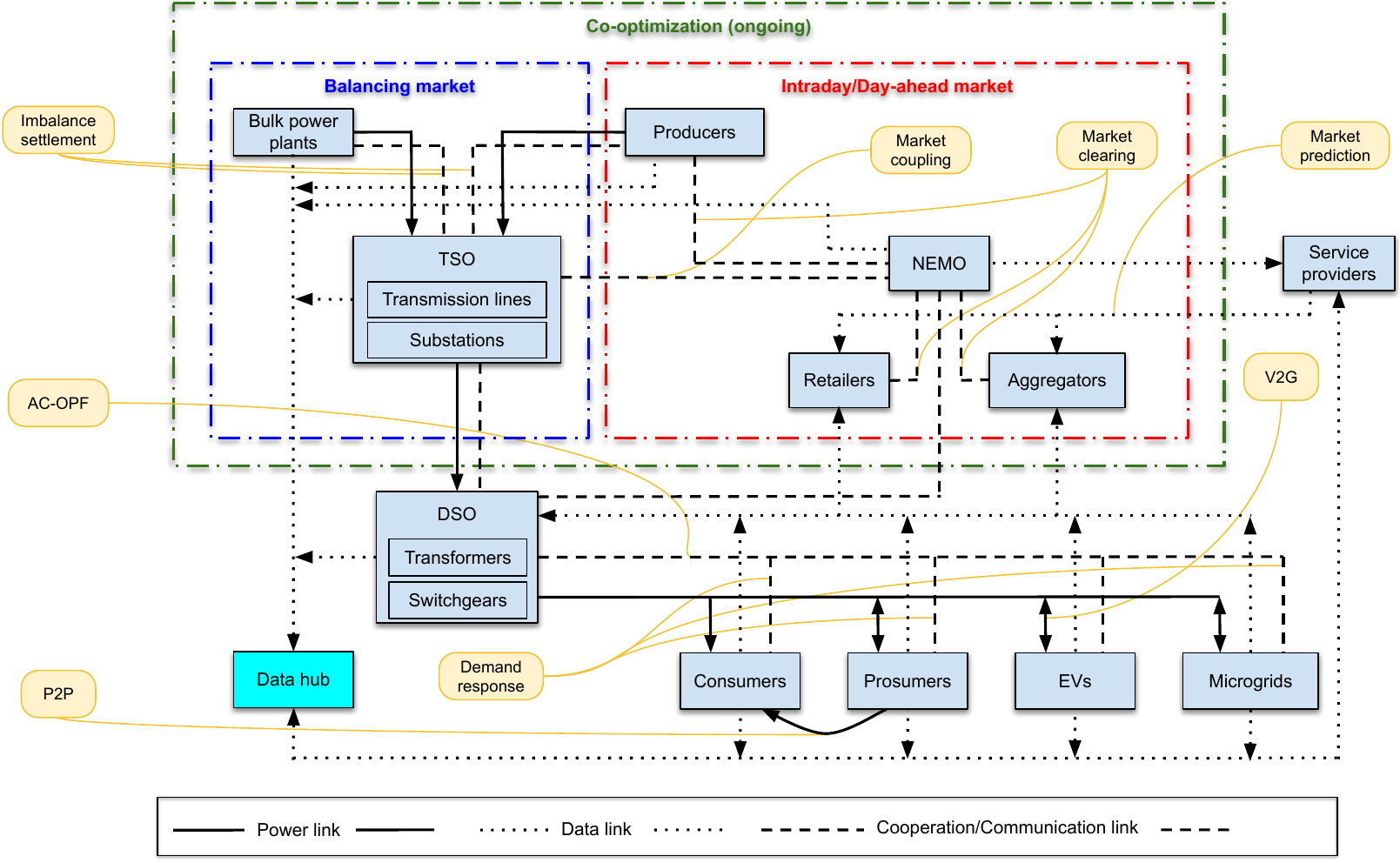}\\
	\caption{\label{fig:energy_ecosystem}A decentralized and distributed energy system (\textit{AC-OPF}: Alternate current-optimal power flow. \textit{P2P}: Peer to peer. \textit{V2G}: Vehicle to grid. \textit{TSO}: Transmission system operator. \textit{DSO}: Distribution system operator. \textit{NEMO}: Nominated energy market operator).}
\end{figure*}

The ongoing transition of the current energy system into a more distributed and participatory energy system heavily depends on data and data communication. Figure~\ref{fig:energy_ecosystem} shows the information exchange and energy flow between involved stakeholders in the energy system. Data sharing and data communication are critical for facilitating functionalities like market operations and demand response, and also underpin the ongoing evolution of the energy system, \textit{e.g.}, the co-optimization~\cite{papavasiliou2023welfare} that integrates the balancing market and intraday/day-ahead market and the optimization of energy flow in the distribution grid~\cite{6624135}. Nevertheless, the increasing data and data communication interfaces lead to vulnerabilities that can affect stakeholders at different levels of the energy system. \textit{E.g.}, the leakage or unauthorized usage of consumer energy consumption data can disclose user privacy~\cite{AHMED2024122403}, hacked communication of energy information~\cite{Ghiasi2023} can lead to severe incidents in energy transmission, and failures in sharing energy statistics by authorities can raise transparency issues and lead to sub-optimal decision-making of energy stakeholders in their business~\cite{ZHANG2024107749}. To address such issues, regulatory provisions and compliance are of critical importance.

Data-related issues in the energy system, \textit{e.g.}, privacy and security, are coupled. Understanding those issues from a technical perspective is the first step towards articulating regulatory initiatives and frameworks. In this section, we review open-access data, data communication, and data-processing techniques adopted in the energy sector, and categorize the issues in data exchange and data processing. We then review studies on data-related issues from a regulatory perspective and analyze the limitations of the existing regulatory studies.

\subsection{Technical landscape of data communication and processing in energy system}

The energy sector involves complex system interactions and generates massive amount of data from various sources, \textit{e.g.}, power plants, renewable energy sources, and consumers/prosumers~\cite{CALEARO2021111518}. Such data are processed and communicated to understand, analyze, and optimize energy usage, services, management, maintenance, and marketing, \textit{etc.}~\cite{CHEN2020109466,KAZMI2021111290,LUO2021111224,SINGH2021100410}. The availability, communication, and processing of data are the basis for contextualizing regulatory data usage in the energy sector. Understanding those technical aspects is crucial to connect data-related activities with corresponding regulations. Reviewing what data exists, stakeholders involved in the data communication and processing and how the processing happens is the necessary context for the data-related regulations and further analysis on the regulations. To this end, we review the open-access data, data communication, and data processing in the energy sector in the following.

\subsubsection{Available data in the energy sector}

Data from the energy sector has been analyzed in different application domains~\cite{ATHANASIADIS2024114151}. Infrastructures and devices like smart meters can generate large amount of data at a high resolution and down to an individual level~\cite{USHAKOVA2020101428}. To gain an understanding of data usage in the energy sector, we summarize the available data resources in Table~\ref{tbl:data_source}. In this table, we categorize the reviewed data into five categories: Residential data, EV data, commercial/industrial data, system and operational data, and market data, as described below.

\begin{enumerate}
    \item Residential data refers to the information of energy consumption/production and energy usage patterns, or energy meta data in residential household/buildings/communities. 
    \item EV data denotes the information related to EV's energy activities like consumption, charging/discharging management, and its interactions with other EVs and energy infrastructures like charging station or the power grid, as well as its mobility data like locations and trails associated with different applications - mainly fuel saving and charging optimization.~\cite{HECHT2020100079}.
    \item Commercial/industrial energy data relates to the information of energy usage and forecast in commercial or industrial buildings and the breakdown of the usage, \textit{e.g.}, HVAC and lighting~\cite{MILLER2017439}. Such buildings generally require high level of secure and stable energy supply and consume more energy than, \textit{e.g.}, residential buildings.
    \item System and operational data relates to the information about monitoring, management, and operation of the energy system, along with the information from the energy demand side and the supply side that is crucial for O\&M of the infrastructure~\cite{TSANAS2012560}.
    \item Market data refers to the purchasing and selling of energy in the market and the information that facilitates the energy transactions, \textit{e.g.}, energy price, energy load and supply forecast~\cite{HIRTH20181054}.
\end{enumerate}
For each category, we provide sub-categories when necessary and give examples as shown in Table~\ref{tbl:data_source}. We also indicate whether the reviewed data resource is related to individuals, devices/environment, or merely statistics, since sharing it can lead to different consequences regarding privacy and cyber security.

\begin{table*}[htb] 
	\centering \caption{\label{tbl:data_source}Data source for information issues in the energy system. Note that we only consider data that can be measured and recorded from devices/infrastructures in the energy system, rather that data from interviews/questionnaires/tests.}
	\resizebox{0.9\textwidth}{!}{%
		\begin{tabular}{|l|l|l|l|l|l|}
			\cline{1-6}
			\tabincell{l}{Data\\domain} & \tabincell{l}{Description} & \tabincell{l}{Example} & \tabincell{l}{Individual\\ related (I)?\\Device/Env.\\ related (D)?\\Statistics (S)?} & \tabincell{l}{Reference with\\data repository} & \tabincell{l}{Applications}  \\

			\cline{1-6}
			\multirow{3}{*}{\tabincell{l}{Residential\\energy\\consumption\\ data}} & \tabincell{l}{Consumption data} &  \tabincell{l}{Household  consumption\\Appliance  consumption\\Current and voltage} &\tabincell{l}{I} & \tabincell{l}{~\cite{DBLP:conf/eenergy/JazizadehABS18,einfalt2009adres,chavat2022ecd}\\~\cite{kolter2011redd,kelly2015uk,DBLP:conf/sensys/BeckelKCSS14,9894187,pereira2022residential}\\~\cite{filip2011blued,DBLP:conf/sensys/GaoGKB14,shin2019enertalk}} & \tabincell{l}{Demand forecast\\Peak demand forecast\\Non intrusive load monitoring}  \\
			\cline{2-6}
			& \tabincell{l}{End-user metadata} &  \tabincell{l}{Demographics information} &\tabincell{l}{S} & \tabincell{l}{~\cite{DBLP:conf/eenergy/NadeemA19}} & \tabincell{l}{Demand forecast\\Peak demand forecast}  \\
			\cline{2-6}
			& \tabincell{l}{Production data} &  \tabincell{l}{Solar energy generation} & \tabincell{l}{I} & \tabincell{l}{~\cite{LOPEZPROL2023106974}} & \tabincell{l}{Curtailment Reduction~\cite{9392581}}  \\
            \cline{2-6}
			& \tabincell{l}{Ambient data} & \tabincell{l}{Indoor/Outdoor temperature\\Indoor humidity\\Weather data} & \tabincell{l}{D} &  \tabincell{l}{~\cite{DBLP:conf/sensys/NambiLP15,DBLP:journals/corr/abs-2110-02166}\\~\cite{CANDANEDO201781}\\~\cite{barker2012smart}\\} & \tabincell{l}{HVAC, demand response, energy\\ generation forecast, load forecast}  \\

            \cline{1-6}
			EV & \tabincell{l}{Charging data} & \tabincell{l}{Battery capacity\\Charging profile, e.g., current\\Charger location\\Plug-in/out time\\Charged energy\\Charging station usage\\Charging transaction} & \tabincell{l}{D\\D\\I\\I\\I\\D\\I} & \tabincell{l}{~\cite{distribution2021western-capacity}\\~\cite{distribution2021western-current}\\~\cite{su132112025}\\~\cite{charger2017}\\~\cite{chargedenergy2018,chargedenergy20181}\\~\cite{HECHT2020100079}\tablefootnote{Data accessible at\url{https://ars.els-cdn.com/content/image/1-s2.0-S2590116820300369-mmc1.xlsx}}~\\~\cite{distribution2021western}\\} & \tabincell{l}{Charging behavior analysis\cite{HECHT2020100079,DBLP:conf/eenergy/LeeLL19}\\Charging impact on distributed \\network\tablefootnote{\url{https://myelectricavenue.info} and \url{https://www.esbnetworks.ie/docs/default-source/publications/ev-pilot-project-report.pdf?Status=Master&sfvrsn=427613c6_6/}}\cite{6038948}\\Charging flexibility analysis\tablefootnote{\url{https://www.sintef.no/en/projects/2016/modelling-flexible-resources-in-smart-distribution/}}\\Charging cost minimization\cite{GONZALEZGARRIDO2019381,DBLP:journals/tii/PertlCTMKK19}}  \\
            \cline{2-6}
			 & \tabincell{l}{Trail data} & \tabincell{l}{Arrival/Departure time\\Consecutive GPS locations\\Travel distance\\Vehicle speed\\Energy consumption\\Charging event\\Parking event } & \tabincell{l}{I} & \tabincell{l}{\\~\cite{zhang2022extended,location2017,c7j010-22}\\~\\~\cite{6bxe-bp52-19}\\~\cite{DBLP:journals/tits/OhLP22}\\~\\~\cite{parking2014}} & \tabincell{l}{Route schedule\\Energy consumption prediction\\Transportation congestion analysis}  \\

            \cline{1-6}
			\tabincell{l}{Commercial\\ or industry\\ energy\\ data} & \tabincell{l}{Energy consumption} & \tabincell{l}{Aggregated consumption\\Appliance consumption} & \tabincell{l}{D} & \tabincell{l}{~\cite{DBLP:journals/corr/BatraPBSR14,Rashid2019,MILLER2017439,doi:10.1080/09613218.2020.1809983}\\~\cite{kriechbaumer2018blond}\\} & \tabincell{l}{Energy Management, cost\\ allocation, demand\\ response, load planning}  \\
            \cline{2-6}
			 & \tabincell{l}{Environmental data} & \tabincell{l}{temperature\\relative humidity\\ambient light } & \tabincell{l}{D} & \tabincell{l}{~\cite{kriechbaumer2018blond}} & \tabincell{l}{Energy efficiency, equipment\\ performance monitoring, energy\\ demand management,\\ fault detection.}  \\

            \cline{1-6}
			 \tabincell{l}{System \&\\ operational\\ data} & \tabincell{l}{Energy efficiency data} & \tabincell{l}{Heating load\\Cooling load\\Building conditions} & \tabincell{l}{S\\S\\D} & \tabincell{l}{~\cite{TSANAS2012560}} & \tabincell{l}{Assessing building energy load \\requirements}  \\
            \cline{2-6}
			 & \tabincell{l}{Regional meta data} & \tabincell{l}{Population\\Population density\\Energy usage sector\\Infrastructural access\\Number of houses} & \tabincell{l}{S} & \tabincell{l}{~\cite{tong2021all}} & \tabincell{l}{Energy planning, load forecasting,\\ infrastructure development\\ planning, energy efficiency,\\ demand response.}  \\

            \cline{1-6}
			\tabincell{l}{Market\\ data} & \tabincell{l}{Pricing data} & \tabincell{l}{Wholesale electricity price,\\ retail price} & \tabincell{l}{S} & \tabincell{l}{\cite{dong2011short,yang2017electricity,peng2018effective}} & \tabincell{l}{Tariff design\tablefootnote{\url{https://energy-stats.uk/octopus-agile-east-midlands/}}\\Demand response}  \\
            \cline{2-6}
			 & \tabincell{l}{Renewable generation \\forecast} & \tabincell{l}{Current forecast\\day ahead forecast\\intraday forecast} & \tabincell{l}{S} & \tabincell{l}{~\cite{HIRTH20181054}} & \tabincell{l}{Energy scheduling}  \\
            \cline{2-6}
			 & \tabincell{l}{Renewable generation \\availability} & \tabincell{l}{Available solar/wind power} & \tabincell{l}{S} & \tabincell{l}{~\cite{WU2023127286}} & \tabincell{l}{Energy planning, grid integration,\\ resource allocation, energy\\ forecasting, demand response}  \\

			\cline{1-6}

		\end{tabular}%
	}
\end{table*}

From Table~\ref{tbl:data_source}, we can observe that energy consumption data (\textit{e.g.}, from EV, residential, or industrial sector) is the most readily available data for studies and applications. Consumption data provides insights into energy usage patterns and trends that are important for energy planning and management. As such, energy consumption data has been used for load forecasting~\cite{https://doi.org/10.1002/er.6745}, demand response~\cite{DBLP:journals/access/LiuSQLX19}, price prediction~\cite{peng2018effective}, market analytics~\cite{WANG2023121550}, \textit{etc}. The consumption data reflects user's activities, and fine-grained consumption data can reveal, \eg energy usage habits and individual patterns. From Table~\ref{tbl:data_source}, We observe less data in the category of system and operation, which is critical to the maintenance of the energy system. Potential reasons behind might be that the system and operation data can be sensitive, the disclosure of which might lead to vulnerabilities not only from a cyber security but also a competition perspective in energy business.

\subsubsection{Data communication in the energy system}

\begin{figure}[tbh]
	\centering
	\includegraphics[width=0.5\textwidth]{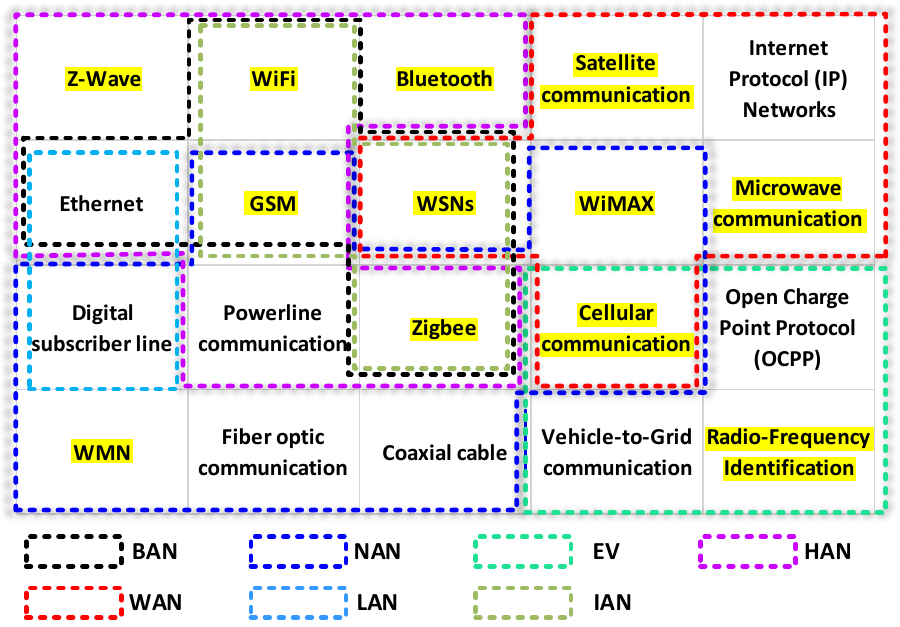}\\
	\caption{\label{fig:communications}The data communication technologies in different part of networks in the energy system. Ellipsoids with colors denote different networks in the energy system. Data communication techniques highlighted by yellow indicate wireless communication.}
\end{figure}

Data exchange in the energy system can expose vulnerable interfaces towards unauthorized data access and cyber security issues like false data injection and data tempering. In the participatory energy system, the prosumers that lack professional level security become the new access points of vulnerability. To understand the issues with data exchange and communication, we review the existing communication technologies for different applications in the energy system, as summarized in Table~\ref{tbl:communication_source} and visualized in Figure~\ref{fig:communications}. 

\begin{table*}[htb]
	\centering \caption{\label{tbl:communication_source}Existing communications in the energy system}
	\resizebox{0.99\textwidth}{!}{%
		\begin{tabular}{|l|l|l|l|l|l|}
			\cline{1-6}
			\tabincell{l}{Communication\\technique} & \tabincell{l}{Network\\domain} & \tabincell{l}{Wireless\\or not} &  \tabincell{l}{Data} & Entities & Functionality  \\

			\cline{1-6}
			\multirow{3}{*}{\tabincell{l}{Wi-Fi\cite{KABALCI2016302}\cite{9023471}\cite{8676311}\\\cite{SHAUKAT20181453}\cite{ABDULSALAM2023100121}\cite{WANG20113604}\cite{Wen_Leung_Li_He_Kuo_2015}\\\cite{MARZAL20183610}\cite{MAHMOOD2015248}\cite{6861946}\cite{8839117}\\\cite{6129368}\cite{9197633}\cite{9759422}\cite{ANCILLOTTI20131665}\\\cite{9806180}\cite{https://doi.org/10.1002/wcm.2258}\cite{REDDY2014180}\cite{8325326}\cite{s22155881}}} & \tabincell{l}{HAN} & \tabincell{l}{Yes} & \tabincell{l}{Energy consumption\\ information. Control\\ commands for\\\ smart devices.} & \tabincell{l}{Smart homes. Smart\\ meters. Smart\\ thermostats} & \tabincell{l}{Remote energy usage monitoring and control.\\Integration of smart home devices with\\ energy management systems.\\Real-time data exchange for energy consumption\\ tracking and optimization.}\\
			\cline{2-6}
			& \tabincell{l}{BAN\\IAN} & \tabincell{l}{Yes} & \tabincell{l}{Energy consumption\\ data. Control\\ signals for\\ heating/cooling.} & \tabincell{l}{Building owners.\\Facility managers.\\Smart meters\\Building automation\\ systems.} & \tabincell{l}{Remote energy monitoring and control.\\Energy usage optimization\\Smart building automation}\\

   			\cline{1-6}
			\multirow{3}{*}{\tabincell{l}{Zigbee(WPAN)\\\cite{KABALCI2016302}\cite{9023471}\cite{6279592}\cite{SHAUKAT20181453}\\\cite{s21238087}\cite{ABDULSALAM2023100121}\cite{WANG20113604}\cite{Wen_Leung_Li_He_Kuo_2015}\\\cite{MARZAL20183610}\cite{MAHMOOD2015248}\cite{6157575}\cite{6861946}\\\cite{8839117}\cite{10.1007/978-3-030-44758-8_48}\cite{6129368}\cite{9197633}\\\cite{9395820}\cite{9759422}\cite{ANCILLOTTI20131665}\cite{9806180}\\\cite{7495234}\cite{https://doi.org/10.1002/wcm.2258}\cite{REDDY2014180}\\\cite{8325326}\cite{s22155881}}} & \tabincell{l}{HAN} & \tabincell{l}{Yes} & \tabincell{l}{Sensor data\\control commands\\status information} & \tabincell{l}{Smart plugs.\\Smart lighting.\\Smart appliances.\\Other energy-efficient\\ devices.} & \tabincell{l}{Seamless connectivity and communication \\between smart energy-efficient devices, such as\\ smart plugs, smart lights, and smart thermostats.\\Support for energy management applications\\ and systems to collect data and control devices\\ within the home area network}\\
			\cline{2-6}
			& \tabincell{l}{BAN\\IAN\\NAN} & \tabincell{l}{Yes} & \tabincell{l}{Appliances energy\\ usage data.\\Lighting/HVAC\\ control signals.} & \tabincell{l}{Smart appliances.\\Building energy\\ management systems.\\Zigbee network\\ controllerss} & \tabincell{l}{Smart lighting control.\\Energy-efficient HVAC management.\\Occupancy sensing.\\Load control}\\

   			\cline{1-6}
			\multirow{3}{*}{\tabincell{l}{Z-Wave\cite{SHAUKAT20181453}\\\cite{s21238087}\cite{MAHMOOD2015248}\cite{8839117}\\\cite{9759422}\cite{https://doi.org/10.1002/wcm.2258}}} & \tabincell{l}{HAN} & \tabincell{l}{Yes} & \tabincell{l}{Energy consumption.\\Device control\\ commands. Status\\ updates for smart\\ home devices} & \tabincell{l}{Smart locks.\\Smart sensors.\\Smart irrigation\\ systems.} & \tabincell{l}{Interoperability and communication between\\ a wide range of smart devices focused on energy\\ efficiency, including smart locks, sensors, and\\ smart appliances.\\Reliable and low-power communication for\\ energy monitoring and control applications}\\

      		\cline{1-6}
			\multirow{3}{*}{\tabincell{l}{Bluetooth\cite{9023471}\\\cite{SHAUKAT20181453}\cite{ABDULSALAM2023100121}\cite{MAHMOOD2015248}\\\cite{8839117}\cite{9806180}\cite{s22155881}}} & \tabincell{l}{HAN} & \tabincell{l}{Yes} & \tabincell{l}{Energy usage.\\Device pairing\\ and control.\\Status information\\ for Bluetooth-enabled\\ devices} & \tabincell{l}{smartphones\\tablets\\smartwatches} & \tabincell{l}{Local and remote control of energy-efficient\\ devices using mobile devices or other\\ Bluetooth-enabled controllers.\\Data exchange for energy management\\ and control functionalities, including\\ scheduling, automation, and energy optimization.}\\

      		\cline{1-6}
			\multirow{3}{*}{\tabincell{l}{Ethernet\cite{EMMANUEL2016133}\\\cite{s21238087}\cite{WANG20113604}\cite{MARZAL20183610}\\\cite{Singh2022}\cite{6861946}\cite{6129368}\cite{9806180}}} & \tabincell{l}{HAN} & \tabincell{l}{No} & \tabincell{l}{Real-time energy\\ consumption metrics.\\System configurations.\\Control commands.} & \tabincell{l}{Smart homes. Smart\\ hubs. Connected\\ appliances and\\ equipment.} & \tabincell{l}{High-speed and reliable connection for energy\\ management systems and smart home devices.\\Efficient data exchange with\\ low latency, enabling real-time monitoring, \\analytics, and control of energy usage}\\
			\cline{2-6}
			& \tabincell{l}{BAN\\IAN} & \tabincell{l}{No} & \tabincell{l}{Real-time monitoring\\ and control data.\\energy usage\\ information.} & \tabincell{l}{industrial control\\ systems supervisory\\ control and data\\ acquisition (SCADA)\\ systems} & \tabincell{l}{Real-time monitoring and control of\\ industrial processes.\\Energy management.\\Predictive maintenance.\\Process optimization}\\
   
      		\cline{1-6}
			\multirow{3}{*}{\tabincell{l}{Powerline \\communication\\\cite{KABALCI2016302}\cite{EMMANUEL2016133}\cite{9023471}\cite{6279592}\cite{SHAUKAT20181453}\\\cite{s21238087}\cite{ABDULSALAM2023100121}\cite{WANG20113604}\cite{Wen_Leung_Li_He_Kuo_2015}\cite{MARZAL20183610}\\\cite{Singh2022}\cite{6157575}\cite{6861946}\cite{8839117}\cite{10.1007/978-3-030-44758-8_48}\\\cite{6129368}\cite{DBLP:journals/winet/JhaAGPGKM21}\cite{9197633}\cite{9395820}\\\cite{9759422}\cite{ANCILLOTTI20131665}\cite{9465005}\cite{9806180}\\\cite{7495234}\cite{https://doi.org/10.1002/wcm.2258}\cite{REDDY2014180}\cite{s22155881}}} & \tabincell{l}{HAN} & \tabincell{l}{No} & \tabincell{l}{energy data\\control signals\\status information} & \tabincell{l}{smart meters\\energy monitors\\powerline adapters} & \tabincell{l}{Utilization of existing electrical wiring for\\ communication, enabling data transmission and\\ control over the powerline network within the home.\\Support for energy monitoring, communication\\ with smart meters, and coordination of\\ energy-efficient devices}\\
			\cline{2-6}
			& \tabincell{l}{NAN} & \tabincell{l}{No} & \tabincell{l}{Energy usage data.\\Control signals.} & \tabincell{l}{Smart meters.\\Utility companies.\\Neighborhood-level \\energy management\\ systems.} & \tabincell{l}{Accurate energy measurement.\\Remote meter reading.\\Demand management.\\Grid balancing}\\

      		\cline{1-6}
			\tabincell{l}{Fiber Optic\\ Communication\\\cite{KABALCI2016302}\cite{9023471}\cite{8676311}\cite{SHAUKAT20181453}\\\cite{s21238087}\cite{ABDULSALAM2023100121}\cite{6861946}\cite{8839117}\\\cite{10.1007/978-3-030-44758-8_48}\cite{6129368}\cite{DBLP:journals/winet/JhaAGPGKM21}\cite{9197633}\\\cite{9395820}\cite{9759422}\cite{ANCILLOTTI20131665}\cite{9806180}\\\cite{7495234}\cite{https://doi.org/10.1002/wcm.2258}\cite{s22155881}} & \tabincell{l}{NAN} & \tabincell{l}{No} & \tabincell{l}{Energy consumption.\\Smart grid\\ control information.} & \tabincell{l}{Utility companies.\\Substations. Distribution\\ system operators.\\ Smart grid\\ management systems} & \tabincell{l}{Real-time monitoring and control.\\Fault detection.\\Load balancing.}\\

			\cline{1-6}

		\end{tabular}
	}
\end{table*}

\begin{table*}[htb]
    \ContinuedFloat
	\centering \caption{\label{tbl:communication_source_1}Existing communications in the energy system (continued)}
	\resizebox{0.99\textwidth}{!}{%
		\begin{tabular}{|l|l|l|l|l|l|}
			\cline{1-6}
			\tabincell{l}{Communication\\technique} & \tabincell{l}{Network\\domain} & \tabincell{l}{Wireless?} &  \tabincell{l}{Data} & Entities & Functionality  \\

      		\cline{1-6}
			\multirow{3}{*}{\tabincell{l}{Cellular\\Communication\\\cite{EMMANUEL2016133}\cite{9023471}\cite{8676311}\cite{SHAUKAT20181453}\\\cite{s21238087}\cite{ABDULSALAM2023100121}\cite{WANG20113604}\cite{MARZAL20183610}\\\cite{MAHMOOD2015248}\cite{6861946}\cite{8839117}\cite{10.1007/978-3-030-44758-8_48}\\\cite{6129368}\cite{DBLP:journals/winet/JhaAGPGKM21}\cite{9197633}\cite{9395820}\\\cite{9759422}\cite{ANCILLOTTI20131665}\cite{9806180}\cite{7495234}\\\cite{https://doi.org/10.1002/wcm.2258}\cite{8325326}\cite{s22155881}}} & \tabincell{l}{NAN} & \tabincell{l}{Yes} & \tabincell{l}{Energy consumption data,\\control signals} & \tabincell{l}{Utility companies, mobile network\\ operators, smart meters,\\smart grid management systems} & \tabincell{l}{Remote meter reading, demand\\ response, grid management, renewable\\ resource monitoring and integration}\\
			\cline{2-6}
			& \tabincell{l}{WAN} & \tabincell{l}{Yes} & \tabincell{l}{Energy consumption data,\\real-time grid monitoring data,\\demand response signals} & \tabincell{l}{Utility companies, mobile network\\ operators, grid operators\\energy management systems} & \tabincell{l}{Remote monitoring, demand\\ response, energy forecasting, outage\\ management, load balancing}\\
            \cline{2-6}
            & \tabincell{l}{EV} & \tabincell{l}{Yes} & \tabincell{l}{Real-time vehicle diagnostics,\\location tracking, battery status,\\software updates, energy\\ consumption information} & \tabincell{l}{EV manufacturers\\telematics service providers\\fleet operators\\energy management platforms} & \tabincell{l}{Remote vehicle monitoring\\Predictive maintenance\\Route optimization\\Energy-efficient driving guidance}\\

      		\cline{1-6}
			\multirow{3}{*}{\tabincell{l}{Wireless Sensor\\ Networks\\(WSNs)\cite{8676311}\cite{6861946}\cite{https://doi.org/10.1002/wcm.2258}}} & \tabincell{l}{NAN} & \tabincell{l}{Yes} & \tabincell{l}{Environmental parameters, energy\\ usage data from sensors, control\\ signals for localized energy\\ management} & \tabincell{l}{sensor nodes\\localized energy management\\ systems\\data aggregators} & \tabincell{l}{Environmental monitoring\\Energy optimization\\Fault detection\\Load control}\\
			\cline{2-6}
			& \tabincell{l}{BAN\\IAN} & \tabincell{l}{Yes} & \tabincell{l}{Environmental parameters\\equipment performance data\\energy usage information} & \tabincell{l}{Industrial sensors, control systems,\\data aggregator/analytic\\ platforms} & \tabincell{l}{Environmental monitoring, equipment\\ condition monitoring, energy efficiency\\ optimization, predictive maintenance}\\

      		\cline{1-6}
			\tabincell{l}{Satellite\\Communication\\\cite{KABALCI2016302}\cite{SHAUKAT20181453}\cite{SHAUKAT20181453}\cite{EMMANUEL2016133}\\\cite{s21238087}\cite{ABDULSALAM2023100121}\cite{8839117}\cite{10.1007/978-3-030-44758-8_48}\\\cite{DBLP:journals/winet/JhaAGPGKM21}\cite{9759422}\cite{ANCILLOTTI20131665}\cite{7495234}\\\cite{https://doi.org/10.1002/wcm.2258}\cite{s22155881}} & \tabincell{l}{WAN} & \tabincell{l}{Yes} & \tabincell{l}{Energy consumption data\\Control signals} & \tabincell{l}{Utility companies, energy grid\\ operators, satellite providers\\regional/grid-level energy\\ management systems} & \tabincell{l}{Remote monitoring of infrastructure\\Wide-scale energy management\\Outage management\\Grid optimization}\\

      		\cline{1-6}
			\tabincell{l}{Internet Protocol \\(IP) Networks\\\cite{9759422}\cite{9465005}\cite{9806180}\cite{https://doi.org/10.1002/wcm.2258}} & \tabincell{l}{WAN} & \tabincell{l}{No} & \tabincell{l}{Energy consumption data,\\real-time monitoring/control\\ information of substations,\\grid operation data} & \tabincell{l}{Utility companies, grid operators,\\distribution system operators,\\regional energy management\\ systems} & \tabincell{l}{Real-time grid monitoring, data\\ analytics, grid management, fault\\ detection, load forecasting}\\ 
   
      		\cline{1-6}
			\tabincell{l}{Microwave \\Communication\\\cite{EMMANUEL2016133}\cite{SHAUKAT20181453}\cite{8839117}\cite{6129368}\\\cite{DBLP:journals/winet/JhaAGPGKM21}\cite{7495234}} & \tabincell{l}{WAN} & \tabincell{l}{Yes} & \tabincell{l}{Energy consumption data,\\wide-scale energy management\\ information} & \tabincell{l}{Utility companies, grid operators,\\transmission system operators\\regional energy management\\ systems} & \tabincell{l}{Wide-scale grid monitoring, fault\\ detection, energy dispatch, grid\\ synchronization, contingency\\ management}\\

            \cline{1-6}
            \tabincell{l}{Digital subscriber \\line\\\cite{KABALCI2016302}\cite{SHAUKAT20181453}\cite{ABDULSALAM2023100121}\cite{WANG20113604}\\\cite{Singh2022}\cite{6861946}\cite{8839117}\cite{9759422}\\\cite{ANCILLOTTI20131665}\cite{9806180}} & \tabincell{l}{LAN\\NAN} & \tabincell{l}{No} & \tabincell{l}{Meter readings, energy\\ consumption data, control\\ commands, status information} & \tabincell{l}{Energy meters, smart home\\ devices, substation equipment,\\ energy management systems,\\ control centers} & \tabincell{l}{Remote meter reading, real-time\\ monitoring of energy consumption,\\control of appliances or devices, fault\\ detection and diagnosis, load balancing,\\energy management, energy system\\ optimization}\\

            \cline{1-6}
            \tabincell{l}{WiMAX\cite{KABALCI2016302}\\\cite{EMMANUEL2016133}\cite{SHAUKAT20181453}\cite{s21238087}\cite{ABDULSALAM2023100121}\\\cite{MAHMOOD2015248}\cite{6157575}\cite{6861946}\cite{8839117}\\\cite{6129368}\cite{9395820}\cite{ANCILLOTTI20131665}\cite{https://doi.org/10.1002/wcm.2258}\\\cite{REDDY2014180}\cite{8325326}\cite{s22155881}} & \tabincell{l}{WAN\\NAN} & \tabincell{l}{Yes} & \tabincell{l}{Meter readings, voltage and\\ current measurements, \\power quality data, equipment\\ status information} & \tabincell{l}{Smart meters, sensors and\\ actuators, distribution automation\\ devices, communication gateways,\\ control centers} & \tabincell{l}{Real-time monitoring and control\\ of distribution equipment,\\outage detection and management,\\ demand response}\\

            \cline{1-6}
            \tabincell{l}{Wireless mesh\\ network(WMN)\\\cite{ABDULSALAM2023100121,WANG20113604,MARZAL20183610,7495234}} & \tabincell{l}{NAN} & \tabincell{l}{Yes} & \tabincell{l}{Meter readings, energy\\ consumption data, power\\ quality measurements, equipment\\ status information} & \tabincell{l}{Smart meters, sensors, actuators,\\ streetlights, distribution automation\\ devices, communication gateways,\\ and energy management systems} & \tabincell{l}{Remote meter reading, real-time\\ monitoring and control of distribution\\ equipment, demand response programs,\\ outage detection and management}\\

            \cline{1-6}
            \tabincell{l}{Coaxial cable\\\cite{WANG20113604,s22155881}} & \tabincell{l}{LAN} & \tabincell{l}{No} &  \tabincell{l}{Meter readings, energy\\ consumption data, control signals,\\ equipment status information} & \tabincell{l}{Energy meters, control devices,\\ substation equipment,\\ energy management systems,\\ data acquisition systems} & \tabincell{l}{Reliable data transfer, real-time\\ monitoring of energy consumption,\\ control of equipment and devices,\\fault detection and diagnosis, remote\\ configuration and programming}\\

            \cline{1-6}
            \multirow{1}{*}{\tabincell{l}{GSM\cite{KABALCI2016302}\cite{WANG20113604}\\\cite{9395820}\cite{9465005}\cite{7495234}\cite{https://doi.org/10.1002/wcm.2258}\cite{REDDY2014180}}} & \tabincell{l}{BAN\\IAN\\HAN\\NAN} & \tabincell{l}{Yes} & \tabincell{l}{Meter readings, energy\\ consumption data, equipment\\ status information, alarms,\\ distribution and management alerts} & \tabincell{l}{Smart meters, sensors, actuators,\\ energy management systems,\\ and control centers} & \tabincell{l}{Remote monitoring of energy\\ consumption, remote control of devices,\\ real-time alerts for system conditions,\\outage detection and reporting}\\

            \cline{1-6}
            \tabincell{l}{Open Charge \\Point Protocol \\(OCPP)\cite{DBLP:conf/iwcmc/VaidyaM18}\cite{hecht2024protocols}} & \multirow{4}{*}{\tabincell{l}{EV}} & \tabincell{l}{Both} & \tabincell{l}{Charging session information\\power and energy measurements\\pricing data, charging station\\ operational status} & \tabincell{l}{Electric Vehicle Supply Equipment\\ (EVSE), manufacturers, charging\\ station operators, management\\ system providers, EV drivers} & \tabincell{l}{Charging station remote monitoring\\ and control, billing and payment\\ processing, energy management,\\ load balancing}\\     
            \cline{1-1}\cline{3-6}
            \tabincell{l}{Radio-Frequency \\Identification \\\cite{EMMANUEL2016133,9759422,7495234}} &  & \tabincell{l}{Yes} & \tabincell{l}{User identification credentials,\\charging authorization information,\\access control data} & \tabincell{l}{EV drivers, charging station\\ operators, back-end systems for\\ authentication and billing} & \tabincell{l}{Secure user access to charging stations,\\ personalized charging profiles, user\\ authentication for billing}\\ 
            \cline{1-1}\cline{3-6}
            \tabincell{l}{Vehicle-to-Grid \\Communication\\\cite{DBLP:journals/tits/UmorenST21}\cite{kilic2023plug}\cite{DBLP:conf/acsac/ZhdanovaUHZHH22}} &  & \tabincell{l}{Both} & \tabincell{l}{Grid demand/response signals,\\charging/discharging commands,\\energy price, vehicle SOC data} & \tabincell{l}{EV manufacturers, grid operators,\\energy aggregators, V2G-enabled\\ EVs} & \tabincell{l}{Peak shaving and frequency regulation,\\optimized EV charging/discharging,\\grid stability enhancement}\\  
			\cline{1-6}

		\end{tabular}%
	}
\end{table*}

Note that data communications in the energy system have been extensively reviewed, \textit{e.g.}, reviews on the communication requirements~\cite{9023471}, standards~\cite{9465005}, architectures~\cite{WANG20113604,ANCILLOTTI20131665}, technologies~\cite{EMMANUEL2016133,SHAUKAT20181453,s21238087,ABDULSALAM2023100121,Singh2022,9806180,7495234}, infrastructures~\cite{6157575,6861946,8839117}, and the review of data communication for Microgrid~\cite{MARZAL20183610,9197633,9395820,s22155881}, renewable energy~\cite{REDDY2014180,8325326}, IoT based smart grid~\cite{9437171}. In comparison with previous reviews, our review aims to contextualize the data communications in the energy system, and show the connection of the communication with the data being exchanged, the entities involved, and the purpose of the communication. Such connections lay the basis for a regulatory analysis on data usage in the energy sector.

We review data communications in home area network (HAN), neighborhood area network (NAN), building area network (BAN), industrial area network (IAN), local area network (LAN) that connects devices in a limited area, wide area network (WAN) that connects components in the energy system across a wide geographic area and the communication network between EVs. In Table~\ref{tbl:communication_source}, we show the network where the communication takes place, and list possible data and entities involved in each communication and the corresponding applications. We summarize the reviewed works in Figure~\ref{fig:communications} in terms of data communication technologies in different part of the networks in the energy system.

From Table~\ref{tbl:communication_source} we observe that most of the communication technologies are for HAN and NAN in household levels for exchange energy usage and maintenance information. HAN and NAN involves tremendous amount of information exchange between households and energy operators/suppliers for purposes of meter reading~\cite{DBLP:journals/cn/KuzluPR14}, energy status monitoring~\cite{DBLP:journals/tii/DuanHSM22}, and control and operation of energy devices~\cite{ELAFIFI2024100634}, necessitating large-scale data communications. We also observe from Figure~\ref{fig:communications} that energy usage data and device control signals are the most communicated data in the energy system, the disclosure of which can lead to privacy issues for the prosumers and for the energy infrastructure. Smart meter and energy management systems are the most involved communication points in data communications, which provide interfaces to monitor and control energy devices for enhanced energy efficiency. Furthermore, Table~\ref{tbl:communication_source} indicates that data communications are mostly involved in the applications of energy monitoring, control, and optimization.

\subsubsection{Data-processing in the energy system}

AI and data processing techniques play a crucial role in gaining valuable insights from data analysis and promoting user experience. Those techniques remand and rely on access to the data and the amount of data, while potential misuse of data by malicious entities can threaten safety and security of the data owner. In this section, we review machine learning and other advanced data-processing approaches used in the energy sector and categorize them in Table~\ref{tbl:technique_source}. 

\begin{figure}[htb]
	\centering
	\includegraphics[width=0.5\textwidth, angle=0]{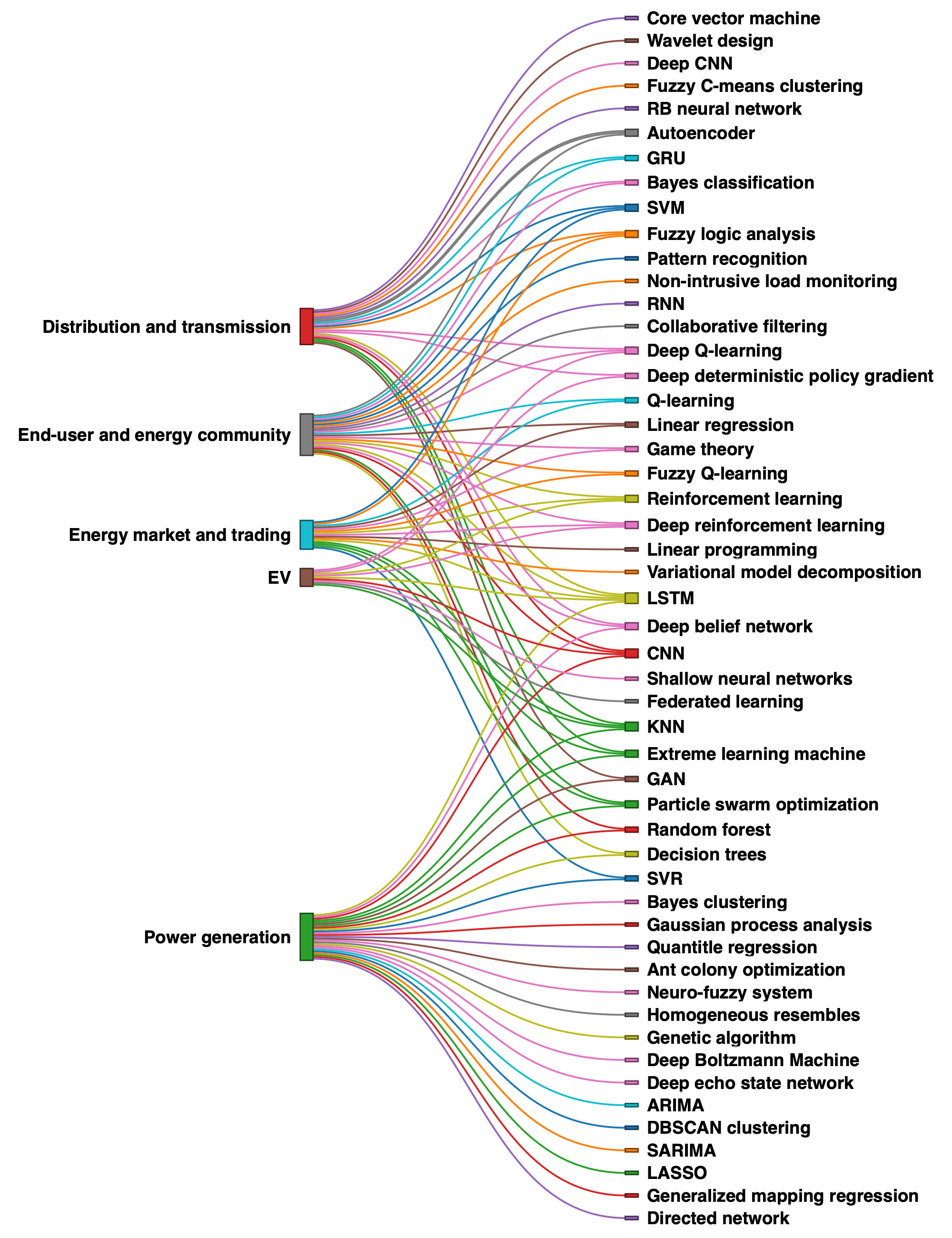}\\
	\caption{\label{fig:ai_techniques}AI techniques leveraged for various applications in the energy system.}
\end{figure}

We note that AI and machine learning approaches related to data processing in the energy system have been extensively reviewed, \textit{e.g.}, machine learning approaches and applications~\cite{DBLP:journals/access/HossainKUSS19,9264022,RANGELMARTINEZ2021414,IFAEI2023126432,HERNANDEZMATHEUS2022112651,MATIJASEVIC202212379,doi:10.1080/15567036.2020.1869867,GAVIRIA2022298,SIERLA2022104174,IBRAHIM2020115237,mocanu2017machine,https://doi.org/10.1002/er.4333,ENTEZARI2023101017,ntakolia2022machine,seyedzadeh2018machine,DBLP:journals/csur/DjenouriLDB19}, deep learning and reinforcement learning approaches and applications in the smart grid~\cite{DBLP:journals/access/MassaoudiARCO21,MISHRA2020104000,WANG2019111799,8468674,DBLP:journals/pieee/LiYSYZC23,9275593,PERERA2021110618,YANG2020145}. In this paper, we provide a structured analysis focusing not only on functionalities and applications, but also the type of data collected and processed for different purposes.

We summarize our review in Table~\ref{tbl:technique_source}. We categorize the reviewed approaches based on specific domains, including distribution and transmission, end user and energy community, energy market and trading, power generation, and EV. For each category, we list the applications involved, and the machine learning and other data-processing techniques used in those applications. We also show what kind of data is processed in each application. In Figure~\ref{fig:ai_techniques}, we cluster and categorize AI and other data-processing approaches for each application domain.

\begin{table*}[htb]
	\centering \caption{\label{tbl:technique_source}Existing AI techniques in the energy system related to data processing}
	\resizebox{0.9\textwidth}{!}{%
		\begin{tabular}{|l|l|l|l|}
			\cline{1-4}
			\tabincell{l}{Domain} & \tabincell{l}{Application} & \tabincell{l}{Technique} & \tabincell{l}{Data involved} \\

			\cline{1-4}
			\multirow{10}{*}{\tabincell{l}{Distribution\\ and \\transmission}} & \tabincell{l}{Grid control\\ and monitoring\cite{DILEEP20202589}} & \tabincell{l}{Extreme machine learning\cite{xu2011real}, Core\\ vector machine\cite{wang2016power}, Wavelet design\cite{DBLP:journals/tsg/AlshareefTM14},\\ Deep CNN\cite{wang2019novel}, Stacked autoencoder\cite{ma2017classification}\\LSTM\cite{rajiv2019long}, Deep Q-learning\cite{duan2019deep}, Gated\\ recurrent unit\cite{DBLP:journals/tii/DengWJTL19}, Deep belief network\cite{li2018classification},\\ Support vector machine\cite{zhang2017data}, Fuzzy C-means\\ clustering\cite{fu2018toward}, ARIMA\cite{DBLP:journals/tii/GangwarMCS20}, distributed\\ optimization\cite{DBLP:journals/iotj/RenLSZGM24}} & \tabincell{l}{Energy generation/consumption profiles, energy prices,\\ market conditions, grid constraints, energy supply and\\ demand, real-time energy demand data, historical\\ consumption patterns, grid status information,\\ operational data} \\
			\cline{2-4}
			& \tabincell{l}{Storage\cite{DILEEP20202589}} & \tabincell{l}{K-means clustering\cite{orgaz2022modeling}} & \tabincell{l}{Historical energy demand patterns, renewable energy\\ generation forecasts, market prices, energy storage\\ system parameters, \textit{e.g.}, capacity, charge or discharge\\ rates, efficiency grid constraints, \textit{e.g.}, grid availability} \\
            \cline{2-4}
			& \tabincell{l}{Energy management\\ and optimization\cite{DILEEP20202589}} & \tabincell{l}{Deep reinforcement learning\cite{hua2019optimal}} & \tabincell{l}{Historical energy consumption and generation data,\\ infrastructure data, market data, weather data,\\ economic and financial indicators} \\
            \cline{2-4}
			& \tabincell{l}{Fault/Anomaly\\ detection\cite{6298960}} & \tabincell{l}{Stacked autoencoders\cite{wang2016deep1}, CNN\cite{guo2017deep},\\ Deep belief network\cite{DBLP:journals/tsg/HeMW17}, SVM\cite{srivastava2021robust},\\ Radial basis function neural network\cite{he2020intelligent}} & \tabincell{l}{Sensor data from energy equipment, maintenance\\ logs, historical failure data} \\
            \cline{2-4}
			& \tabincell{l}{Cyber-security} & \tabincell{l}{GAN\cite{DBLP:conf/globalsip/AhmadianMH18}, Deep deterministic policy\\ gradient\cite{DBLP:journals/tsg/WeiWH20}, Q learning\cite{DBLP:journals/tifs/YanHZT17}, Bayes\\ classifier\cite{DBLP:journals/tsg/CuiWC20}, Fuzzy logic analysis and\\ SVM\cite{patel2017nifty}, K-means clustering\cite{DBLP:journals/tsg/CuiWY19}\\CNN\cite{basumallik2019packet}, Deep auto-encoder\cite{DBLP:journals/tsg/WangSLCDD19},\\ Extreme learning machine\cite{li2019denial}} & \tabincell{l}{Attack behavior data, electricity measurement data,\\ load data, active and reactive power data, phasor\\ measurement unit data} \\

   			\cline{1-4}
			\multirow{14}{*}{\tabincell{l}{End user\\ and energy\\ community}} & \tabincell{l}{Storage\\ optimization\cite{DILEEP20202589}} & \tabincell{l}{LSTM\cite{DBLP:conf/iasam/HafizAQH19}, Reinforcement learning\cite{DBLP:conf/smartgridcomm/MbuwirKD18},\\ Q learning\cite{DBLP:journals/tsg/QiuNC16}} & \tabincell{l}{Historical energy demand patterns, renewable energy\\ generation forecasts market prices, energy storage\\ system parameters, \textit{e.g.}, capacity, charge or discharge\\ rates, efficiency grid constraints, \textit{e.g.}, grid availability} \\
            \cline{2-4}
			& \tabincell{l}{Demand response\cite{TUBALLA2016710}\\ and load control\cite{6298960}} & \tabincell{l}{Fuzzy logic inference\cite{alamaniotis2019elm}, LSTM\cite{DBLP:journals/access/LiuSQLX19},\\ Naive Bayes classifier\cite{DBLP:journals/tsg/HuL13}\\Deep reinforcement learning\cite{DBLP:journals/tsg/LiWH20a}} & \tabincell{l}{data of the appliance states, real-time electricity\\ price, and outdoor temperature} \\
            \cline{2-4}
			& \tabincell{l}{Energy activity\\ analysis and\\ monitoring} & \tabincell{l}{Pattern recognition\cite{li2011predicting}, Non-intrusive\\ load monitoring\cite{DBLP:journals/tsg/MengistuGCAP19}, Deep neural\\ network\cite{DBLP:journals/tsg/CuiLLY19}} & \tabincell{l}{sensor data from energy distribution systems, \\maintenance logs, historical failure data\\energy consumption patterns, demographics,\\ user preferences, time of day/season} \\
            \cline{2-4}
			& \tabincell{l}{Fault detection} & \tabincell{l}{Particle swarm optimization\cite{jiang2017big},\\ Sparse autoencoder\cite{manohar2019enhancing}} & \tabincell{l}{historical/real-time consumption/production data, \\equipment parameters, sensor readings} \\
            \cline{2-4}
			& \tabincell{l}{load forecast} & \tabincell{l}{Neural network and random forest\cite{jurado2015hybrid}\\Recurrent neural network and gated\\ recurrent units\cite{wen2020load}, Deep learning\cite{guo2018deep},\\ Support vector regression\cite{tong2018efficient}\\Deep belief network\cite{qiu2017empirical}, LSTM\cite{DBLP:journals/tsg/KongDJHXZ19},\\ Linear regression\cite{li2020machine}} & \tabincell{l}{Historical energy consumption data, weather data,\\ demographic information, economic indicators} \\
            \cline{2-4}
			& \tabincell{l}{Microgrid energy\\ trading} & \tabincell{l}{Reinforcement learning\cite{boukas2018real}, Deep\\ reinforcement learning and CNN\cite{DBLP:journals/iotj/LuXXDPP19},\\ Game theory\cite{DBLP:conf/gamenets/XiaoDLZX17}} & \tabincell{l}{Individual energy generation profiles, individual\\ energy consumption patterns pricing preferences,\\ available energy surplus grid conditions} \\
            \cline{2-4}
			& \tabincell{l}{Electricity plan\\ recommendation} & \tabincell{l}{Collaborative filtering\cite{DBLP:journals/tii/ZhangMKD19}} & \tabincell{l}{Real-time energy generation data from distributed\\ energy resources (DERs), grid conditions, energy\\ storage state of charge, demand forecasts, operational\\ constraints, dynamic pricing signals} \\
            \cline{2-4}
			& \tabincell{l}{Energy scheduling\\ and pricing} & \tabincell{l}{Reinforcement learning\cite{DBLP:journals/tsg/KimZSL16},\\ Decision tress\cite{huo2021decision}} & \tabincell{l}{Consumer load, forecast load, generation operation\\ cost data, renewable generation profile } \\
            \cline{2-4}
			& \tabincell{l}{Peer-to-peer energy\\ trading} & \tabincell{l}{Reinforcement learning\cite{DBLP:journals/tsg/ChenS19}, Fuzzy Q\\ learning\cite{zhou2019artificial}, Deep Q learning\cite{DBLP:conf/isgteurope/ChenB19}} & \tabincell{l}{Energy prices, market dynamics, individual\\ energy consumption patterns, user preferences} \\

            \cline{1-4}
   
		\end{tabular}%
	}
\end{table*}

\begin{table*}[htb]
    \ContinuedFloat
	\centering \caption{\label{tbl:technique_source_1}Existing AI techniques in the energy system related to data processing (continued)}
	\resizebox{0.9\textwidth}{!}{%
		\begin{tabular}{|l|l|l|l|}
			\cline{1-4}
			\tabincell{l}{Domain} & \tabincell{l}{Application} & \tabincell{l}{Technique} & \tabincell{l}{Data involved} \\
		
            \cline{1-4}
			\multirow{8}{*}{\tabincell{l}{Energy market\\ and trading}} & \tabincell{l}{Retailing\cite{DILEEP20202589}} & \tabincell{l}{Reinforcement learning\cite{peters2013reinforcement}} & \tabincell{l}{Wholesale market price, statistical data on household\\ types, household sizes, appliance saturation, seasonality\\ on multiple timescales, vacations, \textit{etc.}} \\
			\cline{2-4}
			& \tabincell{l}{Aggregation\cite{DILEEP20202589}} & \tabincell{l}{K-means clustering\cite{DBLP:conf/ssci/SpinolaFFV17}, federated learning\cite{10738107}} & \tabincell{l}{Energy price, capacity, and number of resource unit of\\ different type of renewables like hydro, solar, wind, \textit{etc.},\\ energy load of different type of conumsers like domestic,\\ commerce, industry, \textit{etc}.} \\
            \cline{2-4}
			& \tabincell{l}{Demand response\cite{TUBALLA2016710}} & \tabincell{l}{Q-learning\cite{lu2018dynamic}} & \tabincell{l}{Consumer load profiles, wholesale energy price, retailer\\ energy price} \\
            \cline{2-4}
			& \tabincell{l}{Market modeling\\ and simulation} & \tabincell{l}{Game theory\cite{oliveira2009mascem}, Fuzz Q-learning\cite{salehizadeh2016application}\\Extreme learning machine\cite{chi2021data}, Mixed-integer\\ linear programming\cite{heuberger2020ev}, quantile regression\cite{10738058}} & \tabincell{l}{Energy generation capacity and profiles, energy prices,\\ energy demand, historical energy generation and\\ consumption data renewable resource availability data} \\
            \cline{2-4}
			& \tabincell{l}{Trading, bidding\\ and pricing} & \tabincell{l}{Reinforcement learning\cite{DBLP:journals/tcyb/WangHLAC17}, Deep\\ reinforcement learning\cite{DBLP:journals/tsg/XuSNKMH19}} & \tabincell{l}{Consumer demand and bid data, energy seller's supply\\ and reservation price data, initial bid price and expected\\ trading amount data, bid/offer price and quantity,\\ historical load, energy generator capacity } \\

            \cline{2-4}
			& \tabincell{l}{Energy price forecast} & \tabincell{l}{Polynomial regression and support vector\\ regression\cite{luo2019two}, Support vector machine\cite{papadimitriou2014forecasting},\\ LSTM and support vector regression\cite{atef2019comparative}\\Extreme learning machine and variational\\ Mode Decomposition\cite{DBLP:conf/icmla/DeepaNBPST18}, Particle swarm \\optimization\cite{DBLP:journals/tsg/CarriereK19}, Fuzzy inference\cite{DBLP:conf/iisa/BhagatAF19}} & \tabincell{l}{Historical energy market data, including price data,\\ market demand, supply data, economic indicators} \\

            \cline{1-4}
			\multirow{3}{*}{\tabincell{l}{Power\\ generation}} & \tabincell{l}{Power generation\\ forecast\cite{DBLP:journals/access/HossainKUSS19}} & \tabincell{l}{Bayesian clustering\cite{fan2009forecasting}, Gaussian process\\ and neural network\cite{DBLP:journals/tsg/LeeB14a}, Extreme machine\\ learning\cite{salcedo2014feature}, Quantile regression\cite{zhang2016deterministic}\\Swarm optimization\cite{yeh2014forecasting}, Ant colony\\ optimization\cite{rahmani2013hybrid}, Neuro-fuzzy system\cite{chaouachi2010novel},\\ Support vector regression\cite{yang2014weather}, Decision trees\\ and homogeneous resembles\cite{DBLP:conf/aaai/HeinermannK15}, Genetic\\ algorithm and nearest neighbor approach\cite{DBLP:journals/tnn/AkFZ16}\\CNN and LSTM\cite{liu2018smart}, Deep Boltzmann\\ Machine\cite{zhang2015predictive}, Deep belief network\cite{wang2016deep}, Deep\\ echo state network\cite{hu2020forecasting}, Random forest\cite{demolli2019wind},\\ ARIMA\cite{zhang2020adaptive}, DBSCAN Clustering\cite{yan2019uncertainty}\\SARIMA\cite{shadab2019box}, LASSO\cite{DBLP:journals/iotj/TangMWN18}} & \tabincell{l}{Historical power generation data from renewable\\ sources, weather data, solar radiation patterns,\\ wind speed and direction, time of day/week, seasonal\\ patterns, geographical characteristics, \textit{e.g.},\\ location/orientation of solar panels/wind turbines.} \\
            \cline{2-4}
			& \tabincell{l}{Fault detection} & \tabincell{l}{Generalized mapping regressor\cite{marvuglia2012monitoring}} & \tabincell{l}{Wind speed data, power generation data} \\
            \cline{2-4}
			& \tabincell{l}{Generation dispatch} & \tabincell{l}{Directed network\cite{lu2020achieving}} & \tabincell{l}{Energy price, power generation cost, generation\\ capacity, energy demand} \\
            \cline{2-4}
			& \tabincell{l}{Energy profile generation} & \tabincell{l}{GAN\cite{DBLP:journals/corr/ChenWKZ17}} & \tabincell{l}{Power generation data, Power generation forecast data} \\

            \cline{1-4}
			\multirow{3}{*}{\tabincell{l}{EV}} & \tabincell{l}{Charging scheduling\\ and optimization} & \tabincell{l}{Reinforcement learning\cite{DBLP:journals/tsg/VandaelCEHD15}, Deep\\ reinforcement learning\cite{DBLP:journals/tsg/WanLHP19}, Deep Q\\ learning\cite{DBLP:journals/tsg/QianSWS20}, Shallow neural networks and\\ k-nearest neighbors\cite{DBLP:journals/tsg/BedoyaGG19}, LSTM\cite{DBLP:journals/tii/LiHCZHCB22}} & \tabincell{l}{Population density, building data, transportation\\ networks, parking availability, EV usage patterns, grid\\ infrastructure data, user preferences, energy tariff data,\\ grid load data, EV battery data, charging station\\ availability and usage data, real-time energy market data} \\
            \cline{2-4}
			& \tabincell{l}{Energy consumption\\ estimation/optimization} & \tabincell{l}{Federated learning\cite{DBLP:conf/ivs/Zhang21}, Deep reinforcement\\ learning and CNN\cite{DBLP:journals/tii/WangTWP21}, Deep deterministic\\ policy gradient\cite{qi2019deep}} & \tabincell{l}{Battery state of charge, driving patterns, vehicle speed, \\weather data, topographical data like elevation, and slope, \\road network data, historical charging data, cameras\\ sensor data, LiDAR, radar, GPS, road network data, \\real-time traffic data, EV telemetry data, road conditions} \\

            \cline{1-4}
   
		\end{tabular}%
	}
\end{table*}

Table~\ref{tbl:technique_source} shows that the applications of energy usage monitoring, load forecast, energy price forecast, power generation forecast, and cyber security for the power system are the ones where AI and data processing techniques have been extensively used. These applications are directly connected with secure energy supply and efficient operation of the energy system and the energy market. The AI and data processing methods requires data from diverse sources, including data from the environment, aggregated power generation and consumption, energy price, and individual energy usage data. Note that while some of the above data is open and public, unauthenticated transferring or processing of the rest, \textit{e.g.}, information about individual energy consumption or generation, can lead to privacy violation issues or conflicts in business interest.

Table~\ref{tbl:technique_source} shows that most of the data processing takes place in the applications of energy consumption/generation forecast and anomaly detection for energy consumption. Overall, the historical energy usage data, energy price, and weather data (related to solar and wind power generation) are the most widely used in energy data processing.

\subsection{Categorization of data related issues}

In this section, we review the existing studies on issues related to data usage in the energy sector and divide them into the following three categories:

\begin{enumerate}
    \item \textbf{Data-sharing} indicating issues that lead to barriers to transferring and accessing data.
    \item \textbf{Privacy} referring to issues where individual private information is at the risk of disclosure, which is more of a subjective concept from the data owner's perspective.
    \item \textbf{Cyber security} representing issues where any potential circumstance, event or action that could damage, disrupt or otherwise adversely impact network and information systems, the users of such systems and other persons. Cyber security relates closely to the system's capability to secure data storage, processing, and transmission.
\end{enumerate}

For each of the categorized issue, we list the related applications and the critical data that can be processed in the application in Table~\ref{tbl:information_issues}.

We notice that extensive reviews exist on data-related issues for energy system focusing on cyber security and privacy. \textit{E.g.}, there are reviews on cyber security in microgrid~\cite{8333659,inventions8040084} and smart grid communications~\cite{6141833}, smart home technologies~\cite{SOVACOOL2020109663}, advanced metering infrastructures~\cite{CHRISFOREMAN201594}, cloud computing environment for smart grid~\cite{PRIYADARSHINI2021107204}, and reviews on specific cyber attacks like false data injection~\cite{REDA2022112423,BRETAS201943}, and reviews on privacy issues in smart metering~\cite{HORNE201564,RYU2024113851,VONLOESSL2023113645}, distributed energy activities~\cite{NAUS2015125}, e-mobility~\cite{SCHALLEHN2022112756} and EVs~\cite{zhang2020privacy}. Our review in Table~\ref{tbl:information_issues} covers issues with data sharing, which is an critical aspect of data usage in the energy sector, and the unauthorized data sharing can lead to reduced business competence and security concerns. Furthermore, our review also provides the data associated with the reviewed issues, and show the connections between the reviewed issues and applications in the energy system. In Section~\ref{Results of the survey}, we use this review as the basis to connect the three aspects: data sharing, privacy and security with the regulatory analysis on the categorized issues. 

\begin{figure}[htb]
	\centering
	\includegraphics[width=0.27\textwidth]{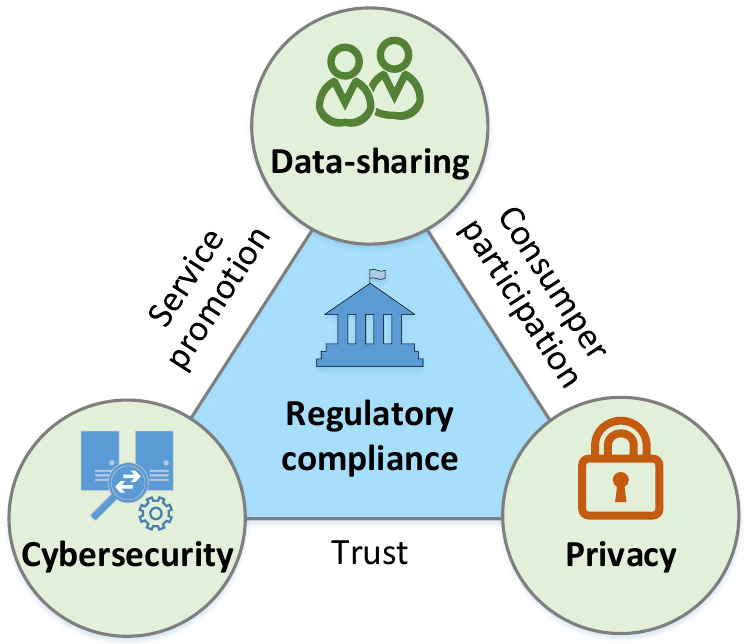}\\
	\caption{\label{fig:categorized_issues}Three challenges for secure data usage in the energy system.}
\end{figure}

We show the relationship and connections between the categorized issues in Figure~\ref{fig:categorized_issues}. We view that data-sharing and privacy connect with each other regarding the usage and regulation of personal information, the sharing of which might lead to the disclosure of individual personal details and thus violation of privacy. A good relationship and balance between data-sharing and privacy can enhance the usage of consumer data and promote consumer participation in energy related applications, services, and new market paradigms like peer-to-peer energy trading~\cite{DBLP:conf/isgteurope/ChenB19}. Data-sharing is also closely connected to cyber security issues, since the storage, transferring, and processing of the shared data requires data security measures that ensure integrity, confidentiality, availability, and authenticity of the data, thus promoting services and business replying on data and data processing. The combination of cyber security and privacy issues concern much about the security of data usage and processing in the energy system, the improvement of which can enhance trust among the stakeholders in the energy system. We provide further details of the listed three issues along with the potential data involved in Table~\ref{tbl:information_issues}.

We observe from Table~\ref{tbl:information_issues} that, the same application can lead to different data-related issues, \eg in the case of vehicle-to-grid (V2G) communication where data-sharing, privacy, and cyber security issues are all involved. This is also true in terms of data, \ie processing of a single type of data can lead to different issues \textit{e.g.}, privacy and cyber security, in the energy system. Due to such coupling, the addressing of one issue might be directly connected with another. 

\begin{table*}[htb]%
	\centering \caption{\label{tbl:information_issues}Summary of data-related issues in the energy system}
	\resizebox{0.9\textwidth}{!}{
		\begin{tabular}{|l|l|l|l|}
			\cline{1-4}
			\tabincell{l}{Domain} & \tabincell{l}{Application} & \tabincell{l}{Description} & \tabincell{l}{Critical data involved} \\

            \cline{1-4}
			\multirow{3}{*}{\tabincell{l}{Data sharing}} & \tabincell{l}{Demand response\cite{snow2022solar}} & \tabincell{l}{Exchange of real-time demand response data between utility\\ companies, grid operators, and consumers may face barriers\\ due to trust, interoperability issues, varying data formats, and\\ the need for standardization across different demand response\\ management systems} & \tabincell{l}{Real-time energy consumption data, peak \\demand information, load profiles, grid conditions, \\and customer-specific usage patterns} \\
			\cline{2-4}
			& \tabincell{l}{Energy system\\ monitoring and\\ maintenance\cite{jaradat2015internet}} & \tabincell{l}{Sharing operational data for grid balancing, frequency regulation,\\ and load forecasting among multiple grid operators and energy\\ market participants can encounter barriers related to data formats,\\ control protocols, and interoperability of grid control systems} & \tabincell{l}{Operational data related to grid stability, energy \\generation and consumption, frequency regulation\\ signals, grid constraints, and system capacity} \\
			\cline{2-4}
			& \tabincell{l}{Vehicle-to-Grid (V2G)\\ Integration\cite{han2016ip2dm}} & \tabincell{l}{Integrating V2G systems with grid infrastructure may face data\\ sharing barriers due to lack of trust or interoperability challenges\\ between EV manufacturers, charging infrastructure providers,\\ and grid operators, impacting the bidirectional exchange of energy\\ flow and operational data} & \tabincell{l}{Charging session records, energy flow data, grid\\ connectivity status, vehicle telematics, battery\\ health information, and vehicle energy demand\\ patterns} \\
            \cline{2-4}
			& \tabincell{l}{Interconnection of\\ Distributed Energy\\ Resources (DERs)\cite{DBLP:conf/chi/WilkinsCL20,photovoltaics2018ieee}} & \tabincell{l}{Connecting and sharing data from various distributed energy\\ resources, such as solar panels, wind turbines, and energy storage\\ systems, presents challenges related to data standardization,\\ system integration, and regulatory frameworks for seamless\\ interconnection with the grid} & \tabincell{l}{Real-time energy production from solar PV, wind\\ turbines, energy storage levels, grid injection data,\\ grid synchronization information, and DER\\ operational parameters} \\
			\cline{2-4}
			& \tabincell{l}{Energy trading\cite{morstyn2018using}} & \tabincell{l}{Data sharing among energy market participants, including\\ independent power producers, aggregators, and utilities, may\\ encounter barriers related to standardizing market data formats,\\ interoperability of trading platforms, and ensuring consistent data \\exchange across diverse energy market systems} & \tabincell{l}{Energy pricing data, market clearing prices,\\ transaction records, supply-demand dynamics,\\ trading agreements, and financial settlement\\ information} \\
			
			\cline{1-4}
			\multirow{3}{*}{\tabincell{l}{Privacy}} & \tabincell{l}{Smart metering and\\ energy monitoring\cite{li2010compressed,efthymiou2010smart}} & \tabincell{l}{reveal sensitive information about occupancy, daily routines, and \\personal habits} & \tabincell{l}{usage patterns, timing, and quantity of \\electricity consumed at specific times} \\
			\cline{2-4}
			& \tabincell{l}{Location tracking \\in EVs\cite{antoun2020detailed}} & \tabincell{l}{continuous tracking of the vehicle's location raises privacy\\ concerns, as it can disclose details about an individual's\\ movements and daily activities} & \tabincell{l}{Real-time and historical location data \\of electric vehicles} \\
			\cline{2-4}
			& \tabincell{l}{Charging services\cite{DBLP:journals/tsg/LiDN17,DBLP:journals/tvt/Pazos-RevillaAG18}} & \tabincell{l}{Charging service providers may collect personal information from\\ EV owners for authentication, billing, and personalized services.\\ The storage and handling of such sensitive data can potentially\\ jeopardize privacy if not adequately protected} & \tabincell{l}{Personal information collected for charging\\ services, including names, addresses, contact\\ details, and payment information} \\
			\cline{2-4}
			& \tabincell{l}{Vehicle-to-Grid (V2G)\\ communication\cite{DBLP:journals/tvt/BansalNCS0G20,DBLP:journals/tsg/LiuNZY12}} & \tabincell{l}{exchange of information between EVs and the grid raises privacy \\concerns} & \tabincell{l}{user identification, energy consumption \\details, charging session records} \\
			\cline{2-4}
			& \tabincell{l}{Decentralized energy\\ trading, peer-to-peer\\ energy sharing\cite{DBLP:journals/tdsc/AitzhanS18,DBLP:conf/isgt/ThandiM22}} & \tabincell{l}{privacy concerns may arise due to the sharing of personal energy\\related information} & \tabincell{l}{Energy production and consumption details,\\ trading transactions, potentially user\\ identification or billing information} \\
			\cline{2-4}
			& \tabincell{l}{Energy system data\\ exchange and\\ integration\cite{kong2019privacy}} & \tabincell{l}{Exchange of data between these entities can raise concerns \\about data privacy} & \tabincell{l}{User energy consumption data, billing\\ information, potentially personally identifiable\\ information} \\

            \cline{1-4}
			\multirow{3}{*}{\tabincell{l}{Cyber security}} & \tabincell{l}{Advanced metering\cite{DBLP:journals/tsg/LiuZZC15}} & \tabincell{l}{Smart meters collect and transmit detailed energy consumption\\ data, and vulnerabilities in the AMI infrastructure can lead to\\ unauthorized access to this data, potential tampering with\\ meter readings, or disruption of metering operations} & \tabincell{l}{Usage patterns, energy demand profiles,\\ potentially identifiable information associated\\ with the customer} \\
			\cline{2-4}
			& \tabincell{l}{Energy management\\ systems (EMS)\cite{DBLP:journals/tsg/MuslehCD20,REDA2022112423}} & \tabincell{l}{Cyber security issues within EMS can lead to unauthorized\\ system control, inaccurate energy optimization, or disruptions\\ to energy supply, affecting grid stability and reliability} & \tabincell{l}{Real-time energy consumption data from various\\ sources, sensitive customer information, energy \\demand forecasts, and grid configuration details} \\
			\cline{2-4}
			& \tabincell{l}{Electric vehicle \\supply equipment\\ (EVSE)\cite{DBLP:journals/tiv/PonnuruRPK22,DBLP:journals/tvt/PonnuruARDSP21}} & \tabincell{l}{charging stations and supporting components may be susceptible\\ to cyber security threats. Vulnerabilities in EVSE can lead to\\ unauthorized access to charging stations, data breaches related\\ to user information, or potential safety risks if the charging\\ process is compromised} & \tabincell{l}{User identification and authentication data,\\ charging session records, billing information,\\ potentially personally identifiable information\\ associated with EV owners} \\
			\cline{2-4}
			& \tabincell{l}{Vehicle-to-Grid \\(V2G) communication\cite{basnet2020deep}} & \tabincell{l}{Threats to V2G communication can result in unauthorized access\\ to vehicles, manipulation of energy flow, or disruptions to grid\\ operations, impacting grid stability and the reliability of EV\\ integration} & \tabincell{l}{energy consumption patterns, charging or\\ discharging schedules, possibly vehicle\\ operational data} \\
            \cline{2-4}
			& \tabincell{l}{Data exchange\\ between vehicles\cite{gawas2021integrative,DBLP:journals/tits/0004OY0022}} & \tabincell{l}{The equipment, \textit{e.g.}, roadside unit, facilitating data exchange\\ between vehicles might be tampered, and the data exchange\\ lacks a trusted environment} & \tabincell{l}{Distance for vehicle communication,\\ communication configurations like encryption\\ settings, packet size during the communication,\\ vehicle GPS trace data, vehicle occupation status} \\
            \cline{2-4}
			& \tabincell{l}{Grid control and \\monitoring systems\cite{DBLP:journals/tac/LiSC18}} & \tabincell{l}{Centralized control systems and monitoring platforms used to\\ oversee grid operations are critical targets for cyber security\\ attacks. Compromising these systems can lead to unauthorized\\ control over grid infrastructure, manipulation of energy\\ distribution, or disruptions in grid operations} & \tabincell{l}{Critical system configurations, control \\commands, operational data, } \\
			\cline{2-4}
			& \tabincell{l}{Decentralized energy \\trading\cite{islam2018impact,kavousi2021effective}} & \tabincell{l}{Platforms that facilitate peer-to-peer energy trading and\\ blockchain-based energy transactions may be vulnerable to\\ cyber security threats. Breaches within decentralized trading\\ systems can compromise the integrity of energy transactions,\\ user authentication, or smart contracts, leading to financial\\ risks and potential system disruptions} & \tabincell{l}{User information, financial details, smart \\contract data, energy production/consumption\\ details, and transaction records} \\

			\cline{1-4}
		
		\end{tabular}
	}
\end{table*}

\subsection{Regulatory frameworks for data-sharing, privacy, and cyber security in the energy system}\label{Information regulation}

The ICT advancements in the evolving energy sector have raised regulatory concerns in preserving human rights and welfare. Regulations and ethics guidelines has been established to mandate secure and rigorous risk management in data processing, sharing, and transferring. Concrete legislative framework is of critical importance to energy system and the stakeholders for ensuring regulatory compliance. There are studies that have explored the regulatory dimension in the context of energy system. In this section, we extensive review the relevant studies, as summarized in Table~\ref{tbl:regulatory_research_information_issues}. In our review, we examine to what level each of those studies cover the regulatory aspects, and we range the reviewed studies by year and the category of data-related issues, so as to check the depth of regulatory studies in data-sharing, privacy, and cyber security, and see how the research efforts have evolved over time. We investigate the following regulatory aspects: Roles and stakeholders, obligations, rights, threats or barriers, objectives, principles, structure or organization, sensitive data examples and definitions, and data usage. We extract those aspects from the reviewed studies, and then we analyze how those aspects are covered in different works.

\begin{table}[htb] 
	\centering \caption{\label{tbl:regulatory_research_information_issues}Summary of studies on data related issues in the energy sector from a regulatory perspective.}
	\resizebox{0.48\textwidth}{!}{%
		\begin{tabular}{|l|l|c|c|c|c|c|c|c|c|c|}
			\cline{1-11}
			\tabincell{l}{Domain} & \tabincell{l}{Reference} & \rotatebox{90}{\tabincell{l}{Roles and Stakeholders}} & \rotatebox{90}{\tabincell{l}{Obligations or Responsibilities}} & \rotatebox{90}{\tabincell{l}{Rights}} & \rotatebox{90}{\tabincell{l}{threats or barriers}} & \rotatebox{90}{\tabincell{l}{Objectives}} & \rotatebox{90}{\tabincell{l}{Principles}} & \rotatebox{90}{\tabincell{l}{Structure or organization}}  & \rotatebox{90}{\tabincell{l}{Data example or definition}} & \rotatebox{90}{\tabincell{l}{Data usage processing}}     \\

			\cline{1-11}
			\multirow{3}{*}{\tabincell{l}{Data-sharing}} & \tabincell{l}{~\cite{forbush2011regulating} (2011)} & \tabincell{l}{} & \tabincell{l}{} & \tabincell{l}{} & \tabincell{l}{\cmark} & \tabincell{l}{} & \tabincell{l}{} & \tabincell{l}{} & \tabincell{l}{\cmark} & \tabincell{l}{}\\
			\cline{2-11}
			& \tabincell{l}{~\cite{10.1093/ijlit/eau001} (2014)} & \tabincell{l}{\cmark} & \tabincell{l}{} & \tabincell{l}{\cmark} & \tabincell{l}{\cmark} & \tabincell{l}{} & \tabincell{l}{} &  \tabincell{l}{} & \tabincell{l}{\cmark} & \tabincell{l}{\cmark}\\   
			\cline{2-11}
			& \tabincell{l}{~\cite{doi:10.1177/1783591719895390} (2019)} & \tabincell{l}{\cmark} & \tabincell{l}{\cmark} & \tabincell{l}{} & \tabincell{l}{} & \tabincell{l}{} & \tabincell{l}{} & \tabincell{l}{} & \tabincell{l}{\cmark} & \tabincell{l}{}\\
            \cline{2-11}
			& \tabincell{l}{~\cite{betti2020share} (2020)} & \tabincell{l}{} & \tabincell{l}{} & \tabincell{l}{} & \tabincell{l}{\cmark} & \tabincell{l}{} & \tabincell{l}{} & \tabincell{l}{} & \tabincell{l}{} & \tabincell{l}{}\\   
			\cline{2-11}
			& \tabincell{l}{~\cite{doi:10.1161/HYPERTENSIONAHA.120.16340} (2021)} & \tabincell{l}{} & \tabincell{l}{} & \tabincell{l}{} & \tabincell{l}{} & \tabincell{l}{} & \tabincell{l}{} & \tabincell{l}{} & \tabincell{l}{\cmark} & \tabincell{l}{\cmark}\\
            \cline{2-11}
            & \tabincell{l}{~\cite{FERNANDES2022113240} (2022)} & \tabincell{l}{} & \tabincell{l}{} & \tabincell{l}{} & \tabincell{l}{\cmark} & \tabincell{l}{} & \tabincell{l}{} & \tabincell{l}{} & \tabincell{l}{\cmark} & \tabincell{l}{\cmark}\\

			\cline{1-11}
			\multirow{3}{*}{\tabincell{l}{Privacy}} & \tabincell{l}{~\cite{10.1093/idpl/ipr004} (2011)} & \tabincell{l}{\cmark} & \tabincell{l}{} & \tabincell{l}{} & \tabincell{l}{\cmark} & \tabincell{l}{} & \tabincell{l}{} & \tabincell{l}{} & \tabincell{l}{} & \tabincell{l}{\cmark}\\
			\cline{2-11}
			& \tabincell{l}{~\cite{havlikova2011smart} (2011)} & \tabincell{l}{\cmark} & \tabincell{l}{\cmark} & \tabincell{l}{\cmark} & \tabincell{l}{} & \tabincell{l}{} & \tabincell{l}{} &  \tabincell{l}{} & \tabincell{l}{} & \tabincell{l}{\cmark}\\   
			\cline{2-11}
			& \tabincell{l}{~\cite{Cuijpers2013} (2012)} & \tabincell{l}{} & \tabincell{l}{} & \tabincell{l}{} & \tabincell{l}{\cmark} & \tabincell{l}{} & \tabincell{l}{} & \tabincell{l}{} & \tabincell{l}{} & \tabincell{l}{\cmark}\\   
			\cline{2-11}
			& \tabincell{l}{~\cite{DBLP:conf/isgt/Pallas12} (2012)} & \tabincell{l}{\cmark} & \tabincell{l}{\cmark} & \tabincell{l}{} & \tabincell{l}{} & \tabincell{l}{} & \tabincell{l}{\cmark} &  \tabincell{l}{} & \tabincell{l}{} & \tabincell{l}{\cmark}\\   
			\cline{2-11}
			& \tabincell{l}{~\cite{Pallas2013} (2012)} & \tabincell{l}{} & \tabincell{l}{} & \tabincell{l}{} & \tabincell{l}{} & \tabincell{l}{} & \tabincell{l}{} & \tabincell{l}{} & \tabincell{l}{} & \tabincell{l}{\cmark}\\   
			\cline{2-11}
			& \tabincell{l}{~\cite{harvey2013smart} (2013)} & \tabincell{l}{} & \tabincell{l}{} & \tabincell{l}{} & \tabincell{l}{\cmark} & \tabincell{l}{} & \tabincell{l}{} & \tabincell{l}{} & \tabincell{l}{\cmark} & \tabincell{l}{}\\   
			\cline{2-11}
			& \tabincell{l}{~\cite{DBLP:conf/ccs/BiselliFC13} (2013)} & \tabincell{l}{\cmark} & \tabincell{l}{} & \tabincell{l}{} & \tabincell{l}{} & \tabincell{l}{} & \tabincell{l}{} & \tabincell{l}{} & \tabincell{l}{} & \tabincell{l}{}\\   
			\cline{2-11}
			& \tabincell{l}{~\cite{Papakonstantinou2015} (2015)} & \tabincell{l}{\cmark} & \tabincell{l}{\cmark} & \tabincell{l}{\cmark} & \tabincell{l}{} & \tabincell{l}{} & \tabincell{l}{\cmark} &  \tabincell{l}{\cmark} & \tabincell{l}{\cmark} & \tabincell{l}{\cmark}\\   
			\cline{2-11}
			& \tabincell{l}{~\cite{Milaj2016} (2016)} & \tabincell{l}{} & \tabincell{l}{} & \tabincell{l}{} & \tabincell{l}{} & \tabincell{l}{} & \tabincell{l}{} & \tabincell{l}{} & \tabincell{l}{\cmark} & \tabincell{l}{\cmark} \\   
			\cline{2-11}
			& \tabincell{l}{~\cite{doi:10.1080/13600869.2017.1371576} (2017)} & \tabincell{l}{} & \tabincell{l}{} & \tabincell{l}{\cmark} & \tabincell{l}{} & \tabincell{l}{} & \tabincell{l}{} & \tabincell{l}{} & \tabincell{l}{} & \tabincell{l}{}\\
            \cline{2-11}
			 & \tabincell{l}{~\cite{10.1093/jwelb/jwx001} (2017)} & \tabincell{l}{} & \tabincell{l}{} & \tabincell{l}{\cmark} & \tabincell{l}{} & \tabincell{l}{} & \tabincell{l}{} & \tabincell{l}{} & \tabincell{l}{\cmark} & \tabincell{l}{}\\   
            \cline{2-11}
            & \tabincell{l}{~\cite{DBLP:conf/isdevel/MartinezRPAM19} (2019)} & \tabincell{l}{} & \tabincell{l}{\cmark} & \tabincell{l}{\cmark} & \tabincell{l}{} & \tabincell{l}{} & \tabincell{l}{\cmark} & \tabincell{l}{} & \tabincell{l}{} & \tabincell{l}{}\\
			\cline{2-11}
			& \tabincell{l}{~\cite{doi:10.1080/02646811.2019.1622244} (2019)} & \tabincell{l}{\cmark} & \tabincell{l}{\cmark} & \tabincell{l}{} & \tabincell{l}{} & \tabincell{l}{} & \tabincell{l}{} & \tabincell{l}{} & \tabincell{l}{\cmark} & \tabincell{l}{\cmark}\\            
			\cline{2-11}
			& \tabincell{l}{~\cite{vojkovic2020iot} (2020)} & \tabincell{l}{\cmark} & \tabincell{l}{} & \tabincell{l}{} & \tabincell{l}{\cmark} & \tabincell{l}{} & \tabincell{l}{} & \tabincell{l}{} & \tabincell{l}{\cmark} & \tabincell{l}{\cmark}\\            
			\cline{2-11}
			& \tabincell{l}{~\cite{9519329} (2021)} & \tabincell{l}{\cmark} & \tabincell{l}{} & \tabincell{l}{\cmark} & \tabincell{l}{\cmark} & \tabincell{l}{} & \tabincell{l}{\cmark} & \tabincell{l}{} & \tabincell{l}{} & \tabincell{l}{\cmark}\\
			\cline{2-11}
			& \tabincell{l}{~\cite{orlando2021smart} (2021)} & \tabincell{l}{} & \tabincell{l}{\cmark} & \tabincell{l}{} & \tabincell{l}{\cmark} & \tabincell{l}{} & \tabincell{l}{} & \tabincell{l}{} & \tabincell{l}{} & \tabincell{l}{}\\

			\cline{1-11}
			\multirow{3}{*}{\tabincell{l}{Cyber security}} & \tabincell{l}{~\cite{Mah2014} (2014)} & \tabincell{l}{} & \tabincell{l}{} & \tabincell{l}{} & \tabincell{l}{\cmark} & \tabincell{l}{} & \tabincell{l}{} & \tabincell{l}{} & \tabincell{l}{} & \tabincell{l}{}\\
			\cline{2-11}
			& \tabincell{l}{~\cite{ANANDAKUMAR2014126} (2014)} & \tabincell{l}{} & \tabincell{l}{} & \tabincell{l}{} & \tabincell{l}{\cmark} & \tabincell{l}{} & \tabincell{l}{} & \tabincell{l}{} & \tabincell{l}{} & \tabincell{l}{}\\
			\cline{2-11}
			& \tabincell{l}{~\cite{DBLP:journals/ei/HolzleitnerR17} (2017)} & \tabincell{l}{\cmark} & \tabincell{l}{} & \tabincell{l}{} & \tabincell{l}{} & \tabincell{l}{} & \tabincell{l}{} & \tabincell{l}{\cmark} & \tabincell{l}{} & \tabincell{l}{}\\
			\cline{2-11}
			& \tabincell{l}{~\cite{draffin2017cybersecurity} (2017)} & \tabincell{l}{} & \tabincell{l}{} & \tabincell{l}{} & \tabincell{l}{\cmark} & \tabincell{l}{\cmark} & \tabincell{l}{} & \tabincell{l}{\cmark} & \tabincell{l}{} & \tabincell{l}{}\\    
			\cline{2-11}
			& \tabincell{l}{~\cite{LESZCZYNA201862} (2018)} & \tabincell{l}{} & \tabincell{l}{} & \tabincell{l}{} & \tabincell{l}{} & \tabincell{l}{} & \tabincell{l}{}  & \tabincell{l}{} & \tabincell{l}{} & \tabincell{l}{}\\
			\cline{2-11}
			& \tabincell{l}{~\cite{LESZCZYNA2018262} (2018)} & \tabincell{l}{} & \tabincell{l}{} & \tabincell{l}{} & \tabincell{l}{} & \tabincell{l}{} & \tabincell{l}{}  & \tabincell{l}{} & \tabincell{l}{} & \tabincell{l}{}\\
			\cline{2-11}
			& \tabincell{l}{~\cite{anderson2018cybersecurity} (2018)} & \tabincell{l}{} & \tabincell{l}{} & \tabincell{l}{} & \tabincell{l}{\cmark} & \tabincell{l}{} & \tabincell{l}{} &  \tabincell{l}{} & \tabincell{l}{} & \tabincell{l}{}\\
			\cline{2-11}
			& \tabincell{l}{~\cite{zukowska2020legal} (2020)} & \tabincell{l}{} & \tabincell{l}{\cmark} & \tabincell{l}{} & \tabincell{l}{} & \tabincell{l}{} & \tabincell{l}{} &  \tabincell{l}{} & \tabincell{l}{} & \tabincell{l}{}\\
			\cline{2-11}
			& \tabincell{l}{~\cite{en14237836} (2021)} & \tabincell{l}{\cmark} & \tabincell{l}{} & \tabincell{l}{} & \tabincell{l}{} & \tabincell{l}{} & \tabincell{l}{}  & \tabincell{l}{} & \tabincell{l}{} & \tabincell{l}{}\\   
            \cline{2-11}
			& \tabincell{l}{~\cite{10.1007/978-3-031-15559-8_38} (2022)} & \tabincell{l}{} & \tabincell{l}{} & \tabincell{l}{} & \tabincell{l}{\cmark} & \tabincell{l}{} & \tabincell{l}{}  & \tabincell{l}{} & \tabincell{l}{} & \tabincell{l}{}\\  
			\cline{2-11}
			& \tabincell{l}{~\cite{DBLP:journals/corr/abs-2210-13119} (2022)} & \tabincell{l}{} & \tabincell{l}{} & \tabincell{l}{} & \tabincell{l}{\cmark} & \tabincell{l}{} & \tabincell{l}{} &  \tabincell{l}{} & \tabincell{l}{} & \tabincell{l}{}\\
            \cline{2-11}
            & \tabincell{l}{~\cite{doi:10.1080/13600869.2022.2094609} (2022)} & \tabincell{l}{} & \tabincell{l}{} & \tabincell{l}{} & \tabincell{l}{\cmark} & \tabincell{l}{} & \tabincell{l}{}  & \tabincell{l}{} & \tabincell{l}{} & \tabincell{l}{}\\

			\cline{1-11}
				
		\end{tabular}%
	}
\end{table}

From Table~\ref{tbl:regulatory_research_information_issues}, we observe that the issues of privacy and cyber security are more investigated than data-sharing from a regulatory perspective. We also observe that there are regulatory aspects that are not investigated across the three data-related issues, \eg roles and stakeholders, rights, obligations, and principles are studied more in the context of privacy than cyber security and data-sharing, while data processing and regulatory definitions about data are investigated more in privacy and data-sharing than in cyber security. There also exists regulatory aspect, \eg organizational structure for the implementation of the regulations, that is insufficiently studied in all the three data-relate issues.

Overall, the research on data related issues in the energy system from a regulatory perspective is sparse. The existence of regulation on data-sharing, privacy and data-protection, cyber-security does not necessarily mean informed and updated knowledge for energy stakeholders. It is not clear to end-users and stakeholders the legal categorization, definition, and boundaries between sensitive/private and non-sensitive data in the energy system. The legal aspects of data-related issues are not well structured and articulated, \eg roles and responsibilities, legislative basis, principles, organizational structures, \textit{etc.}, in the existing regulatory studies. There is a clear gap between the technical research and the regulatory framework in terms of data-sharing, privacy, and cyber security. For energy stakeholders, there is a knowledge gap of pragmatic guidelines, \textit{i.e.}, which regulations to follow and how to follow in their business to avoid regulatory issues in data-sharing, privacy, and cyber security. In addition, there is a lack of knowledge for the energy stakeholders about the organizational structures for regulations compliance, and what kind of responsibilities and interactions are anticipated in their role and responsibility for data related activities.

\section{Articulation of existing regulatory solutions for data-related issues in the energy system: Taking the EU as an example}\label{Results of the survey}

In this section, we review the issues of data-sharing, privacy, and cyber security from a regulatory perspective, to create the bridge between the technical and regulatory dimensions by articulating regulatory frameworks and structures towards those data-related issues. Our review looks into EU regulations and draws analysis upon them. We choose the EU as our scope of regulation review, because the EU has a leading role in decentralization and digitalization of the energy sector, and has established an internal energy market with participants of member states with diversified energy conditions, \eg energy capacity, proportion of energy generation, availability of energy sources, \textit{etc}. Such a leading role and diversified energy sector participants contribute to and also benefit from comprehensive legislative progress regarding data sharing, privacy, and cyber security in the energy sector.

\subsection{Data-sharing with regulatory compliance in the energy system}

In this section, we review and describe the legislative framework for data sharing in energy sector in the EU and how it works with energy stakeholders. Our review reveals the dynamics and interactions between different regulations over time, so as to illustrate the evolution of data-sharing with regulatory compliance in the energy sector. This review of the evolution of the regulatory framework informs which regulations are pertinent to draw the regulatory articulation towards data-sharing. 

From the most pertinent regulations, we extract the critical definitions, concepts, and principles in data-sharing related to the energy sector, clarify the rights to be legally protected regarding data-sharing, and explain roles and responsibilities involved. We present the organizational structure for this regulatory data-sharing framework that navigates an efficient practice of sharing data in the energy sector. Through this, we believe that our regulatory articulation establishes what data-sharing implies in the energy sector from a regulatory perspective, and also that it provides a status review and outlook into the the required path and measures for a conducive regulatory framework to facilitate emerging distributed and participatory energy paradigms.

\subsubsection{Regulations regarding data-sharing and their evolution}

\begin{figure*}[htb]
	\centering
	\includegraphics[width=0.89\textwidth]{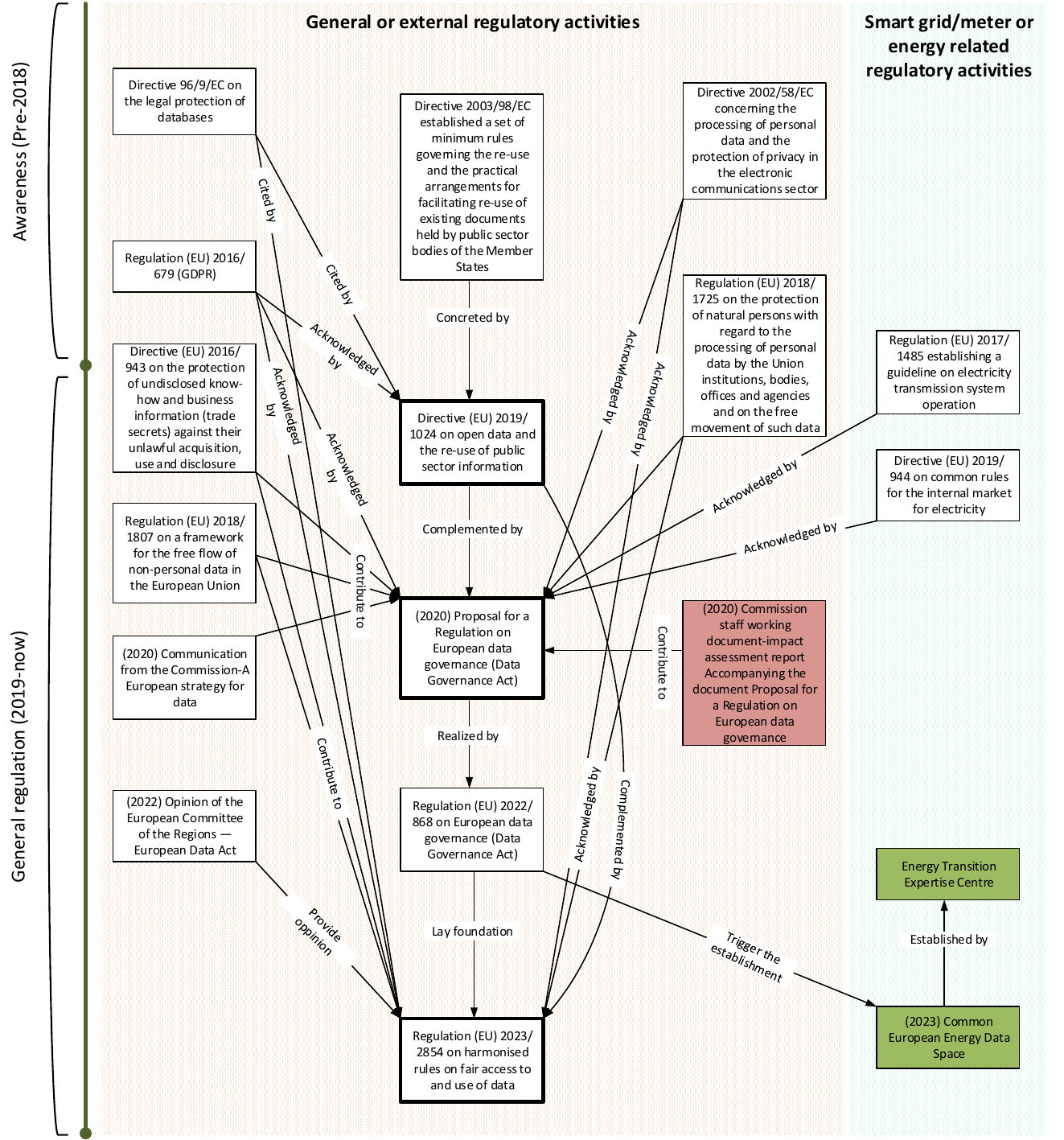}\\
	\caption{\label{fig:legislative process data-sharing}The EU legislative process regarding data-sharing in smart grid. Boxes with thick edge line interplay frequently with other regulations/entities and are the focal point of the legislative process.}
\end{figure*}

In this section, we review EU regulation related to data-sharing in the energy sector from 1996 to 2024. We review relevant regulations from both general regulations applicable to diverse sectors and those dedicated to the energy sector. We also cover critical regulatory entities and documents involved in the evolution of the regulatory framework, and depict the relationships and interactions between the involved elements over time to illustrate how the data-sharing regulations develop, as shown in Figure~\ref{fig:legislative process data-sharing}.

In Figure~\ref{fig:legislative process data-sharing}, we partition the evolution of the regulations into two periods. In the first period (from 1996 to 2018), the regulations mainly concerned data-sharing in the energy sector by promoting awareness of the importance of data re-use, without targeting specific sectors. From 2019, which we call the general regulation period, regulations began to pay attention to the improvement of data-sharing and the data economics through provisions on open data and re-use of data from public sectors. In that period, regulations were established to develop data-sharing frameworks, and regulate fair data access that can lead to commercial insights and potential business profits. In the second period, there was no contextualization of the general regulations in the energy sector, except that the ``Common European Energy Data Space'' was triggered to be established, which aims to empower smooth data exchange and the availability of data access in the energy sector.

From the evolution of the regulatory framework, we observe that data-sharing related regulations interact and intertwine with those related to data protection, \eg Regulation (EU) 2016/679 (GDPR). We also observe the connection between general regulations with energy related regulation, indicating the applicability of those general regulation to the energy sector.

From Figure~\ref{fig:legislative process data-sharing}, we observe that Regulation (EU) 2022/868 and Regulation (EU) 2023/2854 are the most recent laws related to data-sharing. The two regulations are also the most pertinent ones since Regulation (EU) 2022/868 establishes a framework for managing, exchanging, and using data, and Regulation (EU) 2023/2854 deals with fair access to and use of data. In the next section, we will draw the most relevant regulatory definitions, concepts, objectives, and principles from those two regulations for comprehensive understanding of the regulatory framework for data-sharing in the energy sector.

\subsubsection{Objectives, basic definitions, concepts, and principles}

\begin{table*}[htb] 
	\centering \caption{\label{tbl:principles for data sharing}Principles data sharing with regulatory compliance applicable to the energy sector}
	\resizebox{0.7\textwidth}{!}{
		\begin{tabular}{ll}
			\cline{1-2}
			\multicolumn{2}{l}{\tabincell{l}{Principles and descriptions}}  \\

			\cline{1-2}
			\tabincell{l}{Contractual\\ freedom\cite{eckardt2024property}} & \tabincell{l}{Parties should remain free to negotiate the precise conditions for\\ making data available in their contracts within the framework for\\ the general access rules for making data available} \\
            \cline{1-2}
			\tabincell{l}{Open by\\ design and\\ by default\cite{mokobombang2020open}} & \tabincell{l}{Open data as a concept is generally understood to denote data in an\\ open format that can be freely used, re-used and shared by anyone\\ for any purpose. Open data policies which encourage the wide\\ availability and re-use of public sector information for private or\\ commercial purposes, with minimal or no legal, technical or financial\\ constraints, and which promote the circulation of information not only\\ for economic operators but primarily for the public, can play an\\ important role in promoting social engagement, and kick-start and\\ promote the development of new services based on novel ways to\\ combine and make use of such information.} \\
            \cline{1-2}
			\tabincell{l}{Fair, reasonable, and\\ non-discrimination\cite{suh2024non}} & \tabincell{l}{Where, in business-to-business relations, a data holder is obliged to\\ make data available to a data recipient under Article 5 Regulation\\ (EU) 2023/2854 or under other applicable Union law or national\\ legislation adopted in accordance with Union law, it shall agree with\\ a data recipient the arrangements for making the data available} \\
            \cline{1-2}
			\tabincell{l}{Transparency\cite{hacker2024regulating}} & \tabincell{l}{A recognised data altruism organisation shall keep full and accurate\\ records concerning personal data processing, and shall draw up\\ and transmit to the relevant competent authority for the registration\\ of data altruism organisations an annual activity report regarding\\ data processing.\\The data holder should provide to the data recipient sufficiently\\ detailed information for the calculation of the compensation.} \\
            \cline{1-2}
            
			\tabincell{l}{Respect for\\ fundamental rights\\ and freedoms} & \tabincell{l}{It is used as guiding principles in particular the Charter of Fundamental\\ Rights of the European Union~\cite{kellerbauer2024eu}, including the right to privacy, the\\ protection of personal data, the freedom to conduct a business,\\ the right to property and the integration of persons with disabilities} \\
            \cline{1-2}
            \tabincell{l}{Security\cite{DBLP:conf/re/RoulandGJ23}} & \tabincell{l}{Data intermediation services provider shall take necessary measures to\\ ensure an appropriate level of security for the storage, processing\\ and transmission of non-personal data, and the data intermediation\\ services provider shall further ensure the highest level of security\\ for the storage and transmission of competitively sensitive information} \\
            \cline{1-2}
			\tabincell{l}{Interoperability and\\ standardization\cite{hughes2023data}} & \tabincell{l}{Data intermediation services providers should be allowed to adapt the\\ data exchanged in order to improve the usability of the data by the\\ data user where the data user so desires, or to improve interoperability\\ by, for example, converting the data into specific formats.} \\
            \cline{1-2}
			\tabincell{l}{Data sovereignty\\ and user control\cite{von2024data}} & \tabincell{l}{In aggregating access to data so that big data analyses or machine\\ learning can be facilitated, users remain in full control of\\ whether to provide their data to such aggregation and the commercial\\ terms under which their data are to be used.} \\
		 
            \cline{1-2}
					
		\end{tabular}
	}
\end{table*}

The objectives of Regulation (EU) 2022/868 is to regulate the re-use, within the EU, of certain categories of data held by public sector bodies as well as the establishment of a notification and supervisory framework for the provision of data intermediation services, a framework for voluntary registration of entities which make data available for altruistic purposes and a framework for the establishment of a European Data Innovation Board, a commission expert
group that support the implementation of Regulation (EU) 2022/868. The objectives of Regulation (EU) 2023/2854 is to ensure fairness in the allocation of value from data among actors in the data economy and fostering fair access to and use of data in order to contribute to establishing a genuine internal market for data.

In the following, we present the basic definitions, concepts, objectives, and principles related to data-sharing from Regulation (EU) 2022/868 and Regulation (EU) 2023/2854. In Table~\ref{tbl:principles for data sharing}, we list the principles relating to data sharing in the energy sector. In Table~\ref{tbl:concepts/definitions for data sharing}, we list the regulatory definitions and concepts that are applicable to the energy sector, which are the basis for understanding the rights and obligations of different stakeholders regarding data-sharing in the energy system, as shown in Table~\ref{tbl:priciples, rights, obligations}.

\begin{table*}[htb] 
	\centering \caption{\label{tbl:concepts/definitions for data sharing}Basic concepts and definition related to data sharing according to Regulation (EU) 2022/868 and Regulation (EU) 2023/2854}
	\resizebox{0.8\textwidth}{!}{
		\begin{tabular}{ll}
			\cline{1-2}
			\multicolumn{2}{c}{\tabincell{l}{Definition/Concept and description}}  \\ 
		
			\cline{1-2}
			\tabincell{l}{Data} & \tabincell{l}{Any digital representation of acts, facts or information and any compilation of such\\ acts, facts or information, including in the form of sound, visual or audiovisual recording.} \\
			\cline{1-2}
			\tabincell{l}{Data subject} & \tabincell{l}{An identified or identifiable natural person (according to in Article 4, point (1), of\\ Regulation (EU) 2016/679(GDPR)\cite{sharma2019data})} \\   
            \cline{1-2}
			\tabincell{l}{Data holder} & \tabincell{l}{A legal person, including public sector bodies and international organisations, or a natural\\ person who is not a data subject with respect to the specific data in question, which, in\\ accordance with applicable Union or national law, has the right to grant access to or to\\ share certain personal data or non-personal data} \\
			\cline{1-2}
			\tabincell{l}{Data user} & \tabincell{l}{A natural or legal person who has lawful access to certain personal or non-personal data\cite{Irti2022}\\ and has the right, including under Regulation (EU) 2016/679 in the case of personal data,\\ to use that data for commercial or non-commercial purposes.} \\  
            \cline{1-2}
			\tabincell{l}{Data sharing} & \tabincell{l}{The provision of data by a data subject or a data holder to a data user for the purpose of\\ the joint or individual use of such data, based on voluntary agreements or Union or national\\ law, directly or through an intermediary, for example under open or commercial licences\\ subject to a fee or free of charge.} \\
			\cline{1-2}
			\tabincell{l}{Public sector\\ body} & \tabincell{l}{The State, regional or local authorities, bodies governed by public law or associations formed\\ by one or more such authorities, or one or more such bodies governed by public law.} \\  
            \cline{1-2}
			\tabincell{l}{Data intermediation\\ service} & \tabincell{l}{A service which aims to establish commercial relationships for the purposes of data sharing\\ between an undetermined number of data subjects and data holders on the one hand and\\ data users on the other, through technical, legal or other means, including for the purpose of\\ exercising the rights of data subjects in relation to personal data} \\
			\cline{1-2}

		\end{tabular}
	}
\end{table*}

\begin{table*}[htb] 
	\centering \caption{\label{tbl:priciples, rights, obligations}Rights and obligations of different roles involved in data sharing according to Regulation (EU) 2022/868 and Regulation (EU) 2023/2854}
	\resizebox{0.95\textwidth}{!}{
		\begin{tabular}{|l|l|l|l|}
			\cline{1-4}
   
			\tabincell{l}{Data holder} & \tabincell{l}{Data user} & \tabincell{l}{Data intermediaries} & \tabincell{l}{Public sector bodies} \\
    		\cline{1-4}

            \tabincell{l}{\textbf{Obligations}: Define transparent and objective\\ criteria for data access, avoiding exclusion\\ based on irrelevant factors. Offer comparable\\ terms and conditions across different groups.\\\textbf{Rights}: Set reasonable pricing and\\ usage conditions within market boundaries.\\\textbf{Obligations}: Assess the necessity and\\ proportionality of data requests. Provide data\\ access limited to the specific needs\\ and purpose outlined by the user.\\\textbf{Rights}: Seek clarification if requests seem\\ excessive or unclear.\\\textbf{Obligations}: Publish easily understandable\\ information about data access terms, pricing,\\ specifications, and quality in various formats\\ and languages. Communicate updates readily.\\\textbf{Rights}: Clarify ambiguities and address user \\inquiries regarding data access conditions.\\\textbf{Obligations}: Ensure personal data is only\\ shared in compliance with GDPR and other\\ relevant regulations. Implement appropriate\\ safeguards to protect privacy and security.\\\textbf{Rights}: Seek consent and inform\\ individuals about data use purposes\\ when handling personal data.\\\textbf{Obligations}: Implement appropriate technical\\ and organizational measures to ensure data\\ security, integrity, and confidentiality. Regularly\\ assess and update security controls.\\\textbf{Rights}: Choose appropriate technologies and\\ partners for data storage and processing.\\\textbf{Obligations}: Use open standards and\\ formats for data storage and sharing\\ whenever possible. Facilitate seamless data\\ exchange with other platforms and systems.\\\textbf{Rights}: Advocate for adoption of common\\ standards within their sectors or data ecosystems.\\\textbf{Obligations}: Act in a trustworthy manner,\\ fulfilling data access commitments and\\ demonstrating responsible data stewardship.\\ Be transparent about their data governance\\ practices and accountability mechanisms.\\\textbf{Rights}: Hold data intermediaries and\\ users accountable for any misuse or\\ non-compliance with data-sharing agreements.\\\textbf{Obligations}: Make data available in a sustainable\\ manner, considering long-term benefits and\\ potential negative impacts. Invest in data quality\\ and accessibility infrastructure.\\\textbf{Rights}: Benefit from economic returns generated\\ through responsible data-sharing practices.\\\textbf{Obligations}: Respect the data sovereignty of\\ individuals and organizations, providing clear\\ information about data ownership and control \\\textbf{Rights}: Determine the terms and conditions for\\ data access and sharing within the legal framework.} 
            & 
            \tabincell{l}{\textbf{Rights}: Request and access data\\ under defined fair and non\\discriminatory conditions.\\ Challenge unjustified access\\ restrictions.\\\textbf{Obligations}: Clearly specify the\\ intended purpose and scope of\\ data access requests. Avoid\\ requesting more data than strictly\\ necessary.\\\textbf{Rights}: Obtain justifications if\\ access requests are limited on\\ proportionality grounds.\\\textbf{Rights}: Access clear and readily\\ available information about data\\ access conditions before engaging\\ in data-sharing activities.\\\textbf{Obligations}: Only use data for\\ the purposes it was obtained and\\ in accordance with relevant\\ regulations. Respect individual\\ privacy rights and refrain from\\ unauthorized processing.\\\textbf{Rights}: Request access,\\ rectification, or erasure of their\\ personal data held by data\\ holders.\\\textbf{Obligations}: Implement\\ appropriate security measures to\\ protect accessed data from\\ unauthorized access or misuse.\\ Comply with any specific data\\ security requirements stipulated\\ by data holders.\\\textbf{Rights}: Report suspected\\ security vulnerabilities to data\\ holders or intermediaries.\\\textbf{Obligations}: Support adoption\\ of common data standards to\\ ensure smooth data transfer and\\ integration. Communicate\\ compatibility needs to data\\ holders and intermediaries.\\
            \textbf{Obligations}: Use data\\ responsibly and ethically,\\ adhering to the agreed-upon\\ purposes and data access\\ conditions. Report any misuse\\ or non-compliance to relevant\\ authorities.\\
            \textbf{Rights}: Exercise control over\\ their own data and make\\ informed decisions about\\ sharing it. Understand and\\ utilize mechanisms for managing\\ their data privacy and security.} 
            & 
            \tabincell{l}{\textbf{Obligations}: Facilitate fair and\\ non-discriminatory access mechanisms\\ within their platforms.\\
            \textbf{Rights}: Employ neutral algorithms and\\ procedures for user matching and data\\ access determination.\\\textbf{Obligations}: Display transparent\\ information about their platform's\\ terms, fees, and data quality standards.\\ Make clear the roles and responsibilities\\ of different actors involved.\\\textbf{Rights}: Seek additional information\\ from data holders if needed to ensure\\ transparency for users.\\\textbf{Obligations}: Ensure their platforms\\ handle personal data in accordance\\ with data protection regulations.\\ Provide tools and mechanisms for users\\ to exercise their data rights.\\\textbf{Rights}: Report data breaches and\\ privacy violations to relevant\\ authorities.\\\textbf{Obligations}: Provide secure\\ mechanisms for data access and transfer.\\ Implement robust cybersecurity\\ measures to protect data from\\ unauthorized access or breaches.\\\textbf{Rights}: Hold data holders accountable\\ for breaches or non-compliance with\\ security standards.\\\textbf{Obligations}: Develop platforms and\\ services compatible with widely\\ accepted data standards. Facilitate\\ data translation and conversion where\\ necessary.\\\textbf{Rights}: Propose and promote new\\ data standards for emerging use cases.\\\textbf{Obligations}: Establish clear\\ governance structures and operate\\ with transparency. Implement\\ mechanisms for dispute resolution and\\ user redress.\\\textbf{Rights}: Hold data holders\\ accountable for providing accurate\\ and reliable data and complying with\\ agreed-upon terms.\\
            \textbf{Rights}: Participate in initiatives to\\ define and implement sustainable\\ data governance practices.\\\textbf{Obligations}: Implement mechanisms\\ for users to exercise control over their\\ data, including access, rectification,\\ and erasure rights.\\
            } 
            & 
            \tabincell{l}{\textbf{Obligations}: obligated to make\\ high value datasets reusable\\ by potential re-users, following\\ the principle of open data by\\ default; Public sector bodies can\\ only refuse a request to re-use\\ data on specific grounds outlined\\ in the regulation. In such cases,\\ they must provide a clear and\\ reasoned justification for refusal;\\ When re-use of data cannot be\\ allowed due to legal restrictions,\\ the public sector body should try\\ their best to assist the potential\\ re-user in finding alternative\\ solutions, like anonymizing the\\ data or obtaining consent from\\ data subjects.\\\textbf{Obligations}: set clear\\ conditions for re-using their\\ data, but these conditions cannot\\ be discriminatory or excessively\\ restrict re-use. The conditions\\ should be transparent, \\documented, and objective.\\\textbf{Rights}: Public sector bodies can\\ anonymize data before sharing\\ it to address privacy concerns\\ or comply with data protection\\ regulations.\\\textbf{Obligations}: ensure the data\\ they make available for re-use\\ is of high quality, machine\\ readable, and easily accessible\\ through application \\programming interfaces (APIs)\\ or downloads.\\\textbf{Rights}: Public sector bodies\\ retain control over the data\\ they hold. They can decide\\ what data to make available\\ for re-use, set conditions for\\ re-use, and refuse requests\\ in specific situations outlined\\ by the regulation.} \\
            \cline{1-4}

		\end{tabular}
	}
\end{table*}

\subsubsection{Organizational structure for ensuring data-sharing}

\begin{figure}[htb]
	\centering
	\includegraphics[width=0.5\textwidth]{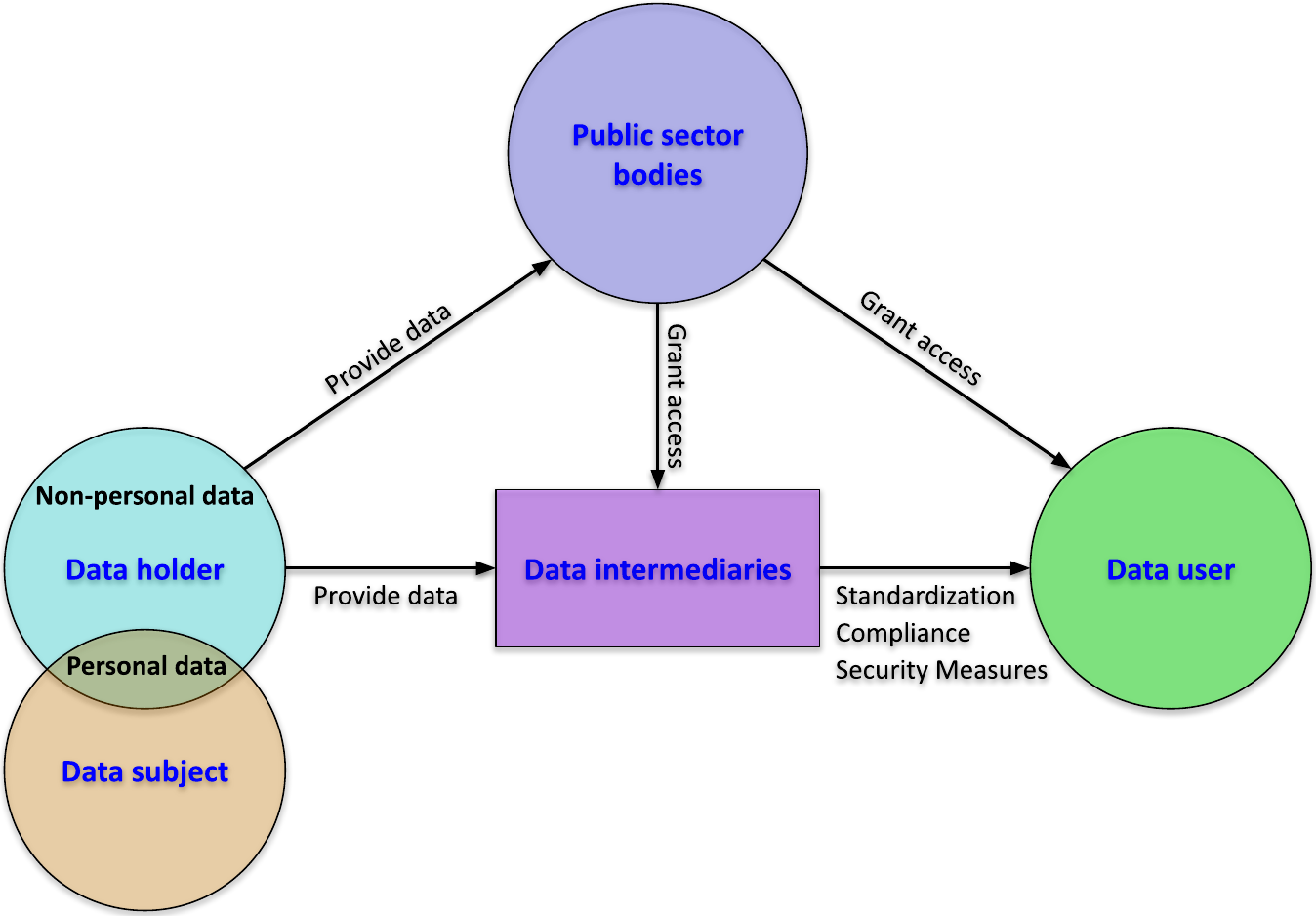}\\
	\caption{\label{fig:organization data-sharing}The organization structure of regulatory data-sharing in the EU according to the Data Governance Act.}
\end{figure}

According to Regulation (EU) 2022/868 (the Data Governance Act), we present the organizational structure for data-sharing in Figure~\ref{fig:organization data-sharing}.

In Figure~\ref{fig:organization data-sharing}, the public sector body plays the leading role in data-sharing by making data more open and accessible through various means, including (i) data-sharing services that offer dedicated platforms or services for data access and reuse (ii) Common European Data Spaces (CEDS) which participates in sector-specific data spaces with harmonized rules and infrastructure, which brings together various stakeholders within a specific sector to establish common technical infrastructure, governance frameworks, and data-sharing practices (iii) data altruism mechanisms that facilitate individuals or organizations to voluntarily contribute data for public interests. Public Sector bodies often take the initiative by making their data available and participating in CEDS. Data holders have rights to control their information and understand how it's being used, \textit{i.e.}, data holder has the legal right to grant access to or share specific data (both personal and non-personal). Public sector bodies might have legal requirements to access certain data from holders, following specific procedures and data protection regulations (\textit{e.g.}, GDPR). A data subject can also be a data holder when it comes to personal data. A data users can be businesses, research institutions, or even individuals interested in discovering, accessing, and utilizing relevant public sector data. Since sharing data can be complex due to technical issues, legal hurdles, and concerns about data security and privacy, data intermediaries are recognized by DGA and act as trusted third parties to facilitate data sharing between public sector bodies and data users. Data intermediaries can help data users find relevant datasets held by public sector bodies, ensure that data is formatted and presented in a way that is usable by different parties, guide both public sector bodies and data users on adhering to DGA regulations and data protection rules (like GDPR), implement technical safeguards to protect data privacy and security during the sharing process, \textit{etc.}, to streamline data-sharing process, make data more accessible to innovations, and promote a transparent and trustworthy data sharing environment.

\subsection{Regulatory framework for privacy in the energy system}

In this section, we describe the regulatory framework for privacy protection in energy sector in the EU and its practical implications. We review EU regulations and regulatory documents related to privacy protection in the energy sector. We reveal the dynamics and connections between different regulations over time to illustrate the evolution of the regulatory framework for privacy protection in the energy sector in the EU. Based on this review, we select the regulations most relevant to privacy protection in the energy system and elaborate the regulatory framework that applies to the evolving energy sector. 

\begin{figure*}[htb]
	\centering
	\includegraphics[width=0.67\textwidth]{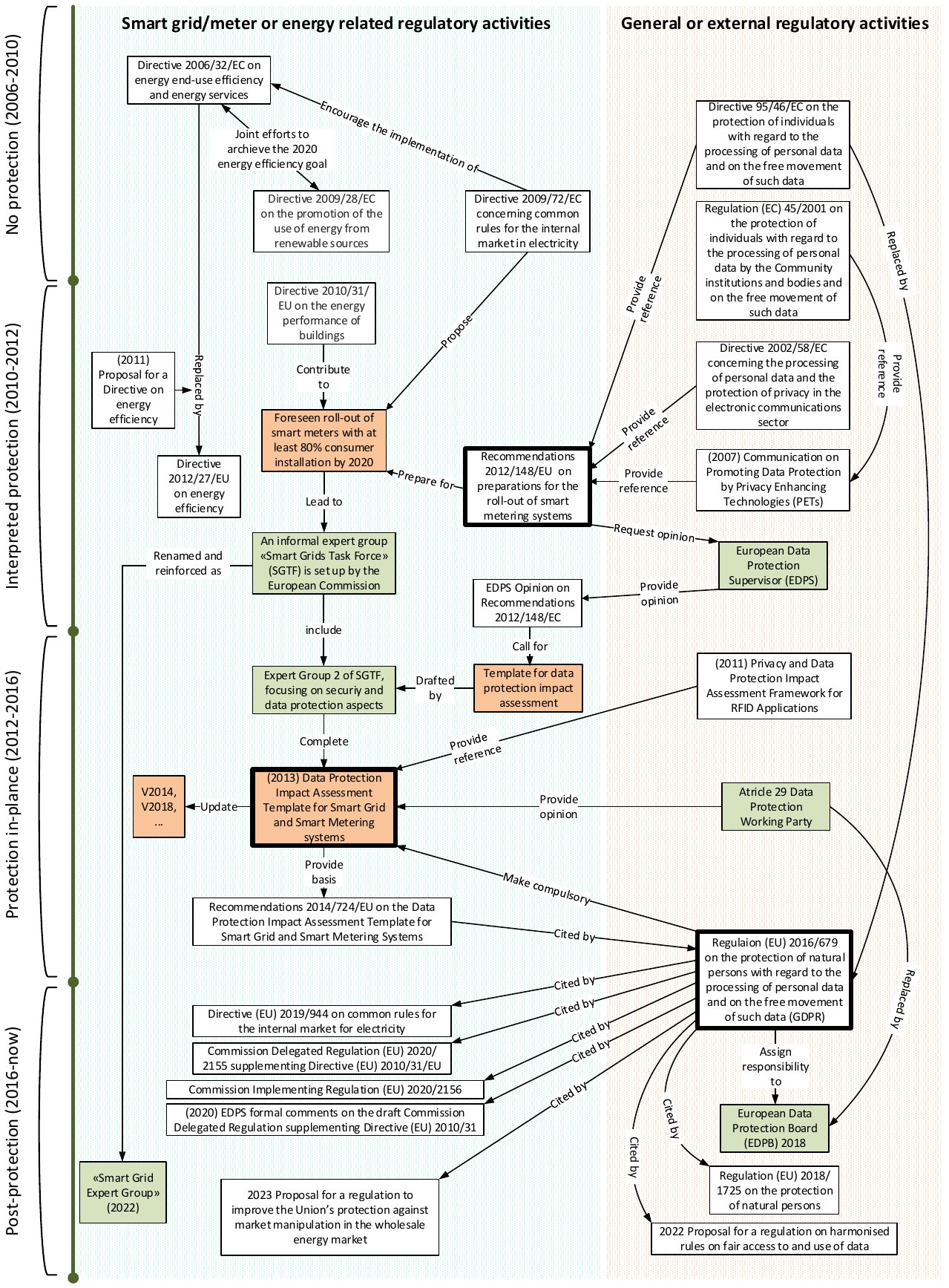}\\
	\caption{\label{fig:privacy_legislative_process}The evolution of the regulatory framework in EU regarding privacy protection in the energy sector, where the boxes in white indicate regulatory documents, the boxes in brown indicate concepts derived from the regulations, the boxes in green represent regulatory entities. The three boxes with thick edge lines are extensively interconnected to other regulations/entities, and thus are the focal point in the regulatory evolution.}
\end{figure*}

From the most pertinent regulations, we extract the critical definitions, concepts, and principles related to privacy preservation in the energy sector, elaborate the rights to be legally protected, and explain roles and responsibilities of the entities involved. We also present the organizational structure for this regulatory privacy protection that navigates the process for privacy protection for the key stakeholders in energy sector. 

\subsubsection{Regulations regarding regulatory privacy and their evolution}

In this section, we review privacy-protection related EU regulations from 2006 to 2024. The review covers both general regulations and those specific to the energy sector. We include critical regulatory groups and documents that contribute to the privacy regulation progress, and reveal the interactions and relationships amongst the involved elements in this progress. We partition the evolution of privacy related regulations over time, and provide a detailed analysis of how privacy regulation landscape has developed in the energy sector in Figure~\ref{fig:privacy_legislative_process}.

We partition the evolution of the regulatory progress in Figure~\ref{fig:privacy_legislative_process} into four stages: (i) no privacy protection from 2006 to 2010 (ii) interpreted protection from 2010 to 2012 (iii) protection in place from 2013 to 2016, and (iv) post-protection from 2017 to 2024. 

In the first stage (2006-2010), the energy sector began to introduce and use smart meters to promote energy usage monitoring and energy efficiency. Privacy issues associated with meter reading was not explicitly considered in this phase. In the second stage (2010-2012), privacy issues in the energy sector and thus challenges related to privacy protection started drawing attention due to the increasing adoption and installation of smart meters. However, this aspect was still not fully analyzed and regulated. Regulatory privacy protection in the energy sector mostly relied on interpreting existing general regulations. Towards the end of this stage, attempts to specify regulatory privacy protection for the energy sector began. In the third stage (2013-2016), Regulation (EU) 2016/679 (GDPR) was established in which regulatory privacy was fully incorporated. While GDPR is not dedicated to the energy sector, it is fully applicable. Moreover, there has been sufficient effort to contextualize GDPR in the energy sector by several regulatory working groups, \eg the Expert Group 2 in the Smart Grid Task Force. Such contextualization effort clarifies what is personal data and related regulatory roles in the energy sector, together with sufficient examples that provide regulatory certainty in privacy practice in the energy sector. Furthermore, GDPR makes the conducting of data protection impact assessment (DPIA) compulsory at specific conditions, and the existing DPIA template\footnote{DPIA by Expert Group 2 of Smart Grid Task Force, accessed Jan. 28, 2025\url{https://energy.ec.europa.eu/document/download/eee93bb8-1bda-4bdc-ac64-7edd6d0e60bc_en?filename=dpia_for_publication_2018.pdf}} by the Expert Group 2 connects GDPR provisions with energy sector with practical guidelines on how such a DPIA should be implemented by energy stakeholders. In the last stage (2017-2024), we note some improvement on regulatory privacy provisions provided by GDPR, and we observe frequent citations by related regulations, \eg on data-sharing, since different issues with secure data usage in the energy sector are interconnected. 

Based on the evolution of privacy regulations above, we focus on Regulation (EU) 2016/679 (GDPR) and the data protection impact assessment template developed by the Expert Group 2 in deriving our articulation of regulatory privacy in the energy sector.

\subsubsection{Objectives, basic definitions, and concepts in Regulation (EU) 2016/679}\label{section: privacy-objectives, definitions, principles}

The objective of Regulation (EU) 2016/679 is to protect the fundamental rights and freedom of natural persons and in particular their right to the protection of personal data and to ensure the free movement of personal data within the EU. There are two rights aimed to be protected: (i) Respect for private and family life\footnote{See interpretation of this right at \url{https://www.echr.coe.int/documents/d/echr/guide_art_8_eng}}, referring to Article 7, TITLE II, Charter of Fundamental Rights of the European Union (2012/C 326/02), and (ii) Protection of personal data\footnote{See interpretation of this right at \url{https://ec.europa.eu/justice/article-29/documentation/opinion-recommendation/files/2007/wp136_en.pdf}}, referring to Article 8, TITLE II, Charter of Fundamental Rights of the European Union (2012/C 326/02).

Table~\ref{tbl:concepts/definitions for privacy} provide the most relevant definitions and concepts of privacy protection to energy end-users and stakeholders. Since there exists a significant level of contextualizing GDPR in the energy sector, we provide exemplifications of those definitions for the energy sector. 

We note that the protection of personal data is at the center of the regulations for preserving individual privacy. While personal data is clearly defined to distinguish it from non-personal data in GDPR and personal data in the energy system is exemplified by regulatory groups, there is also the concept of sensitive data which is a special category of personal data. Special categories of personal data means personal data the processing of which reveals \textit{racial or ethnic origin, political opinions, religious or philosophical beliefs, or trade union membership, and the processing of genetic data, biometric data for the purpose of uniquely identifying a natural person, data concerning health or data concerning a natural person's sex life or sexual orientation (from Article 9(1), GDPR)}. Such data is under high level protection by GDPR, and processing special categories of personal data requires strict conditions. Such data has not been found in the current energy systems according to the DPIA template developed by the Expert Group 2, yet attention should be paid as new technologies and energy trading paradigms are evolving and expanding.

\begin{table*}[htb] 
	\centering \caption{\label{tbl:concepts/definitions for privacy}Basic regulatory concepts and definitions for privacy protection in the energy system according to Regulation (EU) 2016/679 (GDPR) and the DPIA template developed by the Expert Group 2 of Smart Grid Task Force.}
	\resizebox{0.95\textwidth}{!}{
		\begin{tabular}{lll}
			\cline{1-3}
			\multicolumn{2}{l}{\tabincell{l}{Definition/Concept and description}} & Exemplification  \\ 
		
			\cline{1-3}
			\tabincell{l}{Personal\\ data} & \tabincell{l}{Any information relating to an identified or identifiable\\ natural person (‘data subject’); an identifiable natural\\ person is one who can be identified, directly or indirectly,\\ in particular by reference to an identifier such as a name,\\ an identification number, location data, an online\\ identifier or to one or more factors specific to the physical,\\ physiological, genetic, mental, economic, cultural or social\\ identity of that natural person} 
            & 
            \tabincell{l}{Examples of \textbf{data subject}: Energy consumers, house\\ owners. Examples of \textbf{personal data} in the energy\\ sector: Consumer registration data like names, addresses;\\ Energy usage data (like household consumption, demand\\ information); Amount of energy provided to grid (energy\\ production); Profile of consumers; Facility operations\\ profile data (\textit{e.g.}, hours of use, number and type of\\ occupants); Frequency of transmitting data; Billing data\\ and consumer’s payment method; Bank account data (\textit{e.g.},\\ of natural persons running a photovoltaic systems;\\ Geolocation data. Examples of \textbf{non-personal data}: Locally\\ produced weather forecast, consumption prediction/forecasts;\\ Demand forecast of buildings; Anonymized consumption data;\\ Technical meter data, such as voltage curve at the meter.} \\
            \cline{1-3}
			\tabincell{l}{Data\\ controller} & \tabincell{l}{The natural or legal person, public authority, agency or\\ other body which, alone or jointly with others, determines\\ the purposes and means of the processing of personal\\ data; where the purposes and means of such processing\\ are determined by Union or Member State law, the\\ controller or the specific criteria for its nomination may\\ be provided for by Union or Member State law.} & \tabincell{l}{Examples of data controller in the energy sector include:\\ Energy suppliers, DSOs, Energy Regulators, Third party\\ service providers.} \\
            \cline{1-3}
			\tabincell{l}{Data\\ processor} & \tabincell{l}{A natural or legal person, public authority, agency or\\ other body which processes personal data on behalf of\\ the controller.} & \tabincell{l}{Examples of data processors in the energy sector include:\\ entities that the metering business is outsourced to (\eg\\ reading out meters, delivery of meter data to a DSO).} \\
            \cline{1-3}
			\tabincell{l}{Processing} & \tabincell{l}{Any operation or set of operations which is performed\\ on personal data or on sets of personal data, whether or\\ not by automated means, such as collection, recording,\\ organisation, structuring, storage, adaptation or\\ alteration, retrieval, consultation, use, disclosure by\\ transmission, dissemination or otherwise making available,\\ alignment or combination, restriction, erasure or\\ destruction.} & \tabincell{l}{Examples of processing personal data in the energy sector\\ include: Remote readings for billing purposes; Frequent\\ remote readings for network planning; Dynamic and\\ advanced tariffing that provide information to consumer\\ online (\eg Website, mobile App); Remote switching.} \\
            \cline{1-3}
			\tabincell{l}{Data\\ protection\\ officer} & \tabincell{l}{A person with expert knowledge of Data Protection law\\ and practices who advises the Data Controller or Data\\ Processor with the EU Data Protection regulation and\\ monitors internal compliance of the organization. Data\\ Protection officers, whether or not they are an employee\\ of the Data Controller, should be in a position to perform\\ their duties and tasks in an independent manner.} & \tabincell{l}{A data protection officer is designated case by case. It can\\ be a person in, \textit{e.g.}, an energy management company, grid\\ operator, utility company, who fulfills the responsibilities\\ and duties of a data protection officer, including overseeing\\ data protection practices, ensuring compliance with\\ regulations, and acting as a point of contact for data\\ protection matters, \textit{etc.} } \\
            \cline{1-3}
			\tabincell{l}{Supervisory\\ authority} & \tabincell{l}{An independent public authority which is established by\\ a Member State who oversees the operations of\\ energy companies and ensures compliance with energy\\ related regulations and policies} & \tabincell{l}{Examples of supervisory authority for the energy sector in\\ different countries: Norwegian Water Resources and Energy\\ Directorate (NVE)\tablefootnote{\url{https://www.nve.no/english/}, accessed Jan. 28, 2025.} in Norway, the Swedish Energy Agency,\\ a.k.a Energimyndigheten\tablefootnote{\url{https://www.energimyndigheten.se/en/}, accessed Jan. 28, 2025.} in Sweden, the Energy Authority,\\ known as Energiamarkkinavirasto\tablefootnote{\url{https://energiavirasto.fi/en/frontpage}, accessed Jan. 28, 2025.} in Finland, and the Energy\\ Agency (Energistyrelsen\tablefootnote{\url{https://ens.dk/en}, accessed Jan. 28, 2025.}) in Denmark.} \\
            \cline{1-3}
			\tabincell{l}{Data\\ protection\\ impact\\ assessment} & \tabincell{l}{A systematic process for evaluating the potential impact\\ of risks where processing operations are likely to present\\ specific risks to the rights and freedoms of data subjects\\ by virtue of their nature, their scope or their purposes to\\ be to carried by the controller or processor or the\\ processor acting on the controller’s behalf.} & \tabincell{l}{The Smart Grid Task Force maintains a template (Data\\ Protection Impact Assessment Template for Smart Grid\\ and Smart Metering systems\tablefootnote{\url{https://energy.ec.europa.eu/document/download/eee93bb8-1bda-4bdc-ac64-7edd6d0e60bc_en?filename=dpia_for_publication_2018.pdf.}, accessed Jan. 28, 2025.}) that can be used for energy\\ stakeholders.} \\
			\cline{1-3}

		\end{tabular}
	}
\end{table*}

\subsubsection{Principles, roles, rights, and obligations in processing personal data in the energy system}

We list and explain the principles relating to processing of personal data in Table~\ref{tbl:principles for privacy}, and we show the obligations and rights of related roles towards legal person data processing in Table~\ref{tbl:rights and obligations in personal data processing}. Combining Table~\ref{tbl:concepts/definitions for privacy} and Table~\ref{tbl:rights and obligations in personal data processing}, stakeholders in the energy sector might recognize its regulatory role in handling personal data and identify their corresponding rights and obligations as the basic step toward regulation compliance.

\begin{table*}[htb] 
	\centering \caption{\label{tbl:principles for privacy}Principles for personal data processing in the energy sector according to Regulation (EU) 2016/679 (GDPR)}
	\resizebox{0.7\textwidth}{!}{
		\begin{tabular}{ll}
			\cline{1-2}
			\multicolumn{2}{l}{\tabincell{l}{Principles and descriptions}}  \\ 
		
			\cline{1-2}
			\tabincell{l}{Lawfulness, fairness\\ and transparency} & \tabincell{l}{Personal data shall be processed lawfully, fairly and in a transparent manner\\ in relation to the data subject.} \\
			\cline{1-2}
			\tabincell{l}{Purpose limitation} & \tabincell{l}{Personal data shall be collected for specified, explicit and legitimate\\ purposes and not further processed in a manner that is incompatible\\ with those purposes; further processing for archiving purposes\\ in the public interest, scientific or historical research purposes\\ or statistical purposes shall, in accordance with Article 89(1) GDPR,\\ not be considered to be incompatible with the initial purposes.} \\   
            \cline{1-2}
            \tabincell{l}{Data minimisation} & \tabincell{l}{Personal data shall be adequate, relevant and limited to what is necessary\\ in relation to the purposes for which they are processed.} \\   
            \cline{1-2}
            \tabincell{l}{Accuracy} & \tabincell{l}{Personal data shall be accurate and, where necessary, kept up to date; \\every reasonable step must be taken to ensure that personal data that\\ are inaccurate, having regard to the purposes for which they are processed,\\ are erased or rectified without delay.} \\   
            \cline{1-2}
            \tabincell{l}{Storage limitation} & \tabincell{l}{Personal data shall be kept in a form which permits identification of data\\ subjects for no longer than is necessary for the purposes for which\\ the personal data are processed; personal data may be stored for longer\\ periods insofar as the personal data will be processed solely for\\ archiving purposes in the public interest, scientific or historical\\ research purposes or statistical purposes in accordance with Article 89(1)\\ subject to implementation of the appropriate technical and\\ organisational measures required by this Regulation in order to\\ safeguard the rights and freedoms of the data subject.} \\   
            \cline{1-2}
            \tabincell{l}{Integrity and\\ confidentiality} & \tabincell{l}{Personal data shall be processed in a manner that ensures appropriate\\ security of the personal data, including protection against unauthorised\\ or unlawful processing and against accidental loss, destruction or\\ damage, using appropriate technical or organisational measures.} \\   
            \cline{1-2}
            \tabincell{l}{Accountability} & \tabincell{l}{The controller shall be responsible for, and be able to demonstrate\\ compliance with data protection principles.} \\   
            \cline{1-2}
					
		\end{tabular}
	}
\end{table*}

\begin{table*}[htb] 
	\centering \caption{\label{tbl:rights and obligations in personal data processing}Rights and obligations in personal data processing in the energy sector according to Regulation (EU) 2016/679 (GDPR).}
	\resizebox{0.98\textwidth}{!}{
		\begin{tabular}{|l|l|l|l|}
			\cline{1-4}
   
			 \tabincell{l}{Data subject} & \tabincell{l}{Data controller} & \tabincell{l}{Data processor} & \tabincell{l}{Data protection officer} \\
    		\cline{1-4}
      
            \tabincell{l}{\textbf{Rights}:\\Right to be informed (transparency).~\\Right of access.~\\Right to rectification~\\Right to erasure (right to be forgotten).~\\Restriction of processing.~\\Right to data portability (in \eg\\ machine-readable format).~\\Right to object.~\\Right not to be subject to a decision\\ based solely on automated processing} & \tabincell{l}{\textbf{Obligations}:\\Practice data protection by design\\ and by default.~\\Maintain a record of processing\\ activities.~\\Implement appropriate technical\\ and organisational measures\\ to ensure a level of security.~\\Notify a personal data breach to\\ the supervisory authority.~\\Communicate a personal data\\ breach to the data subject.~\\Document any personal data\\ breaches.~\\Carry out data protection impact\\ assessment and prior consultation.~\\Designate a data protection officer} & \tabincell{l}{\textbf{Obligations}:\\Maintain a record of all categories\\ of processing activities carried out\\ on behalf of a controller.~\\Implement appropriate technical\\ and organisational measures to\\ ensure a level of security.~\\Notify the controller without undue\\ delay after becoming aware of a\\ personal data breach.~\\Designate a data protection officer} & \tabincell{l}{\textbf{Obligations}:\\Inform and advise the controller or\\ the processor and the employees who\\ carry out processing of their obligations\\ and to other Union or Member State\\ data protection provisions.~\\Monitor compliance with GDPR.~\\Provide advice where requested as\\ regards the data protection impact\\ assessment and monitor its performance.~\\Cooperate with the supervisory authority.~\\Act as the contact point for the\\ supervisory authority} \\

			\cline{1-4}
		\end{tabular}
	}
\end{table*}

\subsubsection{Organizational structures for implementing privacy protection} 

\begin{figure}[htb]
	\centering
	\includegraphics[width=0.4\textwidth]{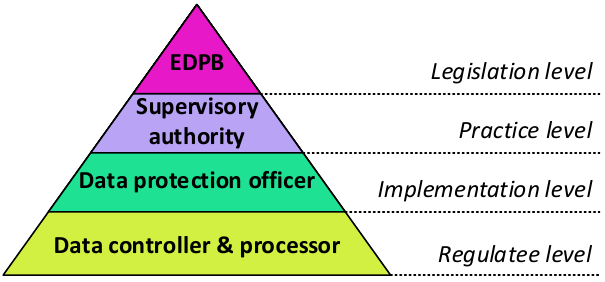}\\
	\caption{\label{fig:organizational structure}Organizational structure of privacy protection (EDPB: the European Data Protection Board, or ``the Board") according to GDPR. Note that this structure is not specified for the energy sector but fully applicable.}
\end{figure}

The organization structure for personal data protection practice involves four roles: the European Data Protection Board (EDPB), supervisory authority, data protection officer, and data controller and processor. Each of those roles has their position in the organization structure as shown in Figure~\ref{fig:organizational structure}. This structure contributes to the systematic compliance of GDPR and an organized implementation of personal data protection. Such an organizational effort is a critical and complimentary element to regulatory and technical efforts towards personal data protection.

In Figure~\ref{fig:organizational structure}, EDPB acts as the central EU body at the legislation level, ensuring consistent application of GDPR across member states. EDPB issues guidelines and recommendations to ensure consistent interpretation and application of GDPR by the Supervisory Authorities. EDPB also resolves disputes between Supervisory Authorities from different EU countries regarding cross-border data processing. Supervisory authorities is at the practice level, who are responsible to enforce GDPR within individual EU member states. They are responsible to monitor relevant developments that have an impact on the protection of personal data. They also establish and maintain a list in relation to the requirement for data protection impact assessment. Data protection officers are at the implementation level to inform, advise, and monitor the compliance of data controllers and data processors with GDPR. Data protection officers are also responsible to monitor the implementation of data protection impact assessment by data controllers or processors. Data controllers and processors are at the regulatee level who are responsible to take measures to secure personal data processing and ensure the rights of data subjects. They are obligated to report personal data breach and document the processing of personal data to offer transparency.

\subsection{Regulatory framework for cyber security in the energy system}

In this section, we describe regulatory cyber security in energy sector in the EU. We review EU regulations and regulatory documents related to cyber security. We reveal the dynamics and interactions between different regulations over time to illustrate the evolution of regulatory cyber security in the energy sector. Based on this review, we select the most relevant regulations, and articulate on the regulatory cyber security in the energy sector pertaining to those regulations.

From the most pertinent regulations, we extract the critical definitions, concepts, and principles in cyber security in the energy sector, clarify the rights and obligations, and explain roles involved in cyber security. We show the organizational structure for regulatory cyber security practice in energy activities.

\subsubsection{Related regulations regarding cyber security and their evolution}

This section reviews cyber security related regulations applicable to the energy sector from 1995 to 2024. From the reviewed regulations we note that beyond general regulations, there are also regulations dedicated specifically to the energy sector. The general and the specific regulations are connected, and they facilitate extensive interactions among the key stakeholders in terms of their roles, rights, responsibilities, \textit{etc}. We sort the relevant regulations and important documents and entities involved in the regulatory progress by timeline, and analyze how this progress evolves. As shown in Figure~\ref{fig:legislative process cybersecurity}, we partition the development of cyber security regulations into three periods as follows:

\begin{figure*}[htb]
	\centering
	\includegraphics[width=0.85\textwidth]{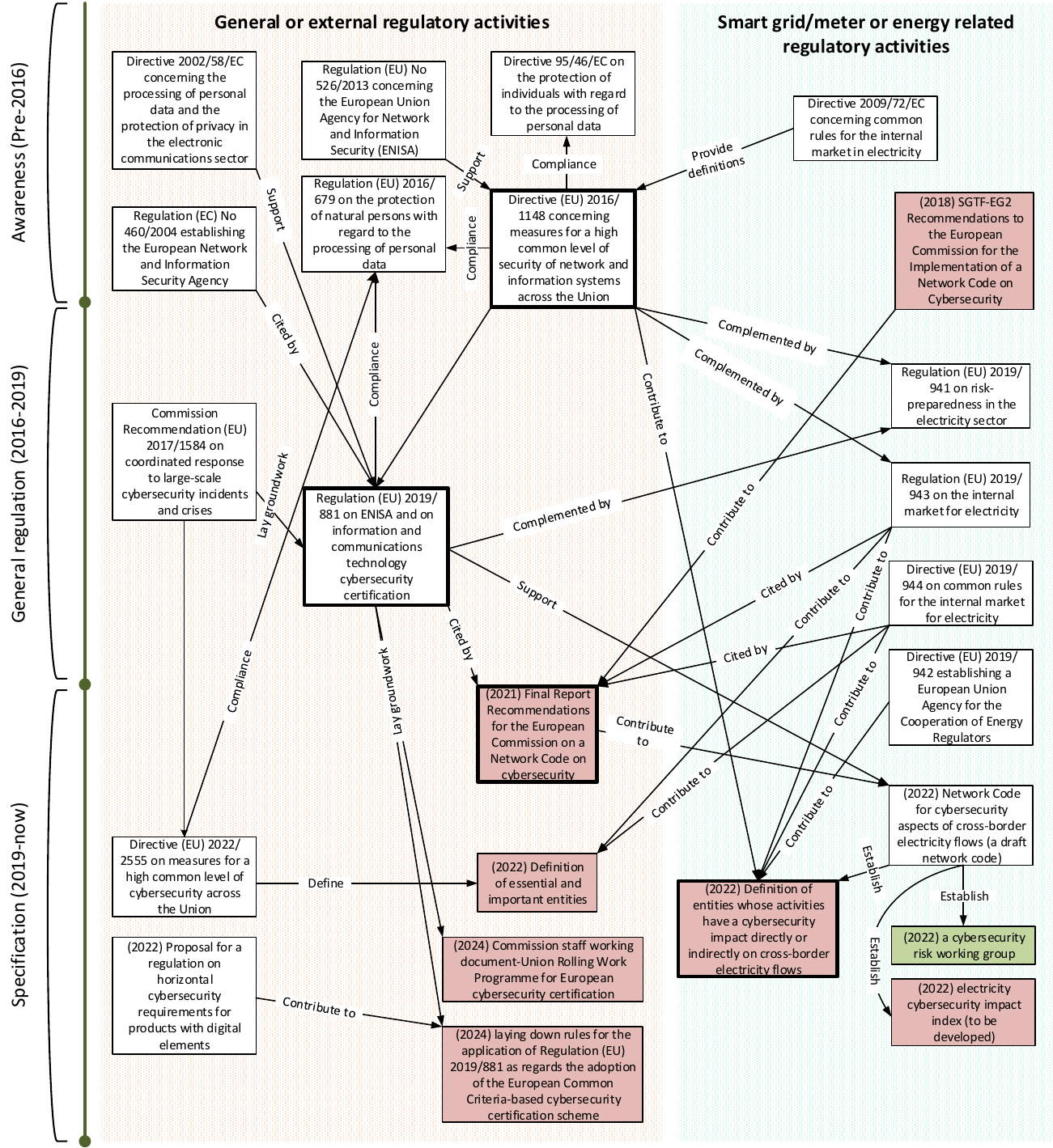}\\
	\caption{\label{fig:legislative process cybersecurity}The EU legislative process regarding cyber security in the energy system. The boxes with thick edge lines are extensively interconnected to other regulations/entities, and thus are the focal point in the regulatory evolution.}
\end{figure*}

\begin{enumerate}
    \item Awareness of cyber security from 1995 to 2016.
    \item Emergence of general regulations towards cyber security from 2016 to 2019
    \item Specification of cyber security in energy regulations from 2019 to 2024
\end{enumerate}

In the first stage (1995-2016), the increasing impact of security issues in network and information system draws attention. Regulations are proposed to lay down measures to achieve a high common level of security of network and information systems. During this period, we also observe the acknowledgement of data protection regulations by cyber security related regulations, implying the interconnection between different data-related issues. In the second stage (2016-2019), the issue of cyber security is explicitly addressed in the regulations side, and general cyber security regulations, \eg Regulation (EU) 2019/881 (Cybersecurity Act) emerge that are applicable to the energy sector. From the beginning of the third stage (2019-2024), the interactions between general cyber security regulations and energy related regulations become intensive, and increasing efforts specific to energy activities trigger the specification of cyber security regulations in the energy sector. We can see from Figure~\ref{fig:legislative process cybersecurity} that the third stage is still evolving, with field-specific regulation proposal available, yet the anticipated regulation and critical document dedicated to the energy cyber security are still under preparation. 

Based on the above analysis, we select and elaborate on Regulation (EU) 2019/881 and Directive (EU) 2022/2555 (a.k.a. NIS 2 Directive), which are the latest and the most pertinent regulations to cyber security, and we use those two regulations to articulate regulatory cyber security in the energy sector.

\subsubsection{Objectives, basic definitions, concepts, principles, rights and obligations for regulatory cyber security}

The objectives of Regulation (EU) 2019/881 are (i) to achieve a high common level of cyber security across the European Union, and (ii) to create a framework for European cyber security certification schemes. The objectives of Directive (EU) 2022/2555 are (i) to establish a high common level of cyber security across the EU by setting out minimum cyber security risk management requirements for relevant entities (ii) to improve cooperation among member states through facilitating information sharing and coordinated incident response (iii) to enhance cyber security risk management for essential and important entities, and (iv) to strengthen incident reporting toward cyber security issues.

Table~\ref{tbl:concepts/definitions for cybersecurity} provides the most relevant definitions and concepts for regulatory cyber security to energy stakeholders. We also provide exemplifications of those regulatory definitions for the energy sector in the table. Table~\ref{tbl:principles for cyber-security} lists the principles relating to regulatory cyber security in the energy sector. Table~\ref{tbl:rights and obligations in regulatory cybersecurity} provides the roles along with their obligations/rights toward regulatory cyber security in the energy sector.

\begin{table*}[htb] 
	\centering \caption{\label{tbl:concepts/definitions for cybersecurity}Basic concepts and definition in regulatory framework for cyber-security in the energy system, according to Regulation (EU) 2019/881 and Directive (EU) 2022/2555.}
	\resizebox{0.95\textwidth}{!}{
		\begin{tabular}{lll}
			\cline{1-3}
			\multicolumn{2}{l}{\tabincell{l}{Definition/Concept and description}} & Exemplification  \\%
		
			\cline{1-3}
			\tabincell{l}{cyber security} & \tabincell{l}{The activities necessary to protect network and\\ information systems, the users of such systems,\\ and other persons affected by cyber threats.} & \tabincell{l}{Implementing access control for information in the\\ energy system, power grid monitoring, \textit{etc}.} \\
            \cline{1-3}
			\tabincell{l}{Cyber threat} & \tabincell{l}{Any potential circumstance, event or action that\\ could damage, disrupt or otherwise adversely impact\\ network and information systems, the users of such\\ systems and other persons.} & \tabincell{l}{Examples of cyber security threat in the energy\\ ystem\footnote{Drawn from the report for the Network Code on cyber security\url{https://energy.ec.europa.eu/system/files/2021-04/nccs_report_network_code_on_cybersecurity_0.pdf}}:\\The unauthorised access, unauthorized modification, or\\ unplanned availability of critical data to key systems,\\ \textit{e.g.}, Energy Management Systems (EMS), Distribution\\ Management Systems (DMS), Day ahead forecast schedules\\ (Market), Common Grid Model (CGM).\\Unauthorized access to and simultaneous control over\\ many the same IOT devices which might potentially\\ impact the security of supply to X\footnote{This threshold might be determined by\url{https://eepublicdownloads.entsoe.eu/clean-documents/SOC\%20documents/Incident_Classification_Scale/IN_USE_FROM_JANUARY_2020_191204_Incident_Classification_Scale.pdf}} households.\\Exploits of a serious vulnerability in common\\ equipment purchased and used by multiple grid\\ participants.} \\
            \cline{1-3}
            \tabincell{l}{National\\ Competent\\ Authority\\ (NCA)} & \tabincell{l}{Organizations designated by individual EU member\\ states to oversee the implementation and enforcement\\ of specific EU regulations within their national\\ borders. These organizations have the legal authority\\ to grant approvals, conduct inspections, and ensure\\ compliance with the regulations.} & \tabincell{l}{NCA for energy sector in Sweden: Energimyndigheten\\ (The Swedish Energy Agency). NCA for energy sector\\ in Finland: Energivirasto (Energy Authority).}\\
            \cline{1-3}
			\tabincell{l}{Essential entities} & \tabincell{l}{Operators of essential services, and entities identified\\ as critical entities under Directive (EU) 2022/2557,\\ and providers that are designated as very large online\\ platforms within the meaning of Article 33 of\\ Regulation (EU) 2022/2065} & \tabincell{l}{Entities exceed the ceilings for medium-sized enterprises.~\\Qualified trust service providers, regardless of their size.~\\Providers of public electronic communications\\ networks or of publicly available electronic communications\\ services which qualify as medium-sized enterprises.~\\Entities identified as critical entities referred\\ to in Article 2(3) of Directive (EU) 2022/2555} \\
            \cline{1-3}
			\tabincell{l}{Important entities} & \tabincell{l}{Entities of a type referred to in Annex I or II of\\ Directive (EU) 2022/2555 which do not qualify as\\ essential entities shall be considered to be important\\ entities.} & \tabincell{l}{The identifying of important entities is still in progress.\\ Possible examples of important entities can be companies\\ that operate power plants and storage facilities, Retailers\\ and energy service companies, Data providers and\\ aggregators, Market operators and balance responsible\\ parties, \textit{etc}.} \\
            \cline{1-3}
			\tabincell{l}{ICT product} & \tabincell{l}{An element or a group of elements of a network or\\ information system} & \tabincell{l}{Smart meters, energy management systems, supervisory\\ control and data acquisition (SCADA) systems, \textit{etc}.} \\
            \cline{1-3}
			\tabincell{l}{ICT service} & \tabincell{l}{A service consisting fully or mainly in the transmission,\\ storing, retrieving or processing of information\\ by means of network and information systems} & \tabincell{l}{Remote monitoring of grid infrastructure, demand\\ response management services, \textit{etc}.} \\
			\cline{1-3}
            \tabincell{l}{ICT process} & \tabincell{l}{A set of activities performed to design, develop,\\ deliver or maintain an ICT product or ICT service} & \tabincell{l}{Development of algorithms for energy trading and market\\ optimization, software development for smart meter data\\ collection and analysis} \\
			\cline{1-3}

		\end{tabular}
	}
\end{table*}

\begin{table*}[htb] 
	\centering \caption{\label{tbl:principles for cyber-security}Principles for regulatory framework for cyber security in the energy sector according to Regulation (EU) 2019/881 and Directive (EU) 2022/2555}
	\resizebox{0.7\textwidth}{!}{%
		\begin{tabular}{ll}
			\cline{1-2}
			\multicolumn{2}{l}{\tabincell{l}{Principles and descriptions}}  \\%
		
            \cline{1-2}
            \tabincell{l}{Security by design} & \tabincell{l}{Organisations, manufacturers or providers involved in the design\\ and development of ICT products, ICT services or ICT\\ processes should be encouraged to implement measures at the\\ earliest stages of design and development to protect the\\ security of those products, services and processes to the\\ highest possible degree, in such a way that the occurrence of\\ cyberattacks is presumed and their impact is anticipated and\\ minimised} \\
            \cline{1-2}
            \tabincell{l}{Security by default} & \tabincell{l}{Undertakings, organisations and the public sector should configure the\\ ICT products, ICT services or ICT processes designed by them\\ in a way that ensures a higher level of security which should\\ enable the first user to receive a default configuration with the\\ most secure settings possible} \\
            \cline{1-2}            
			\tabincell{l}{Data protection principles} & \tabincell{l}{It means the principles of data accuracy, data minimisation, fairness\\ and transparency, and data security} \\
		 
            \cline{1-2}
					
		\end{tabular}%
	}
\end{table*}

\begin{table*}[htb] 
	\centering \caption{\label{tbl:rights and obligations in regulatory cybersecurity}Rights and obligations in cyber security in the energy sector according to Regulation (EU) 2019/881 and Directive (EU) 2022/2555.}
	\resizebox{0.9\textwidth}{!}{%
		\begin{tabular}{|l|l|}
			\cline{1-2}
   
			 \tabincell{l}{Essential entities} & \tabincell{l}{Important entities}  \\
    		\cline{1-2}
      
            \tabincell{l}{\textbf{Obligations}:~\\Subject to on-site inspections and off-site\\ supervision, including random checks conducted by\\ trained professionals.~\\Subject to regular and targeted security audits\\ carried out by an independent body or a competent authority.~\\Subject to ad hoc audits, including where justified\\ on the ground of a significant incident or an\\ infringement of this Directive by the essential\\ entity.~\\Subject to security scans based on objective,\\ non-discriminatory, fair and transparent\\ risk assessment criteria, where necessary with the\\ cooperation of the entity concerned.~\\Subject to requests for information necessary to assess\\ the cyber security risk-management measures adopted by\\ the entity concerned, including documented\\ cyber security policies, as well as compliance\\ with the obligation to submit information to the competent\\ authorities.~\\Subject to requests to access data, documents and\\ information necessary to carry out competent authority's\\ supervisory tasks.~\\Subject to requests for evidence of implementation\\ of cyber security policies, such as the results of\\ security audits carried out by a qualified auditor and\\ the respective underlying evidence.} & \tabincell{l}{\textbf{Obligations}:~\\Subject to on-site inspections and off-site ex post\\ supervision conducted by trained professionals.~\\Subject to targeted security audits carried out\\ by an independent body or a competent authority.~\\Subject to security scans based on objective,\\ non-discriminatory, fair and transparent\\ risk assessment criteria, where necessary with the\\ cooperation of the entity concerned.~\\Subject to requests for information necessary to assess,\\ ex post, the cyber security risk-management measures\\ adopted by the entity concerned, including documented\\ cyber security policies, as well as compliance with the\\ obligation to submit information to the\\ competent authorities.~\\Subject to requests to access data, documents and\\ information necessary to carry out their supervisory\\ tasks.~\\Subject to requests for evidence of implementation\\ of cyber security policies, such as the results of security\\ audits carried out by a qualified auditor and the\\ respective underlying evidence.}  \\

			\cline{1-2}
		\end{tabular}%
	}
\end{table*}

\subsubsection{Organizational structure for ensuring regulatory cyber-security}

\begin{figure}[htb]
	\centering
	\includegraphics[width=0.48\textwidth]{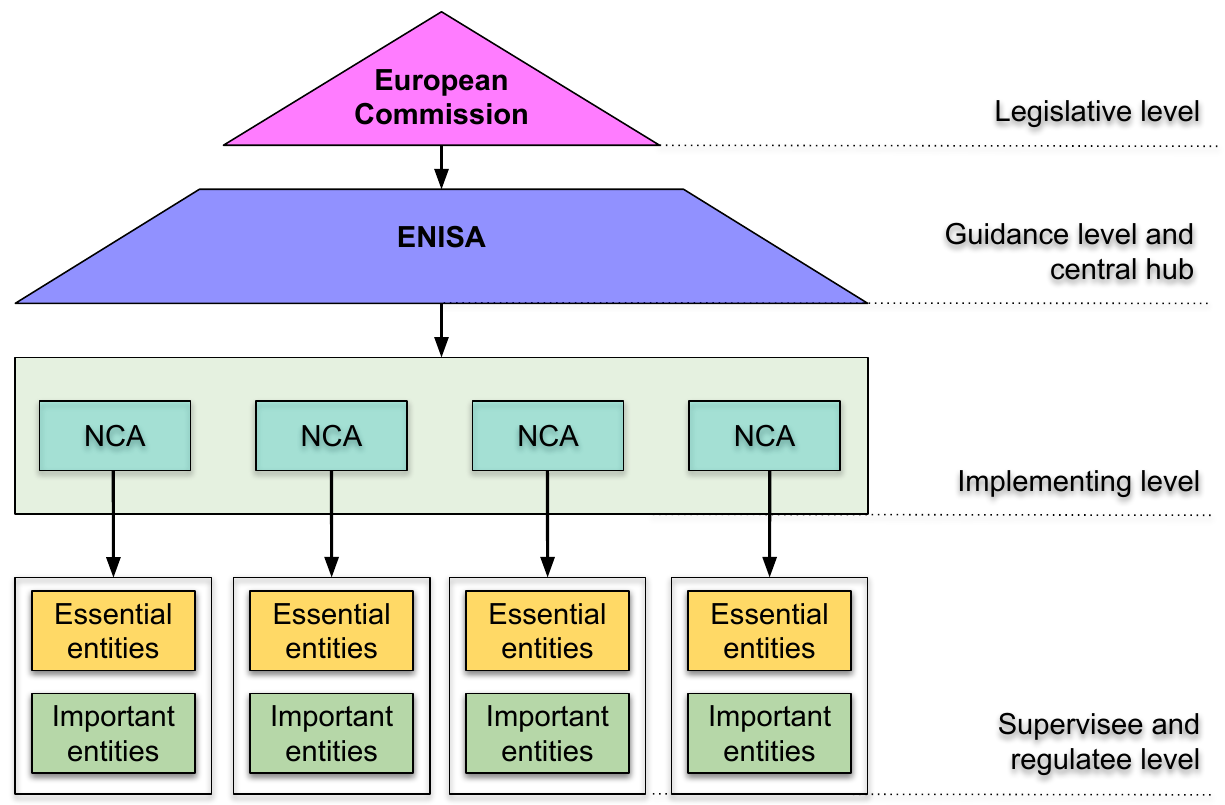}\\
	\caption{\label{fig:cybersecurity organizational structure}Organizational structure of regulatory framework for cyber security. Note that this structure is extracted based on Regulation (EU) 2019/881 and Directive (EU) 2022/2555, which is not specified for the energy sector but fully applicable to the energy sector.}
\end{figure}

According to Regulation (EU) 2019/881 and Directive (EU) 2022/2555, we depict the organizational structure for the regulatory cyber security in Figure~\ref{fig:cybersecurity organizational structure}.

In Figure~\ref{fig:cybersecurity organizational structure}, the European Commission (EC) on the top is responsible for developing and proposing cyber security regulations, and overseeing the implementation of regulations in member states. The European Union Agency for cyber security (ENISA) acts as the central hub for knowledge sharing, expertise, and technical assistance, and develops and disseminates best practices for cyber security measures that member states can leverage. The member states implement the regulations, and each member state appoints a National Competent Authority (NCA) with enforcement powers to oversee compliance with Directive (EU) 2022/2555 in their country. NCAs are responsible for implementing the regulations within their national context, and they report to and collaborate with ENISA on EU-wide initiatives, and work with sectors obligated to implement risk management within their country. The essential entities and important entities, which are organizations in critical sectors designated by Directive (EU) 2022/2555, are responsible for conducting risk assessments, implementing security measures, and reporting incidents. They are directly subject to the regulations on cyber security and responsible for compliance via carrying out their obligations.

\subsubsection{Cybersecurity certification}

\begin{figure}[htb]
	\centering
	\includegraphics[width=0.37\textwidth]{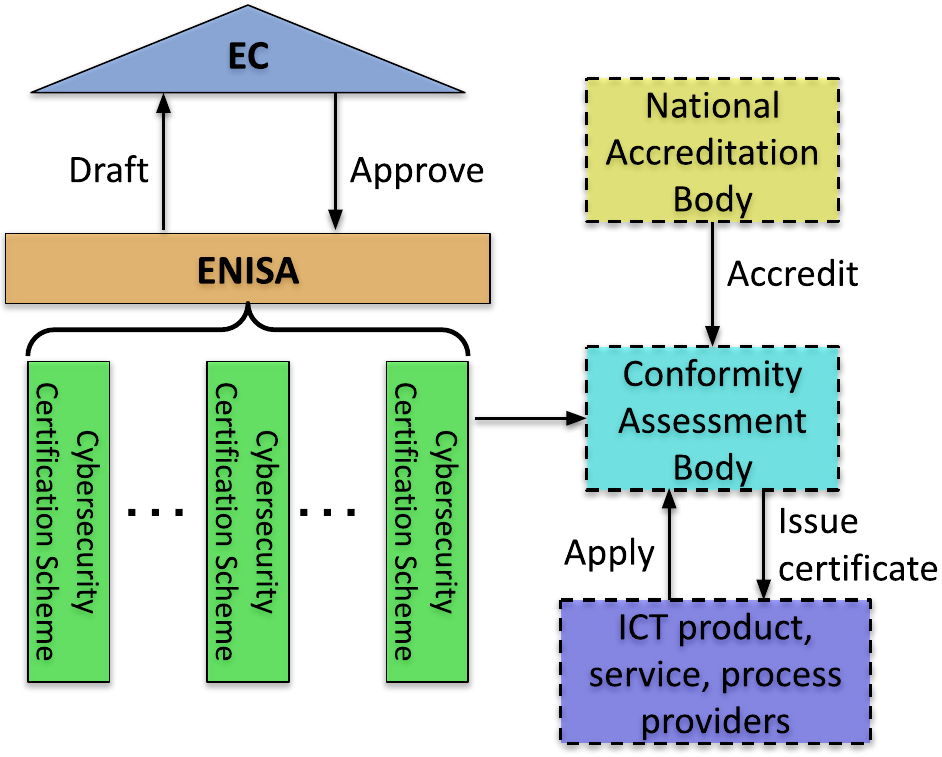}\\
	\caption{\label{fig:cybersecurity_certificate}Organizational structure for the cybersecurity certificate process according to Cyber-security Act.}
\end{figure}

The ``cybersecurity certification'' stipulated in Regulation (EU) 2019/881 also leads to organizational structure that promote cyber security in practice, and we describe the corresponding structure in Figure~\ref{fig:cybersecurity_certificate}.

In Figure~\ref{fig:cybersecurity_certificate}, the European Union Agency for Cybersecurity (ENISA) drafts ``candidate cybersecurity certification schemes'' for the cybersecurity certification. The drafting of the schemes is conducted by a group of experts (Ad-Hoc Working Group) along with the European Commission, EU countries, and relevant stakeholders. The European Commission (EC) defines the requirements and has the authority to adopt the proposed cybersecurity certification schemes. The providers of ICT product, service, or process then have the options to choose a relevant cybersecurity certification schemes which they want to apply for a cybersecurity certificate. The application from the providers will be assessed by a Conformity Assessment Body (CAB), and the provider will be issued the cybersecurity certificate by the CAB if the assessment is approved. Note that the CAB is accredited by a National Accreditation Body (NAB), which is responsible to accredit the CAB to ensure they meet competency standards for conducting assessments.

\section{Reflection: pros-and-cons, barriers, challenges, and guidelines}\label{sec:reflection}

\subsection{Contextualization and specification of regulations in the energy system}

\subsubsection{Data-sharing}

The contextualization of general EU regulations on data-sharing is still at a low level. Their is no regulation specified for data-sharing in the energy sector, nor working groups dedicated to contextualizing data-sharing regulations for the energy sector. We note the establishment of the ``Common European Energy Data Space'' triggered by EU data-sharing related regulation, which is a field-specific effort for data-sharing in the energy sector. This energy data space is anticipated to promote data access and enable smooth data exchange. However, such an effort does not directly specify and interpret how end-users and other stakeholders should behave towards data-sharing from a regulatory perspective.

\subsubsection{Privacy}

While there is no EU regulations specified for privacy protection in the energy sector, general EU regulations exist that are fully applicable. This is achieved through the combination of field legislative documents by dedicated regulatory bodies and general privacy protection regulations. The field legislative documents contextualize the concepts in general regulations, thus translating regulatory privacy and data protection to the energy sector. In return, the general privacy regulations refer those field documents and make them mandatory in privacy practice. In this way, it offers to the stakeholders in the energy sector the steps to follow for clear privacy regulation compliance.

\subsubsection{Cyber security}

The Network Code on Cybersecurity is specified for the energy sector, and other general regulations are also available in EU applicable to cyber security in the energy sector. However, the regulation provisions do not fully address the need of energy stakeholders. Particularly, there is a lack of established cyber security impact risk matrix for the energy sector, though mentioned in regulatory documents. Such a document is of critical importance to bridge regulatory articles and practice, which is expected to provide clarification of exemplified regulatory definitions and concepts and field guidelines on who is responsible to implement what measures under what conditions. While similar document is established for privacy, \textit{i.e.}, the data protection impact assessment template created by the the Expert Group 2 of Smart Grid Task Force, it is still absent in cyber security regulations in the energy sector.

\subsection{Open challenges with emerging scenarios and applications in the evolving energy sector}

\subsubsection{Independent aggregators}

The role of independent aggregators has drawn increasing attention in recent years~\cite{doi:10.1080/09640568.2022.2027233}, and independent aggregators in the energy market will empower more innovations, \textit{e.g.}, peer-to-peer (P2P) energy trading and Blockchain-based transactive energy systems~\cite{berntzen2021blockchain}. In fact, aggregator is viewed as a critical element when designing P2P energy trading mechanisms~\cite{WANG2023121550}. While P2P energy trading and transactive energy systems have mainly been limited to pilot projects so far, these are the emerging distributed paradigms with great potential that can create and facilitate innovative business models and can indeed be lucrative for the end users, the service providers, the energy market and also for the grid. 

However, the aggregators' legal access to the energy market is limited due to various challenges~\cite{bray2019barriers,SCHITTEKATTE2021106971}. \textit{E.g.}, when it comes to the access to the market, several EU countries seem to struggle with
implementing independent aggregators' access to wholesale market. Belgium has enabled such access with limited success in practice. France has facilitated such access but only for demand turndown with restrictions on baselines\footnote{\url{https://pub.norden.org/nordicenergyresearch2022-04/nordicenergyresearch2022-04.pdf}, accessed Jan. 28, 2025.}. Without access to the energy market, aggregators face challenges in terms of data connection and communication, and energy digitalization and decentralization. 

Numerous pilot projects have demonstrated the potential and feasibility of integrating aggregators in the energy market. The recent amendment of Electricity Regulation (2019/943)\footnote{\url{https://eur-lex.europa.eu/legal-content/EN/TXT/PDF/?uri=CELEX:02019R0943-20220623}, accessed Jan. 28, 2025.} and Electricity Directive (2019/944)\footnote{\url{https://eur-lex.europa.eu/legal-content/EN/TXT/PDF/?uri=CELEX:02019L0944-20220623}, accessed Jan. 28, 2025.} allows independent aggregators with specific licenses to directly buy and sell electricity in the wholesale market. This is a significant change and opens up new opportunities for aggregators. In this regard, it is imperative that a regulatory framework that supports a higher level of legal access to aggregators can essentially contribute to empower more data connections and communications, which can encourage the integration of distributed energy system framework and the energy transition at large.

The realisation of independent aggregators can result in more data interfaces that are necessary for operation and maintenance purposes. This also implies new forms of vulnerabilities in data-sharing, privacy, and cyber security. Due awareness and regulatory measures against potential vulnerabilities can help integrate new innovations and mitigate risk, \textit{e.g.}, with the establishment of a regulatory document for impact assessment of integrating aggregators.

\subsubsection{Vehicle to grid (V2G) communications and electricity exchange}

The regulatory framework regarding V2G architecture is in its early stage in the EU, mainly due to the lack of an agreed standard for the usage of V2G across cities and countries\footnote{Regulation (EU) 2023/1804, Preamble (30), \url{https://eur-lex.europa.eu/legal-content/en/TXT/?uri=CELEX\%3A32023R1804}, accessed Jan. 28, 2025}. Even though, we observe that V2G privacy and cyber security have been recognized as an important element in the development of the energy sector, and the framework mainly points to the existing regulatory solutions\footnote{Regulation (EU) 2023/1804, Preamble (75)}. \textit{E.g.}, 

\begin{quote}
\textit{[......] The lack of common technical specifications constitutes a barrier for [......] alternative fuels infrastructure. Therefore, it is necessary to lay down technical specifications [......] In particular, those technical specifications should cover the communication between the electric vehicle and the recharging point, the communication between the recharging point and the recharging software management system (back-end), the communication related to the electric vehicle roaming service and the communication with the electricity grid, while ensuring the highest level of \textbf{cybersecurity} protection and protection of final customers’ \textbf{personal data}. It is also necessary to establish a suitable governance framework and the roles of the different actors involved in the \textbf{vehicle-to-grid} communication sector.} (From \textit{Regulation (EU) 2023/1804 on the Deployment of Alternative Fuels Infrastructure})
\end{quote}

However, the criteria to differentiate personal and non-personal data is not explicitly specified from a regulatory perspective in the context of V2G, nor the roles, responsibilities, obligations in the context of data sharing, personal data protection, and cyber security. The organizational structure that guarantees secure data usage in V2G is unknown. To this end, we emphasize the need of specific stipulations or guidelines towards V2G data-sharing, privacy, and cyber security rather than borrowing/interpreting solutions from general or other existing regulations. Such specified efforts can avoid dependency on interpretation and reduce regulatory ambiguities for EV users and other stakeholders, including aggregators, charging stations, DSOs, service providers \textit{etc.}, that acquire, store, transfer, and process data. To this end, it is critical to provide contextualized regulatory roles, definitions/concepts, rights and obligations, and organizational structures for the V2G framework. This will be a crucial step forward that will play a key role in acceptance and widespread adoption of the V2G technology, and in realizing the tremendous potential to meet the growing demand of electricity globally by exploiting the mobile storage at the energy edge through innovative business models for the energy sector. 

\subsection{Open/Data protection/Security by design and by default}

To ensure that a high level of awareness and measures is in place against data-related issues, EU regulations explicitly encourage the principles of ``open by design and by default'' (Article 5(2) of Directive (EU) 2019/1024), ``data protection by design and by default'' (Regulation (EU) 2016/679), and ``security by design and by default'' (Regulation (EU) 2019/881) and ``cybersecurity by design'' (Network Code for cybersecurity aspects of cross-border electricity flows-draft) toward data-sharing, privacy protection, and cyber security, respectively. While those three principles mean different things considering their context, they can be adopted by the same entity in the energy system. The measures taken by a entity following different principles can overlap or conflict with each other to some level. \textit{E.g.}, a DSO can be responsible to release energy statistics and other data, yet the risk in data protection and security increase with the level of data openness. Nevertheless, there is no investigation on the merge and overlap of the three principles, nor any analysis on how to design measures that simultaneously follow two or three of those principles. Such a gap can lead to ambiguity in understanding the regulation, and thus also in the understanding of the individual roles and responsibilities by different stakeholders, which can naturally lead to conflicts. This can act as a major barrier in realizing the emerging distributed and participatory energy paradigms.

We observed that the principle of ``data protection by design and by default'' has been interpreted in detail to draw explicit understanding in general. However such as interpretation is lacking in the context of the energy sector~\cite{bygrave2021data}. We believe that similar regulatory studies - ideally with an energy context - on the principle of ``open by design and by default'' and ``cybersecurity by design'' can be valuable to align regulatory requirements and stakeholder understanding.

\subsection{Data-sharing, privacy, and cyber security in the energy sector: A Unified Framework}

Several connected but different terminologies have been utilized to address data-sharing, privacy, and cyber security issues for the energy sector. As a result, one entity can have multiple legal roles and obligations/rights under different contexts. \textit{E.g.}, a DSO can be a public sector body under data-sharing regulations, a data controller under data protection regulations, and an essential service provider under cyber security regulations, and different regulatory roles imply different obligations and rights, or lead to redundant efforts to ensure regulation compliance. Such complexity adds to regulatory ambiguity for energy stakeholders, and unaligned contextualization of general regulations deteriorates the situation even further. A clear description of the legal roles and responsibilities for the stakeholders explicitly addressing such overlapping as well as disjoint cases is needed. It is therefore important to conduct role-oriented regulatory analysis to derive explicit and concrete guidelines for targeted energy stakeholders about their role, responsibility, a provisioning of overriding where relevant, \textit{etc.}, for their involvement in data handling. This can be an effective approach towards facilitating regulatory compliance as a whole for specific stakeholders rather than piecemeal tips on separated issues.

\subsection{The European Artificial Intelligence Act and its impact}
The European Artificial Intelligence Act (Regulation (EU) 2024/1689\footnote{\url{https://eur-lex.europa.eu/legal-content/EN/TXT/PDF/?uri=OJ:L_202401689}, accessed Jan. 28, 2025}) or the AI Act came into force on 1st August, 2024, which is the first comprehensive AI regulation. This regulation applies category-based risk identification of AI and anti-risk measures. AI has been progressing in critical systems like the energy system, and it is clear that the AI Act will constrain the use of AI in EU energy sector to limit human and ethical risks~\cite{DBLP:journals/frai/NietEV21,HEYMANN20234538} and to protect fundamental human rights like privacy. However, the bulk of mandatory requirements by the EU AI Act only apply to ``high-risk AI systems''~\cite{https://doi.org/10.1111/reel.12574}. AI system and its applications in the energy system can be connected and interdependent, such that there can be implicit boundaries between different categories of risk levels. The ``all-or-nothing'' approach in the AI Act might risk that user privacy and security might be out of scope and ignored in AI systems that fall beyond the ``high-risk'' category, yet still impact user's energy practice.

\section{Conclusions}\label{Conclusions}

In this work, we reviewed the issues with data-sharing, privacy and cyber security in the energy sector. We started by examining the technical basis for these challenges, \textit{i.e.}, the available data, data communication, and data processing techniques in the energy sector. Then we categorized the data-related issues into data-sharing, privacy protection, and cyber security. We investigated how these issues are addressed in the EU regulatory framework. Based on the review of regulatory studies, we identified the gap between existing regulatory studies and a clear articulation of regulatory frameworks for data usage in the energy sector, and highlighted the importance of clearly and explicitly addressing the aspects important for regulatory compliance, such as role and responsibility of the key stakeholders in the energy sector, for the envisioned transition of the energy sector to a more distributed and participatory system. Moreover, to fill this gap, we review existing EU regulations and related regulatory entities, upon which we articulate the pertinent regulatory provisions regarding data-sharing, privacy, and cyber security in the energy sector. Finally, we highlighted the open challenges yet to be addressed along with emerging techniques and new application scenarios in the evolving energy sector. While we contextualize our study within in the scope of the EU, it is possible to extend this study to other parts of the world and also to compare across regions and countries in order to gain a wider view of this integrated technical and regulatory framework for the data related issues in the energy sector.

\section*{Acknowledgment}

This work was supported via the grant ``Privacy preserving Transactive Energy Management (PriTEM)'' funded by UiO:Energy Convergence Environments.

\bibliographystyle{IEEEtran}
\bibliography{references}

\begin{thebibliography}{100}
\providecommand{\url}[1]{#1}
\csname url@samestyle\endcsname
\providecommand{\newblock}{\relax}
\providecommand{\bibinfo}[2]{#2}
\providecommand{\BIBentrySTDinterwordspacing}{\spaceskip=0pt\relax}
\providecommand{\BIBentryALTinterwordstretchfactor}{4}
\providecommand{\BIBentryALTinterwordspacing}{\spaceskip=\fontdimen2\font plus
\BIBentryALTinterwordstretchfactor\fontdimen3\font minus \fontdimen4\font\relax}
\providecommand{\BIBforeignlanguage}[2]{{%
\expandafter\ifx\csname l@#1\endcsname\relax
\typeout{** WARNING: IEEEtran.bst: No hyphenation pattern has been}%
\typeout{** loaded for the language `#1'. Using the pattern for}%
\typeout{** the default language instead.}%
\else
\language=\csname l@#1\endcsname
\fi
#2}}
\providecommand{\BIBdecl}{\relax}
\BIBdecl

\bibitem{TOOKI2024100596}
\BIBentryALTinterwordspacing
O.~O. Tooki and O.~M. Popoola, ``A comprehensive review on recent advances in transactive energy system: Concepts, models, metrics, technologies, challenges, policies and future,'' \emph{Renewable Energy Focus}, vol.~50, p. 100596, 2024. [Online]. Available: \url{https://www.sciencedirect.com/science/article/pii/S1755008424000607}
\BIBentrySTDinterwordspacing

\bibitem{leal2023electricity}
\BIBentryALTinterwordspacing
R.~Leal-Arcas, \emph{Electricity Decentralization in the European Union: Towards Zero Carbon and Energy Transition}.\hskip 1em plus 0.5em minus 0.4em\relax Elsevier, 2023. [Online]. Available: \url{https://doi.org/10.1016/C2022-0-01938-3}
\BIBentrySTDinterwordspacing

\bibitem{RODRIGUES2023112999}
\BIBentryALTinterwordspacing
S.~D. Rodrigues and V.~J. Garcia, ``Transactive energy in microgrid communities: A systematic review,'' \emph{Renewable and Sustainable Energy Reviews}, vol. 171, p. 112999, 2023. [Online]. Available: \url{https://www.sciencedirect.com/science/article/pii/S1364032122008802}
\BIBentrySTDinterwordspacing

\bibitem{KHALID2024110253}
\BIBentryALTinterwordspacing
M.~Khalid, ``Energy 4.0: Ai-enabled digital transformation for sustainable power networks,'' \emph{Computers \& Industrial Engineering}, vol. 193, p. 110253, 2024. [Online]. Available: \url{https://www.sciencedirect.com/science/article/pii/S0360835224003747}
\BIBentrySTDinterwordspacing

\bibitem{10374424}
F.~Ahsan, N.~H. Dana, S.~K. Sarker, L.~Li, S.~M. Muyeen, M.~F. Ali, Z.~Tasneem, M.~M. Hasan, S.~H. Abhi, M.~R. Islam, M.~H. Ahamed, M.~M. Islam, S.~K. Das, M.~F.~R. Badal, and P.~Das, ``Data-driven next-generation smart grid towards sustainable energy evolution: techniques and technology review,'' \emph{Protection and Control of Modern Power Systems}, vol.~8, no.~3, pp. 1--42, 2023.

\bibitem{daehlen2023twin}
\BIBentryALTinterwordspacing
M.~D{\ae}hlen, ``The twin transition century: The role of digital research for a successful green transition of society?(the guild insight paper no. 5),'' 2023. [Online]. Available: \url{https://doi.org/10.48350/184458}
\BIBentrySTDinterwordspacing

\bibitem{rao2023security}
P.~M. Rao and B.~D. Deebak, ``Security and privacy issues in smart cities/industries: technologies, applications, and challenges,'' \emph{Journal of Ambient Intelligence and Humanized Computing}, vol.~14, no.~8, pp. 10\,517--10\,553, 2023.

\bibitem{Papakonstantinou2015}
\BIBentryALTinterwordspacing
V.~Papakonstantinou and D.~Kloza, \emph{Legal Protection of Personal Data in Smart Grid and Smart Metering Systems from the European Perspective}.\hskip 1em plus 0.5em minus 0.4em\relax London: Springer London, 2015, pp. 41--129. [Online]. Available: \url{https://doi.org/10.1007/978-1-4471-6663-4_2}
\BIBentrySTDinterwordspacing

\bibitem{BUGDEN2019137}
\BIBentryALTinterwordspacing
D.~Bugden and R.~Stedman, ``A synthetic view of acceptance and engagement with smart meters in the {United States},'' \emph{Energy Research \& Social Science}, vol.~47, pp. 137--145, 2019. [Online]. Available: \url{https://doi.org/10.1016/j.erss.2018.08.025}
\BIBentrySTDinterwordspacing

\bibitem{DBLP:journals/tochi/JakobiPRSW19}
\BIBentryALTinterwordspacing
T.~Jakobi, S.~Patil, D.~Randall, G.~Stevens, and V.~Wulf, ``It is about what they could do with the data: {A} user perspective on privacy in smart metering,'' \emph{{ACM} Trans. Comput. Hum. Interact.}, vol.~26, no.~1, pp. 2:1--2:44, 2019. [Online]. Available: \url{https://doi.org/10.1145/3281444}
\BIBentrySTDinterwordspacing

\bibitem{DBLP:journals/tsg/EiblE15}
\BIBentryALTinterwordspacing
G.~Eibl and D.~Engel, ``Influence of data granularity on smart meter privacy,'' \emph{{IEEE} Trans. Smart Grid}, vol.~6, no.~2, pp. 930--939, 2015. [Online]. Available: \url{https://doi.org/10.1109/TSG.2014.2376613}
\BIBentrySTDinterwordspacing

\bibitem{DBLP:conf/wpes/TudorAP13}
\BIBentryALTinterwordspacing
V.~Tudor, M.~Almgren, and M.~Papatriantafilou, ``Analysis of the impact of data granularity on privacy for the smart grid,'' in \emph{Proceedings of the 12th annual {ACM} Workshop on Privacy in the Electronic Society, {WPES} 2013, Berlin, Germany, November 4, 2013}, A.~Sadeghi and S.~Foresti, Eds.\hskip 1em plus 0.5em minus 0.4em\relax {ACM}, 2013, pp. 61--70. [Online]. Available: \url{https://doi.org/10.1145/2517840.2517844}
\BIBentrySTDinterwordspacing

\bibitem{RAZAVI2018312}
\BIBentryALTinterwordspacing
R.~Razavi and A.~Gharipour, ``Rethinking the privacy of the smart grid: What your smart meter data can reveal about your household in ireland,'' \emph{Energy Research \& Social Science}, vol.~44, pp. 312--323, 2018. [Online]. Available: \url{https://doi.org/10.1016/j.erss.2018.06.005}
\BIBentrySTDinterwordspacing

\bibitem{BALTAOZKAN201465}
\BIBentryALTinterwordspacing
N.~Balta-Ozkan, B.~Boteler, and O.~Amerighi, ``European smart home market development: Public views on technical and economic aspects across the united kingdom, germany and italy,'' \emph{Energy Research \& Social Science}, vol.~3, pp. 65--77, 2014. [Online]. Available: \url{https://doi.org/10.1016/j.erss.2014.07.007}
\BIBentrySTDinterwordspacing

\bibitem{DBLP:journals/taasm/FellSHE14}
\BIBentryALTinterwordspacing
M.~J. Fell, D.~Shipworth, G.~M. Huebner, and C.~A. Elwell, ``Exploring perceived control in domestic electricity demand-side response,'' \emph{Technol. Anal. Strateg. Manag.}, vol.~26, no.~10, pp. 1118--1130, 2014. [Online]. Available: \url{https://doi.org/10.1080/09537325.2014.974530}
\BIBentrySTDinterwordspacing

\bibitem{WILSON201772}
\BIBentryALTinterwordspacing
C.~Wilson, T.~Hargreaves, and R.~Hauxwell-Baldwin, ``Benefits and risks of smart home technologies,'' \emph{Energy Policy}, vol. 103, pp. 72--83, 2017. [Online]. Available: \url{https://doi.org/10.1016/j.enpol.2016.12.047}
\BIBentrySTDinterwordspacing

\bibitem{HMIELOWSKI2019189}
\BIBentryALTinterwordspacing
J.~D. Hmielowski, A.~D. Boyd, G.~Harvey, and J.~Joo, ``The social dimensions of smart meters in the united states: Demographics, privacy, and technology readiness,'' \emph{Energy Research \& Social Science}, vol.~55, pp. 189--197, 2019. [Online]. Available: \url{https://doi.org/10.1016/j.erss.2019.05.003}
\BIBentrySTDinterwordspacing

\bibitem{HORNE201564}
\BIBentryALTinterwordspacing
C.~Horne, B.~Darras, E.~Bean, A.~Srivastava, and S.~Frickel, ``Privacy, technology, and norms: The case of smart meters,'' \emph{Social Science Research}, vol.~51, pp. 64--76, 2015. [Online]. Available: \url{https://doi.org/10.1016/j.ssresearch.2014.12.003}
\BIBentrySTDinterwordspacing

\bibitem{NAUS2015125}
\BIBentryALTinterwordspacing
J.~Naus, B.~J. {van Vliet}, and A.~Hendriksen, ``Households as change agents in a dutch smart energy transition: On power, privacy and participation,'' \emph{Energy Research \& Social Science}, vol.~9, pp. 125--136, 2015, special Issue on Smart Grids and the Social Sciences. [Online]. Available: \url{https://doi.org/10.1016/j.erss.2015.08.025}
\BIBentrySTDinterwordspacing

\bibitem{HANSEN2017112}
\BIBentryALTinterwordspacing
M.~Hansen and B.~Hauge, ``Scripting, control, and privacy in domestic smart grid technologies: Insights from a danish pilot study,'' \emph{Energy Research \& Social Science}, vol.~25, pp. 112--123, 2017. [Online]. Available: \url{https://doi.org/10.1016/j.erss.2017.01.005}
\BIBentrySTDinterwordspacing

\bibitem{KRISHNAMURTI2012790}
\BIBentryALTinterwordspacing
T.~Krishnamurti, D.~Schwartz, A.~Davis, B.~Fischhoff, W.~B. {de Bruin}, L.~Lave, and J.~Wang, ``Preparing for smart grid technologies: A behavioral decision research approach to understanding consumer expectations about smart meters,'' \emph{Energy Policy}, vol.~41, pp. 790--797, 2012, modeling Transport (Energy) Demand and Policies. [Online]. Available: \url{https://doi.org/10.1016/j.enpol.2011.11.047}
\BIBentrySTDinterwordspacing

\bibitem{uppuluri2023secure}
\BIBentryALTinterwordspacing
S.~Uppuluri and G.~Lakshmeeswari, ``Secure user authentication and key agreement scheme for iot device access control based smart home communications,'' \emph{Wireless Networks}, vol.~29, no.~3, pp. 1333--1354, 2023. [Online]. Available: \url{https://doi.org/10.1007/s11276-022-03197-1}
\BIBentrySTDinterwordspacing

\bibitem{GHIASI2023108975}
\BIBentryALTinterwordspacing
M.~Ghiasi, T.~Niknam, Z.~Wang, M.~Mehrandezh, M.~Dehghani, and N.~Ghadimi, ``A comprehensive review of cyber-attacks and defense mechanisms for improving security in smart grid energy systems: Past, present and future,'' \emph{Electric Power Systems Research}, vol. 215, p. 108975, 2023. [Online]. Available: \url{https://www.sciencedirect.com/science/article/pii/S0378779622010240}
\BIBentrySTDinterwordspacing

\bibitem{10.1145/3655693.3656546}
\BIBentryALTinterwordspacing
U.~Cali, S.~N.~G. Gourisetti, D.~J. Sebastian-Cardenas, F.~O. Catak, A.~Lee, L.~M. Zeger, T.~S. Ustun, M.~F. Dynge, S.~Rao, and J.~E. Ramirez, ``Emerging technologies for privacy preservation in energy systems,'' in \emph{Proceedings of the 2024 European Interdisciplinary Cybersecurity Conference}, ser. EICC '24.\hskip 1em plus 0.5em minus 0.4em\relax New York, NY, USA: Association for Computing Machinery, 2024, p. 163–170. [Online]. Available: \url{https://doi.org/10.1145/3655693.3656546}
\BIBentrySTDinterwordspacing

\bibitem{DBLP:journals/tse/IwayaBR23}
\BIBentryALTinterwordspacing
L.~H. Iwaya, M.~A. Babar, and A.~Rashid, ``Privacy engineering in the wild: Understanding the practitioners' mindset, organizational aspects, and current practices,'' \emph{{IEEE} Trans. Software Eng.}, vol.~49, no.~9, pp. 4324--4348, 2023. [Online]. Available: \url{https://doi.org/10.1109/TSE.2023.3290237}
\BIBentrySTDinterwordspacing

\bibitem{ogunniye2023survey}
\BIBentryALTinterwordspacing
G.~Ogunniye and N.~Kokciyan, ``A survey on understanding and representing privacy requirements in the internet-of-things,'' \emph{Journal of Artificial Intelligence Research}, vol.~76, pp. 163--192, 2023. [Online]. Available: \url{https://doi.org/10.1613/jair.1.14000}
\BIBentrySTDinterwordspacing

\bibitem{MACEDO2023138102}
\BIBentryALTinterwordspacing
D.~P. Macedo and A.~C. Marques, ``Is the energy transition ready for declining budgets in rd\&d for fossil fuels? evidence from a panel of european countries,'' \emph{Journal of Cleaner Production}, vol. 417, p. 138102, 2023. [Online]. Available: \url{https://www.sciencedirect.com/science/article/pii/S0959652623022606}
\BIBentrySTDinterwordspacing

\bibitem{10.1093/ijlit/eau001}
\BIBentryALTinterwordspacing
N.~J. King and P.~W. Jessen, ``{Smart metering systems and data sharing: why getting a smart meter should also mean getting strong information privacy controls to manage data sharing},'' \emph{International Journal of Law and Information Technology}, vol.~22, no.~3, pp. 215--253, 03 2014. [Online]. Available: \url{https://doi.org/10.1093/ijlit/eau001}
\BIBentrySTDinterwordspacing

\bibitem{harvey2013smart}
\BIBentryALTinterwordspacing
S.~J. Harvey, ``Smart meters, smarter regulation: Balancing privacy and innovation in the electric grid,'' \emph{UCLA L. Rev.}, vol.~61, p. 2068, 2013. [Online]. Available: \url{https://heinonline.org/HOL/Page?handle=hein.journals/uclalr61&id=2057&collection=journals&index=#}
\BIBentrySTDinterwordspacing

\bibitem{LESZCZYNA201862}
\BIBentryALTinterwordspacing
R.~Leszczyna, ``Cybersecurity and privacy in standards for smart grids – a comprehensive survey,'' \emph{Computer Standards \& Interfaces}, vol.~56, pp. 62--73, 2018. [Online]. Available: \url{https://www.sciencedirect.com/science/article/pii/S0920548917301277}
\BIBentrySTDinterwordspacing

\bibitem{ANANDAKUMAR2014126}
\BIBentryALTinterwordspacing
V.~{Ananda Kumar}, K.~K. Pandey, and D.~K. Punia, ``Cyber security threats in the power sector: Need for a domain specific regulatory framework in india,'' \emph{Energy Policy}, vol.~65, pp. 126--133, 2014. [Online]. Available: \url{https://www.sciencedirect.com/science/article/pii/S0301421513010471}
\BIBentrySTDinterwordspacing

\bibitem{10.1145/3645091}
\BIBentryALTinterwordspacing
K.~Salehzadeh~Niksirat, L.~Velykoivanenko, N.~Zufferey, M.~Cherubini, K.~Huguenin, and M.~Humbert, ``Wearable activity trackers: A survey on utility, privacy, and security,'' \emph{ACM Comput. Surv.}, vol.~56, no.~7, Apr. 2024. [Online]. Available: \url{https://doi.org/10.1145/3645091}
\BIBentrySTDinterwordspacing

\bibitem{DBLP:journals/misq/SmithDX11}
\BIBentryALTinterwordspacing
H.~J. Smith, T.~Dinev, and H.~Xu, ``Information privacy research: An interdisciplinary review,'' \emph{{MIS} Q.}, vol.~35, no.~4, pp. 989--1015, 2011. [Online]. Available: \url{http://misq.org/catalog/product/view/id/1518/s/information-privacy-research-an-interdisciplinary-review/}
\BIBentrySTDinterwordspacing

\bibitem{XIA2023104771}
\BIBentryALTinterwordspacing
L.~Xia, D.~Semirumi, and R.~Rezaei, ``A thorough examination of smart city applications: Exploring challenges and solutions throughout the life cycle with emphasis on safeguarding citizen privacy,'' \emph{Sustainable Cities and Society}, vol.~98, p. 104771, 2023. [Online]. Available: \url{https://www.sciencedirect.com/science/article/pii/S2210670723003827}
\BIBentrySTDinterwordspacing

\bibitem{VEZZONI2023103134}
\BIBentryALTinterwordspacing
R.~Vezzoni, ``Green growth for whom, how and why? the repowereu plan and the inconsistencies of european union energy policy,'' \emph{Energy Research \& Social Science}, vol. 101, p. 103134, 2023. [Online]. Available: \url{https://www.sciencedirect.com/science/article/pii/S2214629623001949}
\BIBentrySTDinterwordspacing

\bibitem{HAJIBASHI2023113055}
\BIBentryALTinterwordspacing
M.~{Haji Bashi}, L.~{De Tommasi}, A.~{Le Cam}, L.~S. Relaño, P.~Lyons, J.~Mundó, I.~Pandelieva-Dimova, H.~Schapp, K.~Loth-Babut, C.~Egger, M.~Camps, B.~Cassidy, G.~Angelov, and C.~E. Stancioff, ``A review and mapping exercise of energy community regulatory challenges in european member states based on a survey of collective energy actors,'' \emph{Renewable and Sustainable Energy Reviews}, vol. 172, p. 113055, 2023. [Online]. Available: \url{https://www.sciencedirect.com/science/article/pii/S1364032122009364}
\BIBentrySTDinterwordspacing

\bibitem{FRILINGOU2023102934}
\BIBentryALTinterwordspacing
N.~Frilingou, G.~Xexakis, K.~Koasidis, A.~Nikas, L.~Campagnolo, E.~Delpiazzo, A.~Chiodi, M.~Gargiulo, B.~McWilliams, T.~Koutsellis, and H.~Doukas, ``Navigating through an energy crisis: Challenges and progress towards electricity decarbonisation, reliability, and affordability in italy,'' \emph{Energy Research \& Social Science}, vol.~96, p. 102934, 2023. [Online]. Available: \url{https://www.sciencedirect.com/science/article/pii/S2214629622004376}
\BIBentrySTDinterwordspacing

\bibitem{MCCLEAN2023113378}
\BIBentryALTinterwordspacing
A.~McClean and O.~Pedersen, ``The role of regulation in geothermal energy in the uk,'' \emph{Energy Policy}, vol. 173, p. 113378, 2023. [Online]. Available: \url{https://www.sciencedirect.com/science/article/pii/S0301421522005973}
\BIBentrySTDinterwordspacing

\bibitem{LEE2021101188}
\BIBentryALTinterwordspacing
D.~Lee and D.~J. Hess, ``Data privacy and residential smart meters: Comparative analysis and harmonization potential,'' \emph{Utilities Policy}, vol.~70, p. 101188, 2021. [Online]. Available: \url{https://doi.org/10.1016/j.jup.2021.101188}
\BIBentrySTDinterwordspacing

\bibitem{HEUNINCKX2023103040}
\BIBentryALTinterwordspacing
S.~Heuninckx, M.~Meitern, G.~{te Boveldt}, and T.~Coosemans, ``Practical problems before privacy concerns: How european energy community initiatives struggle with data collection,'' \emph{Energy Research \& Social Science}, vol.~98, p. 103040, 2023. [Online]. Available: \url{https://www.sciencedirect.com/science/article/pii/S2214629623001007}
\BIBentrySTDinterwordspacing

\bibitem{CARLANDER2023100097}
\BIBentryALTinterwordspacing
J.~Carlander and P.~Thollander, ``Barriers to implementation of energy-efficient technologies in building construction projects — results from a swedish case study,'' \emph{Resources, Environment and Sustainability}, vol.~11, p. 100097, 2023. [Online]. Available: \url{https://www.sciencedirect.com/science/article/pii/S266691612200041X}
\BIBentrySTDinterwordspacing

\bibitem{papavasiliou2023welfare}
\BIBentryALTinterwordspacing
A.~Papavasiliou and D.~Avila, ``Welfare benefits of co-optimising energy and reserves,'' 2023. [Online]. Available: \url{https://www.acer.europa.eu/sites/default/files/documents/Publications/ACER_Cooptimisation_Benefits_Study_2024.pdf}
\BIBentrySTDinterwordspacing

\bibitem{6624135}
J.~Lavaei, D.~Tse, and B.~Zhang, ``Geometry of power flows and optimization in distribution networks,'' \emph{IEEE Transactions on Power Systems}, vol.~29, no.~2, pp. 572--583, 2014.

\bibitem{AHMED2024122403}
\BIBentryALTinterwordspacing
S.~A. Ahmed, Q.~Huang, Z.~Zhang, J.~Li, W.~Amin, M.~Afzal, J.~Hussain, and F.~Hussain, ``Optimization of social welfare and mitigating privacy risks in p2p energy trading: Differential privacy for secure data reporting,'' \emph{Applied Energy}, vol. 356, p. 122403, 2024. [Online]. Available: \url{https://www.sciencedirect.com/science/article/pii/S0306261923017671}
\BIBentrySTDinterwordspacing

\bibitem{Ghiasi2023}
\BIBentryALTinterwordspacing
M.~Ghiasi, Z.~Wang, T.~Niknam, M.~Dehghani, and H.~R. Ansari, \emph{Cyber-Physical Security in Smart Power Systems from a Resilience Perspective: Concepts and Possible Solutions}.\hskip 1em plus 0.5em minus 0.4em\relax Cham: Springer International Publishing, 2023, pp. 67--89. [Online]. Available: \url{https://doi.org/10.1007/978-3-031-20360-2_3}
\BIBentrySTDinterwordspacing

\bibitem{ZHANG2024107749}
\BIBentryALTinterwordspacing
D.~Zhang, Y.~He, and M.~Lu, ``Is energy firms' investment behavior more sensitive on corporate perception of monetary policy?'' \emph{Energy Economics}, vol. 136, p. 107749, 2024. [Online]. Available: \url{https://www.sciencedirect.com/science/article/pii/S0140988324004572}
\BIBentrySTDinterwordspacing

\bibitem{CALEARO2021111518}
\BIBentryALTinterwordspacing
L.~Calearo, M.~Marinelli, and C.~Ziras, ``A review of data sources for electric vehicle integration studies,'' \emph{Renewable and Sustainable Energy Reviews}, vol. 151, p. 111518, 2021. [Online]. Available: \url{https://www.sciencedirect.com/science/article/pii/S1364032121007966}
\BIBentrySTDinterwordspacing

\bibitem{CHEN2020109466}
\BIBentryALTinterwordspacing
M.~Chen, C.~Gao, M.~Song, S.~Chen, D.~Li, and Q.~Liu, ``Internet data centers participating in demand response: A comprehensive review,'' \emph{Renewable and Sustainable Energy Reviews}, vol. 117, p. 109466, 2020. [Online]. Available: \url{https://www.sciencedirect.com/science/article/pii/S1364032119306744}
\BIBentrySTDinterwordspacing

\bibitem{KAZMI2021111290}
\BIBentryALTinterwordspacing
H.~Kazmi, Íngrid Munné-Collado, F.~Mehmood, T.~A. Syed, and J.~Driesen, ``Towards data-driven energy communities: A review of open-source datasets, models and tools,'' \emph{Renewable and Sustainable Energy Reviews}, vol. 148, p. 111290, 2021. [Online]. Available: \url{https://www.sciencedirect.com/science/article/pii/S1364032121005773}
\BIBentrySTDinterwordspacing

\bibitem{LUO2021111224}
\BIBentryALTinterwordspacing
N.~Luo, M.~Pritoni, and T.~Hong, ``An overview of data tools for representing and managing building information and performance data,'' \emph{Renewable and Sustainable Energy Reviews}, vol. 147, p. 111224, 2021. [Online]. Available: \url{https://www.sciencedirect.com/science/article/pii/S1364032121005116}
\BIBentrySTDinterwordspacing

\bibitem{SINGH2021100410}
\BIBentryALTinterwordspacing
N.~K. Singh and V.~Mahajan, ``End-user privacy protection scheme from cyber intrusion in smart grid advanced metering infrastructure,'' \emph{International Journal of Critical Infrastructure Protection}, vol.~34, p. 100410, 2021. [Online]. Available: \url{https://www.sciencedirect.com/science/article/pii/S1874548221000020}
\BIBentrySTDinterwordspacing

\bibitem{ATHANASIADIS2024114151}
\BIBentryALTinterwordspacing
C.~Athanasiadis, T.~Papadopoulos, G.~Kryonidis, and D.~Doukas, ``A review of distribution network applications based on smart meter data analytics,'' \emph{Renewable and Sustainable Energy Reviews}, vol. 191, p. 114151, 2024. [Online]. Available: \url{https://www.sciencedirect.com/science/article/pii/S1364032123010092}
\BIBentrySTDinterwordspacing

\bibitem{USHAKOVA2020101428}
\BIBentryALTinterwordspacing
A.~Ushakova and S.~{Jankin Mikhaylov}, ``Big data to the rescue? challenges in analysing granular household electricity consumption in the united kingdom,'' \emph{Energy Research \& Social Science}, vol.~64, p. 101428, 2020. [Online]. Available: \url{https://www.sciencedirect.com/science/article/pii/S2214629620300050}
\BIBentrySTDinterwordspacing

\bibitem{HECHT2020100079}
\BIBentryALTinterwordspacing
C.~Hecht, S.~Das, C.~Bussar, and D.~U. Sauer, ``Representative, empirical, real-world charging station usage characteristics and data in {G}ermany,'' \emph{eTransportation}, vol.~6, p. 100079, 2020. [Online]. Available: \url{https://www.sciencedirect.com/science/article/pii/S2590116820300369}
\BIBentrySTDinterwordspacing

\bibitem{MILLER2017439}
\BIBentryALTinterwordspacing
C.~Miller and F.~Meggers, ``The building data genome project: An open, public data set from non-residential building electrical meters,'' \emph{Energy Procedia}, vol. 122, pp. 439--444, 2017, cISBAT 2017 International ConferenceFuture Buildings \& Districts – Energy Efficiency from Nano to Urban Scale. [Online]. Available: \url{https://www.sciencedirect.com/science/article/pii/S1876610217330047}
\BIBentrySTDinterwordspacing

\bibitem{TSANAS2012560}
\BIBentryALTinterwordspacing
A.~Tsanas and A.~Xifara, ``Accurate quantitative estimation of energy performance of residential buildings using statistical machine learning tools,'' \emph{Energy and Buildings}, vol.~49, pp. 560--567, 2012. [Online]. Available: \url{https://www.sciencedirect.com/science/article/pii/S037877881200151X}
\BIBentrySTDinterwordspacing

\bibitem{HIRTH20181054}
\BIBentryALTinterwordspacing
L.~Hirth, J.~Mühlenpfordt, and M.~Bulkeley, ``The {ENTSO-E} transparency platform – a review of {E}urope’s most ambitious electricity data platform,'' \emph{Applied Energy}, vol. 225, pp. 1054--1067, 2018. [Online]. Available: \url{https://www.sciencedirect.com/science/article/pii/S0306261918306068}
\BIBentrySTDinterwordspacing

\bibitem{DBLP:conf/eenergy/JazizadehABS18}
\BIBentryALTinterwordspacing
F.~Jazizadeh, M.~Afzalan, B.~Becerik{-}Gerber, and L.~Soibelman, ``{EMBED:} {A} dataset for energy monitoring through building electricity disaggregation,'' in \emph{Proceedings of the Ninth International Conference on Future Energy Systems, e-Energy 2018, Karlsruhe, Germany, June 12-15, 2018}, H.~Schmeck and V.~Hagenmeyer, Eds.\hskip 1em plus 0.5em minus 0.4em\relax {ACM}, 2018, pp. 230--235. [Online]. Available: \url{https://doi.org/10.1145/3208903.3208939}
\BIBentrySTDinterwordspacing

\bibitem{einfalt2009adres}
\BIBentryALTinterwordspacing
A.~Einfalt, C.~Leitinger, D.~Tiefgraber, and S.~Ghaemi, ``Adres concept--micro grids in {\"o}sterreich,'' \emph{Internationale Energiewirtschaftstagung an der TU Wien. Wien: IEWT}, 2009. [Online]. Available: \url{https://www.researchgate.net/profile/Sara-Ghaemi-2/publication/267412641_ADRES_Concept_-_Micro_Grids_in_Osterreich/links/547f03bc0cf2c1e3d2dc38e2/ADRES-Concept-Micro-Grids-in-Oesterreich.pdf}
\BIBentrySTDinterwordspacing

\bibitem{chavat2022ecd}
\BIBentryALTinterwordspacing
J.~Chavat, S.~Nesmachnow, J.~Graneri, and G.~Alvez, ``{ECD-UY}, detailed household electricity consumption dataset of {U}ruguay,'' \emph{Scientific Data}, vol.~9, no.~1, p.~21, 2022. [Online]. Available: \url{https://doi.org/10.1038/s41597-022-01122-x}
\BIBentrySTDinterwordspacing

\bibitem{kolter2011redd}
\BIBentryALTinterwordspacing
J.~Z. Kolter and M.~J. Johnson, ``Redd: A public data set for energy disaggregation research,'' in \emph{Workshop on data mining applications in sustainability (SIGKDD), San Diego, CA}, vol.~25, no. Citeseer.\hskip 1em plus 0.5em minus 0.4em\relax Citeseer, 2011, pp. 59--62. [Online]. Available: \url{https://citeseerx.ist.psu.edu/document?repid=rep1&type=pdf&doi=d85a51e2978f4563ee74bf9a09d3219e03799819}
\BIBentrySTDinterwordspacing

\bibitem{kelly2015uk}
\BIBentryALTinterwordspacing
J.~Kelly and W.~Knottenbelt, ``The uk-dale dataset, domestic appliance-level electricity demand and whole-house demand from five uk homes,'' \emph{Scientific data}, vol.~2, no.~1, pp. 1--14, 2015. [Online]. Available: \url{https://doi.org/10.1038/sdata.2015.7}
\BIBentrySTDinterwordspacing

\bibitem{DBLP:conf/sensys/BeckelKCSS14}
\BIBentryALTinterwordspacing
C.~Beckel, W.~Kleiminger, R.~Cicchetti, T.~Staake, and S.~Santini, ``The {ECO} data set and the performance of non-intrusive load monitoring algorithms,'' in \emph{Proceedings of the 1st {ACM} Conference on Embedded Systems for Energy-Efficient Buildings, BuildSys 2014, Memphis, TN, USA, November 3-6, 2014}, M.~Srivastava, Ed.\hskip 1em plus 0.5em minus 0.4em\relax {ACM}, 2014, pp. 80--89. [Online]. Available: \url{https://doi.org/10.1145/2674061.2674064}
\BIBentrySTDinterwordspacing

\bibitem{9894187}
\BIBentryALTinterwordspacing
M.~Zaman, S.~Saha, N.~Zohrabi, and S.~Abdelwahed, ``Uncertainty estimation in power consumption of a smart home using bayesian lstm networks,'' in \emph{2022 IEEE International Symposium on Advanced Control of Industrial Processes (AdCONIP)}, 2022, pp. 120--125. [Online]. Available: \url{https://doi.org/10.1109/AdCONIP55568.2022.9894187}
\BIBentrySTDinterwordspacing

\bibitem{pereira2022residential}
\BIBentryALTinterwordspacing
L.~Pereira, D.~Costa, and M.~Ribeiro, ``A residential labeled dataset for smart meter data analytics,'' \emph{Scientific Data}, vol.~9, no.~1, p. 134, 2022. [Online]. Available: \url{https://doi.org/10.1038/s41597-022-01252-2}
\BIBentrySTDinterwordspacing

\bibitem{filip2011blued}
\BIBentryALTinterwordspacing
A.~Filip \emph{et~al.}, ``Blued: A fully labeled public dataset for event-based nonintrusive load monitoring research,'' in \emph{2nd workshop on data mining applications in sustainability (SustKDD)}, vol. 2012, 2011. [Online]. Available: \url{https://www.inferlab.org/wp-content/uploads/2012/08/2012_anderson_SustKDD.pdf}
\BIBentrySTDinterwordspacing

\bibitem{DBLP:conf/sensys/GaoGKB14}
\BIBentryALTinterwordspacing
J.~Gao, S.~Giri, E.~C. Kara, and M.~Berges, ``{PLAID:} a public dataset of high-resoultion electrical appliance measurements for load identification research: demo abstract,'' in \emph{Proceedings of the 1st {ACM} Conference on Embedded Systems for Energy-Efficient Buildings, BuildSys 2014, Memphis, TN, USA, November 3-6, 2014}, M.~Srivastava, Ed.\hskip 1em plus 0.5em minus 0.4em\relax {ACM}, 2014, pp. 198--199. [Online]. Available: \url{https://doi.org/10.1145/2674061.2675032}
\BIBentrySTDinterwordspacing

\bibitem{shin2019enertalk}
\BIBentryALTinterwordspacing
C.~Shin, E.~Lee, J.~Han, J.~Yim, W.~Rhee, and H.~Lee, ``The enertalk dataset, 15 hz electricity consumption data from 22 houses in korea,'' \emph{Scientific data}, vol.~6, no.~1, p. 193, 2019. [Online]. Available: \url{https://doi.org/10.1038/s41597-019-0212-5}
\BIBentrySTDinterwordspacing

\bibitem{DBLP:conf/eenergy/NadeemA19}
\BIBentryALTinterwordspacing
A.~Nadeem and N.~Arshad, ``{PRECON:} pakistan residential electricity consumption dataset,'' in \emph{Proceedings of the Tenth {ACM} International Conference on Future Energy Systems, e-Energy 2019, Phoenix, AZ, USA, June 25-28, 2019}.\hskip 1em plus 0.5em minus 0.4em\relax {ACM}, 2019, pp. 52--57. [Online]. Available: \url{https://doi.org/10.1145/3307772.3328317}
\BIBentrySTDinterwordspacing

\bibitem{LOPEZPROL2023106974}
\BIBentryALTinterwordspacing
J.~{López Prol} and D.~Zilberman, ``No alarms and no surprises: Dynamics of renewable energy curtailment in california,'' \emph{Energy Economics}, vol. 126, p. 106974, 2023. [Online]. Available: \url{https://www.sciencedirect.com/science/article/pii/S0140988323004723}
\BIBentrySTDinterwordspacing

\bibitem{9392581}
S.~Saha, V.~J. Sriprasath, M.~R. Ashwin, B.~Shivram, and A.~G. Kumar, ``Renewable energy curtailment reduction for california,'' in \emph{2021 International Conference on Advances in Electrical, Computing, Communication and Sustainable Technologies (ICAECT)}, 2021, pp. 1--7.

\bibitem{DBLP:conf/sensys/NambiLP15}
\BIBentryALTinterwordspacing
S.~N. A.~U. Nambi, A.~R. Lua, and R.~V. Prasad, ``Loced: Location-aware energy disaggregation framework,'' in \emph{Proceedings of the 2nd {ACM} International Conference on Embedded Systems for Energy-Efficient Built Environments, BuildSys 2015, Seoul, South Korea, November 4-5, 2015}, D.~E. Culler, Y.~Agarwal, and R.~Mangharam, Eds.\hskip 1em plus 0.5em minus 0.4em\relax {ACM}, 2015, pp. 45--54. [Online]. Available: \url{https://doi.org/10.1145/2821650.2821659}
\BIBentrySTDinterwordspacing

\bibitem{DBLP:journals/corr/abs-2110-02166}
\BIBentryALTinterwordspacing
O.~Mey, A.~Schneider, O.~Enge{-}Rosenblatt, Y.~Bravo, and P.~Stenzel, ``Prediction of energy consumption for variable customer portfolios including aleatoric uncertainty estimation,'' \emph{CoRR}, vol. abs/2110.02166, 2021. [Online]. Available: \url{https://arxiv.org/abs/2110.02166}
\BIBentrySTDinterwordspacing

\bibitem{CANDANEDO201781}
\BIBentryALTinterwordspacing
L.~M. Candanedo, V.~Feldheim, and D.~Deramaix, ``Data driven prediction models of energy use of appliances in a low-energy house,'' \emph{Energy and Buildings}, vol. 140, pp. 81--97, 2017. [Online]. Available: \url{https://www.sciencedirect.com/science/article/pii/S0378778816308970}
\BIBentrySTDinterwordspacing

\bibitem{barker2012smart}
\BIBentryALTinterwordspacing
S.~Barker, A.~Mishra, D.~Irwin, E.~Cecchet, P.~Shenoy, J.~Albrecht \emph{et~al.}, ``Smart*: An open data set and tools for enabling research in sustainable homes,'' \emph{SustKDD, August}, vol. 111, no. 112, p. 108, 2012. [Online]. Available: \url{https://odysseus.informatik.uni-oldenburg.de/download/Data/Smart%20Home/sustkdd-smart.pdf}
\BIBentrySTDinterwordspacing

\bibitem{distribution2021western-capacity}
\BIBentryALTinterwordspacing
W.~P. Distribution, ``Electric nation data-charger install,'' 2021. [Online]. Available: \url{https://www.nationalgrid.co.uk/downloads/81655/chargerinstall.xlsx}
\BIBentrySTDinterwordspacing

\bibitem{distribution2021western-current}
\BIBentryALTinterwordspacing
W.~P. D.-M. values, ``Electric nation data-charger install,'' 2021. [Online]. Available: \url{https://www.nationalgrid.co.uk/downloads-view-reciteme/279901}
\BIBentrySTDinterwordspacing

\bibitem{su132112025}
\BIBentryALTinterwordspacing
B.~Dutta and H.-G. Hwang, ``Consumers purchase intentions of green electric vehicles: The influence of consumers technological and environmental considerations,'' \emph{Sustainability}, vol.~13, no.~21, 2021. [Online]. Available: \url{https://www.mdpi.com/2071-1050/13/21/12025}
\BIBentrySTDinterwordspacing

\bibitem{charger2017}
\BIBentryALTinterwordspacing
U.~D. for Transport, ``Electric chargepoint analysis 2017: Domestics,'' 2017. [Online]. Available: \url{https://www.gov.uk/government/statistics/electric-chargepoint-analysis-2017-domestics}
\BIBentrySTDinterwordspacing

\bibitem{chargedenergy2018}
\BIBentryALTinterwordspacing
------, ``Electric chargepoint analysis 2017: Local authority rapids (revised),'' 2018. [Online]. Available: \url{https://www.gov.uk/government/statistics/electric-chargepoint-analysis-2017-local-authority-rapids}
\BIBentrySTDinterwordspacing

\bibitem{chargedenergy20181}
\BIBentryALTinterwordspacing
------, ``Electric chargepoint analysis 2017: Public sector fasts,'' 2018. [Online]. Available: \url{https://www.gov.uk/government/statistics/electric-chargepoint-analysis-2017-public-sector-fasts}
\BIBentrySTDinterwordspacing

\bibitem{distribution2021western}
\BIBentryALTinterwordspacing
W.~P. Distribution, ``Electric nation data-crowd charge transactions,'' 2021. [Online]. Available: \url{https://www.nationalgrid.co.uk/downloads-view/81646}
\BIBentrySTDinterwordspacing

\bibitem{DBLP:conf/eenergy/LeeLL19}
\BIBentryALTinterwordspacing
Z.~J. Lee, T.~Li, and S.~H. Low, ``Acn-data: Analysis and applications of an open {EV} charging dataset,'' in \emph{Proceedings of the Tenth {ACM} International Conference on Future Energy Systems, e-Energy 2019, Phoenix, AZ, USA, June 25-28, 2019}.\hskip 1em plus 0.5em minus 0.4em\relax {ACM}, 2019, pp. 139--149. [Online]. Available: \url{https://doi.org/10.1145/3307772.3328313}
\BIBentrySTDinterwordspacing

\bibitem{6038948}
\BIBentryALTinterwordspacing
A.~Walsh and T.~Fallon, ``Esb networks smart grid demonstration pilot,'' in \emph{2011 IEEE Power and Energy Society General Meeting}, 2011, pp. 1--6. [Online]. Available: \url{http://dx.doi.org/10.1109/PES.2011.6038948}
\BIBentrySTDinterwordspacing

\bibitem{GONZALEZGARRIDO2019381}
\BIBentryALTinterwordspacing
A.~González-Garrido, A.~Thingvad, H.~Gaztañaga, and M.~Marinelli, ``Full-scale electric vehicles penetration in the danish island of bornholm—optimal scheduling and battery degradation under driving constraints,'' \emph{Journal of Energy Storage}, vol.~23, pp. 381--391, 2019. [Online]. Available: \url{https://www.sciencedirect.com/science/article/pii/S2352152X18308405}
\BIBentrySTDinterwordspacing

\bibitem{DBLP:journals/tii/PertlCTMKK19}
\BIBentryALTinterwordspacing
M.~Pertl, F.~Carducci, M.~D. Tabone, M.~Marinelli, S.~Kiliccote, and E.~C. Kara, ``An equivalent time-variant storage model to harness {EV} flexibility: Forecast and aggregation,'' \emph{{IEEE} Trans. Ind. Informatics}, vol.~15, no.~4, pp. 1899--1910, 2019. [Online]. Available: \url{https://doi.org/10.1109/TII.2018.2865433}
\BIBentrySTDinterwordspacing

\bibitem{zhang2022extended}
\BIBentryALTinterwordspacing
S.~Zhang, D.~Fatih, F.~Abdulqadir, T.~Schwarz, and X.~Ma, ``Extended vehicle energy dataset (eved): an enhanced large-scale dataset for deep learning on vehicle trip energy consumption,'' \emph{arXiv preprint arXiv:2203.08630}, 2022. [Online]. Available: \url{https://arxiv.org/pdf/2203.08630.pdf}
\BIBentrySTDinterwordspacing

\bibitem{location2017}
\BIBentryALTinterwordspacing
T.~S.~D. Center, ``2004–2006 puget sound traffic choices study,'' 2017. [Online]. Available: \url{https://www.nrel.gov/transportation/secure-transportation-data/tsdc-puget-sound-traffic-study.html}
\BIBentrySTDinterwordspacing

\bibitem{c7j010-22}
\BIBentryALTinterwordspacing
M.~Piorkowski, N.~Sarafijanovic-Djukic, and M.~Grossglauser, ``Crawdad epfl/mobility,'' 2022. [Online]. Available: \url{https://dx.doi.org/10.15783/C7J010}
\BIBentrySTDinterwordspacing

\bibitem{6bxe-bp52-19}
\BIBentryALTinterwordspacing
G.~Storti~Gajani, G.~Gruosso, and J.-D. Valladolid, ``Electric vehicle speed, pedal, accel, voltage and current data over four different roads,'' 2019. [Online]. Available: \url{https://dx.doi.org/10.21227/6bxe-bp52}
\BIBentrySTDinterwordspacing

\bibitem{DBLP:journals/tits/OhLP22}
\BIBentryALTinterwordspacing
G.~Oh, D.~J. LeBlanc, and H.~Peng, ``Vehicle energy dataset (ved), {A} large-scale dataset for vehicle energy consumption research,'' \emph{{IEEE} Trans. Intell. Transp. Syst.}, vol.~23, no.~4, pp. 3302--3312, 2022. [Online]. Available: \url{https://doi.org/10.1109/TITS.2020.3035596}
\BIBentrySTDinterwordspacing

\bibitem{parking2014}
\BIBentryALTinterwordspacing
J.~Zmud, ``St. louis household travel survey, 2002,'' \emph{Inter-university Consortium for Political and Social Research [distributor]}, July 2014. [Online]. Available: \url{https://doi.org/10.3886/ICPSR35265.v1}
\BIBentrySTDinterwordspacing

\bibitem{DBLP:journals/corr/BatraPBSR14}
\BIBentryALTinterwordspacing
N.~Batra, O.~Parson, M.~Berges, A.~Singh, and A.~Rogers, ``A comparison of non-intrusive load monitoring methods for commercial and residential buildings,'' \emph{CoRR}, vol. abs/1408.6595, 2014. [Online]. Available: \url{http://arxiv.org/abs/1408.6595}
\BIBentrySTDinterwordspacing

\bibitem{Rashid2019}
\BIBentryALTinterwordspacing
H.~Rashid, P.~Singh, and A.~Singh, ``Data descriptor: I-blend, a campus-scale commercial and residential buildings electrical energy dataset,'' \emph{Scientific Data}, vol.~6, 2019. [Online]. Available: \url{https://www.scopus.com/inward/record.uri?eid=2-s2.0-85061859740&doi=10.1038%2fsdata.2019.15&partnerID=40&md5=a3e61a7fa95661e13b7a6eb5a40d9832}
\BIBentrySTDinterwordspacing

\bibitem{doi:10.1080/09613218.2020.1809983}
\BIBentryALTinterwordspacing
C.~S. Sathishkumar V~E and Y.~Cho, ``Efficient energy consumption prediction model for a data analytic-enabled industry building in a smart city,'' \emph{Building Research \& Information}, vol.~49, no.~1, pp. 127--143, 2021. [Online]. Available: \url{https://doi.org/10.1080/09613218.2020.1809983}
\BIBentrySTDinterwordspacing

\bibitem{kriechbaumer2018blond}
\BIBentryALTinterwordspacing
T.~Kriechbaumer and H.-A. Jacobsen, ``Blond, a building-level office environment dataset of typical electrical appliances,'' \emph{Scientific data}, vol.~5, no.~1, pp. 1--14, 2018. [Online]. Available: \url{https://doi.org/10.1038/sdata.2018.48}
\BIBentrySTDinterwordspacing

\bibitem{tong2021all}
\BIBentryALTinterwordspacing
K.~Tong, A.~S. Nagpure, and A.~Ramaswami, ``All urban areas’ energy use data across 640 districts in india for the year 2011,'' \emph{Scientific Data}, vol.~8, no.~1, p. 104, 2021. [Online]. Available: \url{https://doi.org/10.1038/s41597-021-00853-7}
\BIBentrySTDinterwordspacing

\bibitem{dong2011short}
Y.~Dong, J.~Wang, H.~Jiang, and J.~Wu, ``Short-term electricity price forecast based on the improved hybrid model,'' \emph{Energy Conversion and Management}, vol.~52, no. 8-9, pp. 2987--2995, 2011.

\bibitem{yang2017electricity}
Z.~Yang, L.~Ce, and L.~Lian, ``Electricity price forecasting by a hybrid model, combining wavelet transform, arma and kernel-based extreme learning machine methods,'' \emph{Applied Energy}, vol. 190, pp. 291--305, 2017.

\bibitem{peng2018effective}
L.~Peng, S.~Liu, R.~Liu, and L.~Wang, ``Effective long short-term memory with differential evolution algorithm for electricity price prediction,'' \emph{Energy}, vol. 162, pp. 1301--1314, 2018.

\bibitem{WU2023127286}
\BIBentryALTinterwordspacing
S.~Wu, Z.~Ding, J.~Wang, and D.~Shi, ``Unveiling bidding uncertainties in electricity markets: A bayesian deep learning framework based on accurate variational inference,'' \emph{Energy}, vol. 276, p. 127286, 2023. [Online]. Available: \url{https://www.sciencedirect.com/science/article/pii/S0360544223006801}
\BIBentrySTDinterwordspacing

\bibitem{https://doi.org/10.1002/er.6745}
\BIBentryALTinterwordspacing
M.~F. Elahe, M.~Jin, and P.~Zeng, ``Review of load data analytics using deep learning in smart grids: Open load datasets, methodologies, and application challenges,'' \emph{International Journal of Energy Research}, vol.~45, no.~10, pp. 14\,274--14\,305, 2021. [Online]. Available: \url{https://onlinelibrary.wiley.com/doi/abs/10.1002/er.6745}
\BIBentrySTDinterwordspacing

\bibitem{DBLP:journals/access/LiuSQLX19}
\BIBentryALTinterwordspacing
D.~Liu, Y.~Sun, Y.~Qu, B.~Li, and Y.~Xu, ``Analysis and accurate prediction of user's response behavior in incentive-based demand response,'' \emph{{IEEE} Access}, vol.~7, pp. 3170--3180, 2019. [Online]. Available: \url{https://doi.org/10.1109/ACCESS.2018.2889500}
\BIBentrySTDinterwordspacing

\bibitem{WANG2023121550}
\BIBentryALTinterwordspacing
Z.~Wang, L.~Dong, M.~Shi, J.~Qiao, H.~Jia, Y.~Mu, and T.~Pu, ``Market power modeling and restraint of aggregated prosumers in peer-to-peer energy trading: A game-theoretic approach,'' \emph{Applied Energy}, vol. 348, p. 121550, 2023. [Online]. Available: \url{https://www.sciencedirect.com/science/article/pii/S0306261923009145}
\BIBentrySTDinterwordspacing

\bibitem{KABALCI2016302}
\BIBentryALTinterwordspacing
Y.~Kabalci, ``A survey on smart metering and smart grid communication,'' \emph{Renewable and Sustainable Energy Reviews}, vol.~57, pp. 302--318, 2016. [Online]. Available: \url{https://www.sciencedirect.com/science/article/pii/S1364032115014975}
\BIBentrySTDinterwordspacing

\bibitem{9023471}
I.~Serban, S.~Céspedes, C.~Marinescu, C.~A. Azurdia-Meza, J.~S. Gómez, and D.~S. Hueichapan, ``Communication requirements in microgrids: A practical survey,'' \emph{IEEE Access}, vol.~8, pp. 47\,694--47\,712, 2020.

\bibitem{8676311}
M.~H. Rehmani, A.~Davy, B.~Jennings, and C.~Assi, ``Software defined networks-based smart grid communication: A comprehensive survey,'' \emph{IEEE Communications Surveys \& Tutorials}, vol.~21, no.~3, pp. 2637--2670, 2019.

\bibitem{SHAUKAT20181453}
\BIBentryALTinterwordspacing
N.~Shaukat, S.~Ali, C.~Mehmood, B.~Khan, M.~Jawad, U.~Farid, Z.~Ullah, S.~Anwar, and M.~Majid, ``A survey on consumers empowerment, communication technologies, and renewable generation penetration within smart grid,'' \emph{Renewable and Sustainable Energy Reviews}, vol.~81, pp. 1453--1475, 2018. [Online]. Available: \url{https://www.sciencedirect.com/science/article/pii/S1364032117308420}
\BIBentrySTDinterwordspacing

\bibitem{ABDULSALAM2023100121}
\BIBentryALTinterwordspacing
K.~A. Abdulsalam, J.~Adebisi, M.~Emezirinwune, and O.~Babatunde, ``An overview and multicriteria analysis of communication technologies for smart grid applications,'' \emph{e-Prime - Advances in Electrical Engineering, Electronics and Energy}, vol.~3, p. 100121, 2023. [Online]. Available: \url{https://www.sciencedirect.com/science/article/pii/S2772671123000165}
\BIBentrySTDinterwordspacing

\bibitem{WANG20113604}
\BIBentryALTinterwordspacing
W.~Wang, Y.~Xu, and M.~Khanna, ``A survey on the communication architectures in smart grid,'' \emph{Computer Networks}, vol.~55, no.~15, pp. 3604--3629, 2011. [Online]. Available: \url{https://www.sciencedirect.com/science/article/pii/S138912861100260X}
\BIBentrySTDinterwordspacing

\bibitem{Wen_Leung_Li_He_Kuo_2015}
M.~H. Wen, K.-C. Leung, V.~O. Li, X.~He, and C.-C.~J. Kuo, ``A survey on smart grid communication system,'' \emph{APSIPA Transactions on Signal and Information Processing}, vol.~4, p.~e5, 2015.

\bibitem{MARZAL20183610}
\BIBentryALTinterwordspacing
S.~Marzal, R.~Salas, R.~González-Medina, G.~Garcerá, and E.~Figueres, ``Current challenges and future trends in the field of communication architectures for microgrids,'' \emph{Renewable and Sustainable Energy Reviews}, vol.~82, pp. 3610--3622, 2018. [Online]. Available: \url{https://www.sciencedirect.com/science/article/pii/S1364032117314703}
\BIBentrySTDinterwordspacing

\bibitem{MAHMOOD2015248}
\BIBentryALTinterwordspacing
A.~Mahmood, N.~Javaid, and S.~Razzaq, ``A review of wireless communications for smart grid,'' \emph{Renewable and Sustainable Energy Reviews}, vol.~41, pp. 248--260, 2015. [Online]. Available: \url{https://www.sciencedirect.com/science/article/pii/S1364032114007126}
\BIBentrySTDinterwordspacing

\bibitem{6861946}
M.~Erol-Kantarci and H.~T. Mouftah, ``Energy-efficient information and communication infrastructures in the smart grid: A survey on interactions and open issues,'' \emph{IEEE Communications Surveys \& Tutorials}, vol.~17, no.~1, pp. 179--197, 2015.

\bibitem{8839117}
M.~Ghorbanian, S.~H. Dolatabadi, M.~Masjedi, and P.~Siano, ``Communication in smart grids: A comprehensive review on the existing and future communication and information infrastructures,'' \emph{IEEE Systems Journal}, vol.~13, no.~4, pp. 4001--4014, 2019.

\bibitem{6129368}
Z.~Fan, P.~Kulkarni, S.~Gormus, C.~Efthymiou, G.~Kalogridis, M.~Sooriyabandara, Z.~Zhu, S.~Lambotharan, and W.~H. Chin, ``Smart grid communications: Overview of research challenges, solutions, and standardization activities,'' \emph{IEEE Communications Surveys \& Tutorials}, vol.~15, no.~1, pp. 21--38, 2013.

\bibitem{9197633}
Q.~Zhou, M.~Shahidehpour, A.~Paaso, S.~Bahramirad, A.~Alabdulwahab, and A.~Abusorrah, ``Distributed control and communication strategies in networked microgrids,'' \emph{IEEE Communications Surveys \& Tutorials}, vol.~22, no.~4, pp. 2586--2633, 2020.

\bibitem{9759422}
N.~Suhaimy, N.~A.~M. Radzi, W.~S. H. M.~W. Ahmad, K.~H.~M. Azmi, and M.~A. Hannan, ``Current and future communication solutions for smart grids: A review,'' \emph{IEEE Access}, vol.~10, pp. 43\,639--43\,668, 2022.

\bibitem{ANCILLOTTI20131665}
\BIBentryALTinterwordspacing
E.~Ancillotti, R.~Bruno, and M.~Conti, ``The role of communication systems in smart grids: Architectures, technical solutions and research challenges,'' \emph{Computer Communications}, vol.~36, no.~17, pp. 1665--1697, 2013. [Online]. Available: \url{https://www.sciencedirect.com/science/article/pii/S0140366413002090}
\BIBentrySTDinterwordspacing

\bibitem{9806180}
D.~Said, ``A survey on information communication technologies in modern demand-side management for smart grids: Challenges, solutions, and opportunities,'' \emph{IEEE Engineering Management Review}, vol.~51, no.~1, pp. 76--107, 2023.

\bibitem{https://doi.org/10.1002/wcm.2258}
\BIBentryALTinterwordspacing
N.~Kayastha, D.~Niyato, E.~Hossain, and Z.~Han, ``Smart grid sensor data collection, communication, and networking: a tutorial,'' \emph{Wireless Communications and Mobile Computing}, vol.~14, no.~11, pp. 1055--1087, 2014. [Online]. Available: \url{https://onlinelibrary.wiley.com/doi/abs/10.1002/wcm.2258}
\BIBentrySTDinterwordspacing

\bibitem{REDDY2014180}
\BIBentryALTinterwordspacing
K.~Reddy, M.~Kumar, T.~Mallick, H.~Sharon, and S.~Lokeswaran, ``A review of integration, control, communication and metering (iccm) of renewable energy based smart grid,'' \emph{Renewable and Sustainable Energy Reviews}, vol.~38, pp. 180--192, 2014. [Online]. Available: \url{https://www.sciencedirect.com/science/article/pii/S1364032114003748}
\BIBentrySTDinterwordspacing

\bibitem{8325326}
M.~H. Rehmani, M.~Reisslein, A.~Rachedi, M.~Erol-Kantarci, and M.~Radenkovic, ``Integrating renewable energy resources into the smart grid: Recent developments in information and communication technologies,'' \emph{IEEE Transactions on Industrial Informatics}, vol.~14, no.~7, pp. 2814--2825, 2018.

\bibitem{s22155881}
\BIBentryALTinterwordspacing
G.~P. Reddy, Y.~V.~P. Kumar, and M.~K. Chakravarthi, ``Communication technologies for interoperable smart microgrids in urban energy community: A broad review of the state of the art, challenges, and research perspectives,'' \emph{Sensors}, vol.~22, no.~15, 2022. [Online]. Available: \url{https://www.mdpi.com/1424-8220/22/15/5881}
\BIBentrySTDinterwordspacing

\bibitem{6279592}
W.~Saad, Z.~Han, H.~V. Poor, and T.~Basar, ``Game-theoretic methods for the smart grid: An overview of microgrid systems, demand-side management, and smart grid communications,'' \emph{IEEE Signal Processing Magazine}, vol.~29, no.~5, pp. 86--105, 2012.

\bibitem{s21238087}
\BIBentryALTinterwordspacing
F.~E. Abrahamsen, Y.~Ai, and M.~Cheffena, ``Communication technologies for smart grid: A comprehensive survey,'' \emph{Sensors}, vol.~21, no.~23, 2021. [Online]. Available: \url{https://www.mdpi.com/1424-8220/21/23/8087}
\BIBentrySTDinterwordspacing

\bibitem{6157575}
Y.~Yan, Y.~Qian, H.~Sharif, and D.~Tipper, ``A survey on smart grid communication infrastructures: Motivations, requirements and challenges,'' \emph{IEEE Communications Surveys \& Tutorials}, vol.~15, no.~1, pp. 5--20, 2013.

\bibitem{10.1007/978-3-030-44758-8_48}
V.~Tiwari, S.~M. Dubey, H.~M. Dubey, and M.~Pandit, ``Smart grid communication: A survey of state-of-the-art,'' in \emph{Intelligent Computing Applications for Sustainable Real-World Systems}.\hskip 1em plus 0.5em minus 0.4em\relax Cham: Springer International Publishing, 2020, pp. 524--534.

\bibitem{9395820}
S.~Ghosh, C.~K. Chanda, and J.~K. Das, ``A comprehensive survey on communication technologies for a grid connected microgrid system,'' in \emph{2021 International Conference on Artificial Intelligence and Smart Systems (ICAIS)}, 2021, pp. 1525--1528.

\bibitem{7495234}
F.~Khan, A.~u. Rehman, M.~Arif, M.~Aftab, and B.~K. Jadoon, ``A survey of communication technologies for smart grid connectivity,'' in \emph{2016 International Conference on Computing, Electronic and Electrical Engineering (ICE Cube)}, 2016, pp. 256--261.

\bibitem{EMMANUEL2016133}
\BIBentryALTinterwordspacing
M.~Emmanuel and R.~Rayudu, ``Communication technologies for smart grid applications: A survey,'' \emph{Journal of Network and Computer Applications}, vol.~74, pp. 133--148, 2016. [Online]. Available: \url{https://www.sciencedirect.com/science/article/pii/S1084804516301734}
\BIBentrySTDinterwordspacing

\bibitem{Singh2022}
\BIBentryALTinterwordspacing
S.~V. Singh, A.~Khursheed, and Z.~Alam, \emph{Wired Communication Technologies and Networks for Smart Grid---A Review}.\hskip 1em plus 0.5em minus 0.4em\relax Singapore: Springer Singapore, 2022, pp. 183--195. [Online]. Available: \url{https://doi.org/10.1007/978-981-16-8012-0_15}
\BIBentrySTDinterwordspacing

\bibitem{DBLP:journals/winet/JhaAGPGKM21}
\BIBentryALTinterwordspacing
A.~V. Jha, B.~Appasani, A.~N. Ghazali, P.~Pattanayak, D.~S. Gurjar, E.~Kabalci, and D.~K. Mohanta, ``Smart grid cyber-physical systems: communication technologies, standards and challenges,'' \emph{Wirel. Networks}, vol.~27, no.~4, pp. 2595--2613, 2021. [Online]. Available: \url{https://doi.org/10.1007/s11276-021-02579-1}
\BIBentrySTDinterwordspacing

\bibitem{9465005}
G.~P. Reddy and Y.~V. Pavan~Kumar, ``Smart grid communication and networking: Review of standards,'' in \emph{2021 International Conference on Applied and Theoretical Electricity (ICATE)}, 2021, pp. 1--6.

\bibitem{DBLP:conf/iwcmc/VaidyaM18}
\BIBentryALTinterwordspacing
B.~Vaidya and H.~T. Mouftah, ``Deployment of secure {EV} charging system using open charge point protocol,'' in \emph{14th International Wireless Communications {\&} Mobile Computing Conference, {IWCMC} 2018, Limassol, Cyprus, June 25-29, 2018}.\hskip 1em plus 0.5em minus 0.4em\relax {IEEE}, 2018, pp. 922--927. [Online]. Available: \url{https://doi.org/10.1109/IWCMC.2018.8450489}
\BIBentrySTDinterwordspacing

\bibitem{hecht2024protocols}
C.~Hecht, J.~Figgener, and D.~U. Sauer, ``Protocols and interfaces for ev charging,'' in \emph{Next Generation Electrified Vehicles Optimised for the Infrastructure}.\hskip 1em plus 0.5em minus 0.4em\relax Springer, 2024, pp. 77--89.

\bibitem{DBLP:journals/tits/UmorenST21}
\BIBentryALTinterwordspacing
I.~A. Umoren, M.~Z. Shakir, and H.~Tabassum, ``Resource efficient vehicle-to-grid {(V2G)} communication systems for electric vehicle enabled microgrids,'' \emph{{IEEE} Trans. Intell. Transp. Syst.}, vol.~22, no.~7, pp. 4171--4180, 2021. [Online]. Available: \url{https://doi.org/10.1109/TITS.2020.3023899}
\BIBentrySTDinterwordspacing

\bibitem{kilic2023plug}
A.~Kilic, ``Plug and charge solutions with vehicle-to-grid communication,'' \emph{Electric Power Components and Systems}, vol.~51, no.~16, pp. 1786--1814, 2023.

\bibitem{DBLP:conf/acsac/ZhdanovaUHZHH22}
\BIBentryALTinterwordspacing
M.~Zhdanova, J.~Urbansky, A.~Hagemeier, D.~Zelle, I.~Herrmann, and D.~H{\"{o}}ffner, ``Local power grids at risk - an experimental and simulation-based analysis of attacks on vehicle-to-grid communication,'' in \emph{Annual Computer Security Applications Conference, {ACSAC} 2022, Austin, TX, USA, December 5-9, 2022}.\hskip 1em plus 0.5em minus 0.4em\relax {ACM}, 2022, pp. 42--55. [Online]. Available: \url{https://doi.org/10.1145/3564625.3568136}
\BIBentrySTDinterwordspacing

\bibitem{9437171}
S.~Ahmadzadeh, G.~Parr, and W.~Zhao, ``A review on communication aspects of demand response management for future 5g iot- based smart grids,'' \emph{IEEE Access}, vol.~9, pp. 77\,555--77\,571, 2021.

\bibitem{DBLP:journals/cn/KuzluPR14}
\BIBentryALTinterwordspacing
M.~Kuzlu, M.~Pipattanasomporn, and S.~Rahman, ``Communication network requirements for major smart grid applications in han, {NAN} and {WAN},'' \emph{Comput. Networks}, vol.~67, pp. 74--88, 2014. [Online]. Available: \url{https://doi.org/10.1016/j.comnet.2014.03.029}
\BIBentrySTDinterwordspacing

\bibitem{DBLP:journals/tii/DuanHSM22}
\BIBentryALTinterwordspacing
N.~Duan, C.~Huang, C.~Sun, and L.~Min, ``Smart meters enabling voltage monitoring and control: The last-mile voltage stability issue,'' \emph{{IEEE} Trans. Ind. Informatics}, vol.~18, no.~1, pp. 677--687, 2022. [Online]. Available: \url{https://doi.org/10.1109/TII.2021.3062628}
\BIBentrySTDinterwordspacing

\bibitem{ELAFIFI2024100634}
\BIBentryALTinterwordspacing
M.~I. El-Afifi, B.~E. Sedhom, S.~Padmanaban, and A.~A. Eladl, ``A review of iot-enabled smart energy hub systems: Rising, applications, challenges, and future prospects,'' \emph{Renewable Energy Focus}, vol.~51, p. 100634, 2024. [Online]. Available: \url{https://www.sciencedirect.com/science/article/pii/S175500842400098X}
\BIBentrySTDinterwordspacing

\bibitem{DBLP:journals/access/HossainKUSS19}
\BIBentryALTinterwordspacing
E.~Hossain, I.~Khan, F.~Un{-}Noor, S.~S. Sikander, and M.~S.~H. Sunny, ``Application of big data and machine learning in smart grid, and associated security concerns: {A} review,'' \emph{{IEEE} Access}, vol.~7, pp. 13\,960--13\,988, 2019. [Online]. Available: \url{https://doi.org/10.1109/ACCESS.2019.2894819}
\BIBentrySTDinterwordspacing

\bibitem{9264022}
\BIBentryALTinterwordspacing
S.~Mohammadi, M.~R. Hesamzadeh, A.~Vafamehr, and F.~Ferdowsi, ``A review of machine learning applications in electricity market studies,'' in \emph{2020 3rd International Colloquium on Intelligent Grid Metrology (SMAGRIMET)}, 2020, pp. 1--8. [Online]. Available: \url{https://doi.org/10.23919/SMAGRIMET48809.2020.9264022}
\BIBentrySTDinterwordspacing

\bibitem{RANGELMARTINEZ2021414}
\BIBentryALTinterwordspacing
D.~Rangel-Martinez, K.~Nigam, and L.~A. Ricardez-Sandoval, ``Machine learning on sustainable energy: A review and outlook on renewable energy systems, catalysis, smart grid and energy storage,'' \emph{Chemical Engineering Research and Design}, vol. 174, pp. 414--441, 2021. [Online]. Available: \url{https://www.sciencedirect.com/science/article/pii/S0263876221003312}
\BIBentrySTDinterwordspacing

\bibitem{IFAEI2023126432}
\BIBentryALTinterwordspacing
P.~Ifaei, M.~Nazari-Heris, A.~S. {Tayerani Charmchi}, S.~Asadi, and C.~Yoo, ``Sustainable energies and machine learning: An organized review of recent applications and challenges,'' \emph{Energy}, vol. 266, p. 126432, 2023. [Online]. Available: \url{https://www.sciencedirect.com/science/article/pii/S0360544222033187}
\BIBentrySTDinterwordspacing

\bibitem{HERNANDEZMATHEUS2022112651}
\BIBentryALTinterwordspacing
A.~Hernandez-Matheus, M.~Löschenbrand, K.~Berg, I.~Fuchs, M.~Aragüés-Peñalba, E.~Bullich-Massagué, and A.~Sumper, ``A systematic review of machine learning techniques related to local energy communities,'' \emph{Renewable and Sustainable Energy Reviews}, vol. 170, p. 112651, 2022. [Online]. Available: \url{https://www.sciencedirect.com/science/article/pii/S1364032122005433}
\BIBentrySTDinterwordspacing

\bibitem{MATIJASEVIC202212379}
\BIBentryALTinterwordspacing
T.~Matijašević, T.~Antić, and T.~Capuder, ``A systematic review of machine learning applications in the operation of smart distribution systems,'' \emph{Energy Reports}, vol.~8, pp. 12\,379--12\,407, 2022. [Online]. Available: \url{https://www.sciencedirect.com/science/article/pii/S2352484722017929}
\BIBentrySTDinterwordspacing

\bibitem{doi:10.1080/15567036.2020.1869867}
\BIBentryALTinterwordspacing
Y.~Deng, X.~Tang, Z.~Zhou, Y.~Yang, and F.~Niu, ``Application of machine learning algorithms in wind power: a review,'' \emph{Energy Sources, Part A: Recovery, Utilization, and Environmental Effects}, vol.~0, no.~0, pp. 1--22, 2021. [Online]. Available: \url{https://doi.org/10.1080/15567036.2020.1869867}
\BIBentrySTDinterwordspacing

\bibitem{GAVIRIA2022298}
\BIBentryALTinterwordspacing
J.~F. Gaviria, G.~Narváez, C.~Guillen, L.~F. Giraldo, and M.~Bressan, ``Machine learning in photovoltaic systems: A review,'' \emph{Renewable Energy}, vol. 196, pp. 298--318, 2022. [Online]. Available: \url{https://www.sciencedirect.com/science/article/pii/S0960148122009454}
\BIBentrySTDinterwordspacing

\bibitem{SIERLA2022104174}
\BIBentryALTinterwordspacing
S.~Sierla, M.~Pourakbari-Kasmaei, and V.~Vyatkin, ``A taxonomy of machine learning applications for virtual power plants and home/building energy management systems,'' \emph{Automation in Construction}, vol. 136, p. 104174, 2022. [Online]. Available: \url{https://www.sciencedirect.com/science/article/pii/S0926580522000474}
\BIBentrySTDinterwordspacing

\bibitem{IBRAHIM2020115237}
\BIBentryALTinterwordspacing
M.~S. Ibrahim, W.~Dong, and Q.~Yang, ``Machine learning driven smart electric power systems: Current trends and new perspectives,'' \emph{Applied Energy}, vol. 272, p. 115237, 2020. [Online]. Available: \url{https://www.sciencedirect.com/science/article/pii/S0306261920307492}
\BIBentrySTDinterwordspacing

\bibitem{mocanu2017machine}
\BIBentryALTinterwordspacing
E.~Mocanu, ``Machine learning applied to smart grids,'' \emph{({PhD} thesis) Technische Universiteit Eindhoven}, 2017. [Online]. Available: \url{https://pure.tue.nl/ws/portalfiles/portal/77436888/20171009_Mocanu.pdf}
\BIBentrySTDinterwordspacing

\bibitem{https://doi.org/10.1002/er.4333}
\BIBentryALTinterwordspacing
L.~Cheng and T.~Yu, ``A new generation of ai: A review and perspective on machine learning technologies applied to smart energy and electric power systems,'' \emph{International Journal of Energy Research}, vol.~43, no.~6, pp. 1928--1973, 2019. [Online]. Available: \url{https://onlinelibrary.wiley.com/doi/abs/10.1002/er.4333}
\BIBentrySTDinterwordspacing

\bibitem{ENTEZARI2023101017}
\BIBentryALTinterwordspacing
A.~Entezari, A.~Aslani, R.~Zahedi, and Y.~Noorollahi, ``Artificial intelligence and machine learning in energy systems: A bibliographic perspective,'' \emph{Energy Strategy Reviews}, vol.~45, p. 101017, 2023. [Online]. Available: \url{https://www.sciencedirect.com/science/article/pii/S2211467X22002115}
\BIBentrySTDinterwordspacing

\bibitem{ntakolia2022machine}
\BIBentryALTinterwordspacing
C.~Ntakolia, A.~Anagnostis, S.~Moustakidis, and N.~Karcanias, ``Machine learning applied on the district heating and cooling sector: A review,'' \emph{Energy Systems}, vol.~13, no.~1, pp. 1--30, 2022. [Online]. Available: \url{https://link.springer.com/content/pdf/10.1007/s12667-020-00405-9.pdf}
\BIBentrySTDinterwordspacing

\bibitem{seyedzadeh2018machine}
\BIBentryALTinterwordspacing
S.~Seyedzadeh, F.~P. Rahimian, I.~Glesk, and M.~Roper, ``Machine learning for estimation of building energy consumption and performance: a review,'' \emph{Visualization in Engineering}, vol.~6, pp. 1--20, 2018. [Online]. Available: \url{https://link.springer.com/content/pdf/10.1186/s40327-018-0064-7.pdf}
\BIBentrySTDinterwordspacing

\bibitem{DBLP:journals/csur/DjenouriLDB19}
\BIBentryALTinterwordspacing
D.~Djenouri, R.~Laidi, Y.~Djenouri, and I.~Balasingham, ``Machine learning for smart building applications: Review and taxonomy,'' \emph{{ACM} Comput. Surv.}, vol.~52, no.~2, pp. 24:1--24:36, 2019. [Online]. Available: \url{https://doi.org/10.1145/3311950}
\BIBentrySTDinterwordspacing

\bibitem{DBLP:journals/access/MassaoudiARCO21}
\BIBentryALTinterwordspacing
M.~Massaoudi, H.~Abu{-}Rub, S.~S. Refaat, I.~Chihi, and F.~S. Oueslati, ``Deep learning in smart grid technology: {A} review of recent advancements and future prospects,'' \emph{{IEEE} Access}, vol.~9, pp. 54\,558--54\,578, 2021. [Online]. Available: \url{https://doi.org/10.1109/ACCESS.2021.3071269}
\BIBentrySTDinterwordspacing

\bibitem{MISHRA2020104000}
\BIBentryALTinterwordspacing
M.~Mishra, J.~Nayak, B.~Naik, and A.~Abraham, ``Deep learning in electrical utility industry: A comprehensive review of a decade of research,'' \emph{Engineering Applications of Artificial Intelligence}, vol.~96, p. 104000, 2020. [Online]. Available: \url{https://www.sciencedirect.com/science/article/pii/S0952197620302943}
\BIBentrySTDinterwordspacing

\bibitem{WANG2019111799}
\BIBentryALTinterwordspacing
H.~Wang, Z.~Lei, X.~Zhang, B.~Zhou, and J.~Peng, ``A review of deep learning for renewable energy forecasting,'' \emph{Energy Conversion and Management}, vol. 198, p. 111799, 2019. [Online]. Available: \url{https://www.sciencedirect.com/science/article/pii/S0196890419307812}
\BIBentrySTDinterwordspacing

\bibitem{8468674}
\BIBentryALTinterwordspacing
D.~Zhang, X.~Han, and C.~Deng, ``Review on the research and practice of deep learning and reinforcement learning in smart grids,'' \emph{CSEE Journal of Power and Energy Systems}, vol.~4, no.~3, pp. 362--370, 2018. [Online]. Available: \url{https://doi.org/10.17775/CSEEJPES.2018.00520}
\BIBentrySTDinterwordspacing

\bibitem{DBLP:journals/pieee/LiYSYZC23}
\BIBentryALTinterwordspacing
Y.~Li, C.~Yu, M.~Shahidehpour, T.~Yang, Z.~Zeng, and T.~Chai, ``Deep reinforcement learning for smart grid operations: Algorithms, applications, and prospects,'' \emph{Proc. {IEEE}}, vol. 111, no.~9, pp. 1055--1096, 2023. [Online]. Available: \url{https://doi.org/10.1109/JPROC.2023.3303358}
\BIBentrySTDinterwordspacing

\bibitem{9275593}
\BIBentryALTinterwordspacing
D.~Cao, W.~Hu, J.~Zhao, G.~Zhang, B.~Zhang, Z.~Liu, Z.~Chen, and F.~Blaabjerg, ``Reinforcement learning and its applications in modern power and energy systems: A review,'' \emph{Journal of Modern Power Systems and Clean Energy}, vol.~8, no.~6, pp. 1029--1042, 2020. [Online]. Available: \url{10.35833/MPCE.2020.000552}
\BIBentrySTDinterwordspacing

\bibitem{PERERA2021110618}
\BIBentryALTinterwordspacing
A.~Perera and P.~Kamalaruban, ``Applications of reinforcement learning in energy systems,'' \emph{Renewable and Sustainable Energy Reviews}, vol. 137, p. 110618, 2021. [Online]. Available: \url{https://www.sciencedirect.com/science/article/pii/S1364032120309023}
\BIBentrySTDinterwordspacing

\bibitem{YANG2020145}
\BIBentryALTinterwordspacing
T.~Yang, L.~Zhao, W.~Li, and A.~Y. Zomaya, ``Reinforcement learning in sustainable energy and electric systems: a survey,'' \emph{Annual Reviews in Control}, vol.~49, pp. 145--163, 2020. [Online]. Available: \url{https://www.sciencedirect.com/science/article/pii/S1367578820300079}
\BIBentrySTDinterwordspacing

\bibitem{DILEEP20202589}
\BIBentryALTinterwordspacing
G.~Dileep, ``A survey on smart grid technologies and applications,'' \emph{Renewable Energy}, vol. 146, pp. 2589--2625, 2020. [Online]. Available: \url{https://www.sciencedirect.com/science/article/pii/S0960148119312790}
\BIBentrySTDinterwordspacing

\bibitem{xu2011real}
Y.~Xu, Z.~Dong, K.~Meng, R.~Zhang, and K.~Wong, ``Real-time transient stability assessment model using extreme learning machine,'' \emph{IET generation, transmission \& distribution}, vol.~5, no.~3, pp. 314--322, 2011.

\bibitem{wang2016power}
B.~Wang, B.~Fang, Y.~Wang, H.~Liu, and Y.~Liu, ``Power system transient stability assessment based on big data and the core vector machine,'' \emph{IEEE Transactions on Smart Grid}, vol.~7, no.~5, pp. 2561--2570, 2016.

\bibitem{DBLP:journals/tsg/AlshareefTM14}
\BIBentryALTinterwordspacing
S.~Alshareef, S.~Talwar, and W.~G. Morsi, ``A new approach based on wavelet design and machine learning for islanding detection of distributed generation,'' \emph{{IEEE} Trans. Smart Grid}, vol.~5, no.~4, pp. 1575--1583, 2014. [Online]. Available: \url{https://doi.org/10.1109/TSG.2013.2296598}
\BIBentrySTDinterwordspacing

\bibitem{wang2019novel}
S.~Wang and H.~Chen, ``A novel deep learning method for the classification of power quality disturbances using deep convolutional neural network,'' \emph{Applied energy}, vol. 235, pp. 1126--1140, 2019.

\bibitem{ma2017classification}
J.~Ma, J.~Zhang, L.~Xiao, K.~Chen, and J.~Wu, ``Classification of power quality disturbances via deep learning,'' \emph{IETE Technical Review}, vol.~34, no.~4, pp. 408--415, 2017.

\bibitem{rajiv2019long}
K.~Rajiv and M.~Tripathi, ``Long short-term memory-convolution neural network based hybrid deep learning approach for power quality events classification,'' \emph{Innovations in electronics and communication engineering. Lecture notes in networks and systems. Springer, Singapore}, 2019.

\bibitem{duan2019deep}
J.~Duan, D.~Shi, R.~Diao, H.~Li, Z.~Wang, B.~Zhang, D.~Bian, and Z.~Yi, ``Deep-reinforcement-learning-based autonomous voltage control for power grid operations,'' \emph{IEEE Transactions on Power Systems}, vol.~35, no.~1, pp. 814--817, 2019.

\bibitem{DBLP:journals/tii/DengWJTL19}
\BIBentryALTinterwordspacing
Y.~Deng, L.~Wang, H.~Jia, X.~Tong, and F.~Li, ``A sequence-to-sequence deep learning architecture based on bidirectional {GRU} for type recognition and time location of combined power quality disturbance,'' \emph{{IEEE} Trans. Ind. Informatics}, vol.~15, no.~8, pp. 4481--4493, 2019. [Online]. Available: \url{https://doi.org/10.1109/TII.2019.2895054}
\BIBentrySTDinterwordspacing

\bibitem{li2018classification}
C.-M. Li, Z.-X. Li, N.~Jia, Z.-L. Qi, and J.-H. Wu, ``Classification of power-quality disturbances using deep belief network,'' in \emph{2018 International Conference on Wavelet Analysis and Pattern Recognition (ICWAPR)}.\hskip 1em plus 0.5em minus 0.4em\relax IEEE, 2018, pp. 231--237.

\bibitem{zhang2017data}
S.~Zhang, Y.~Wang, M.~Liu, and Z.~Bao, ``Data-based line trip fault prediction in power systems using lstm networks and svm,'' \emph{Ieee Access}, vol.~6, pp. 7675--7686, 2017.

\bibitem{fu2018toward}
Y.-Y. Fu and H.-D. Chiang, ``Toward optimal multiperiod network reconfiguration for increasing the hosting capacity of distribution networks,'' \emph{IEEE Transactions on Power Delivery}, vol.~33, no.~5, pp. 2294--2304, 2018.

\bibitem{DBLP:journals/tii/GangwarMCS20}
\BIBentryALTinterwordspacing
P.~Gangwar, A.~Mallick, S.~Chakrabarti, and S.~N. Singh, ``Short-term forecasting-based network reconfiguration for unbalanced distribution systems with distributed generators,'' \emph{{IEEE} Trans. Ind. Informatics}, vol.~16, no.~7, pp. 4378--4389, 2020. [Online]. Available: \url{https://doi.org/10.1109/TII.2019.2946423}
\BIBentrySTDinterwordspacing

\bibitem{DBLP:journals/iotj/RenLSZGM24}
\BIBentryALTinterwordspacing
R.~Ren, Y.~Li, Q.~Sun, S.~Zhang, D.~W. Gao, and S.~Maharjan, ``Switched surplus-based distributed security dispatch for smart grid with persistent packet loss,'' \emph{{IEEE} Internet Things J.}, vol.~11, no.~4, pp. 6185--6198, 2024. [Online]. Available: \url{https://doi.org/10.1109/JIOT.2023.3311758}
\BIBentrySTDinterwordspacing

\bibitem{orgaz2022modeling}
A.~Orgaz, A.~Bello, and J.~Reneses, ``Modeling storage systems in electricity markets with high shares of renewable generation: A daily clustering approach,'' \emph{International Journal of Electrical Power \& Energy Systems}, vol. 137, p. 107706, 2022.

\bibitem{hua2019optimal}
H.~Hua, Y.~Qin, C.~Hao, and J.~Cao, ``Optimal energy management strategies for energy internet via deep reinforcement learning approach,'' \emph{Applied energy}, vol. 239, pp. 598--609, 2019.

\bibitem{6298960}
V.~C. Gungor, D.~Sahin, T.~Kocak, S.~Ergut, C.~Buccella, C.~Cecati, and G.~P. Hancke, ``A survey on smart grid potential applications and communication requirements,'' \emph{IEEE Transactions on Industrial Informatics}, vol.~9, no.~1, pp. 28--42, 2013.

\bibitem{wang2016deep1}
Y.~Wang, M.~Liu, and Z.~Bao, ``Deep learning neural network for power system fault diagnosis,'' in \emph{2016 35th Chinese control conference (CCC)}.\hskip 1em plus 0.5em minus 0.4em\relax IEEE, 2016, pp. 6678--6683.

\bibitem{guo2017deep}
M.-F. Guo, X.-D. Zeng, D.-Y. Chen, and N.-C. Yang, ``Deep-learning-based earth fault detection using continuous wavelet transform and convolutional neural network in resonant grounding distribution systems,'' \emph{IEEE Sensors Journal}, vol.~18, no.~3, pp. 1291--1300, 2017.

\bibitem{DBLP:journals/tsg/HeMW17}
\BIBentryALTinterwordspacing
Y.~He, G.~J. Mendis, and J.~Wei, ``Real-time detection of false data injection attacks in smart grid: {A} deep learning-based intelligent mechanism,'' \emph{{IEEE} Trans. Smart Grid}, vol.~8, no.~5, pp. 2505--2516, 2017. [Online]. Available: \url{https://doi.org/10.1109/TSG.2017.2703842}
\BIBentrySTDinterwordspacing

\bibitem{srivastava2021robust}
A.~Srivastava and S.~Parida, ``A robust fault detection and location prediction module using support vector machine and gaussian process regression for ac microgrid,'' \emph{IEEE Transactions on Industry Applications}, vol.~58, no.~1, pp. 930--939, 2021.

\bibitem{he2020intelligent}
L.~He, S.~Rong, and C.~Liu, ``An intelligent overcurrent protection algorithm of distribution systems with inverter based distributed energy resources,'' in \emph{2020 IEEE Energy Conversion Congress and Exposition (ECCE)}.\hskip 1em plus 0.5em minus 0.4em\relax IEEE, 2020, pp. 2746--2751.

\bibitem{DBLP:conf/globalsip/AhmadianMH18}
\BIBentryALTinterwordspacing
S.~Ahmadian, H.~Malki, and Z.~Han, ``Cyber attacks on smart energy grids using generative adverserial networks,'' in \emph{2018 {IEEE} Global Conference on Signal and Information Processing, GlobalSIP 2018, Anaheim, CA, USA, November 26-29, 2018}.\hskip 1em plus 0.5em minus 0.4em\relax {IEEE}, 2018, pp. 942--946. [Online]. Available: \url{https://doi.org/10.1109/GlobalSIP.2018.8646424}
\BIBentrySTDinterwordspacing

\bibitem{DBLP:journals/tsg/WeiWH20}
\BIBentryALTinterwordspacing
F.~Wei, Z.~Wan, and H.~He, ``Cyber-attack recovery strategy for smart grid based on deep reinforcement learning,'' \emph{{IEEE} Trans. Smart Grid}, vol.~11, no.~3, pp. 2476--2486, 2020. [Online]. Available: \url{https://doi.org/10.1109/TSG.2019.2956161}
\BIBentrySTDinterwordspacing

\bibitem{DBLP:journals/tifs/YanHZT17}
\BIBentryALTinterwordspacing
J.~Yan, H.~He, X.~Zhong, and Y.~Tang, ``Q-learning-based vulnerability analysis of smart grid against sequential topology attacks,'' \emph{{IEEE} Trans. Inf. Forensics Secur.}, vol.~12, no.~1, pp. 200--210, 2017. [Online]. Available: \url{https://doi.org/10.1109/TIFS.2016.2607701}
\BIBentrySTDinterwordspacing

\bibitem{DBLP:journals/tsg/CuiWC20}
\BIBentryALTinterwordspacing
M.~Cui, J.~Wang, and B.~Chen, ``Flexible machine learning-based cyberattack detection using spatiotemporal patterns for distribution systems,'' \emph{{IEEE} Trans. Smart Grid}, vol.~11, no.~2, pp. 1805--1808, 2020. [Online]. Available: \url{https://doi.org/10.1109/TSG.2020.2965797}
\BIBentrySTDinterwordspacing

\bibitem{patel2017nifty}
A.~Patel, H.~Alhussian, J.~M. Pedersen, B.~Bounabat, J.~C. J{\'u}nior, and S.~Katsikas, ``A nifty collaborative intrusion detection and prevention architecture for smart grid ecosystems,'' \emph{Computers \& Security}, vol.~64, pp. 92--109, 2017.

\bibitem{DBLP:journals/tsg/CuiWY19}
\BIBentryALTinterwordspacing
M.~Cui, J.~Wang, and M.~Yue, ``Machine learning-based anomaly detection for load forecasting under cyberattacks,'' \emph{{IEEE} Trans. Smart Grid}, vol.~10, no.~5, pp. 5724--5734, 2019. [Online]. Available: \url{https://doi.org/10.1109/TSG.2018.2890809}
\BIBentrySTDinterwordspacing

\bibitem{basumallik2019packet}
S.~Basumallik, R.~Ma, and S.~Eftekharnejad, ``Packet-data anomaly detection in pmu-based state estimator using convolutional neural network,'' \emph{International Journal of Electrical Power \& Energy Systems}, vol. 107, pp. 690--702, 2019.

\bibitem{DBLP:journals/tsg/WangSLCDD19}
\BIBentryALTinterwordspacing
J.~Wang, D.~Shi, Y.~Li, J.~Chen, H.~Ding, and X.~Duan, ``Distributed framework for detecting {PMU} data manipulation attacks with deep autoencoders,'' \emph{{IEEE} Trans. Smart Grid}, vol.~10, no.~4, pp. 4401--4410, 2019. [Online]. Available: \url{https://doi.org/10.1109/TSG.2018.2859339}
\BIBentrySTDinterwordspacing

\bibitem{li2019denial}
Y.~Li, P.~Zhang, and L.~Ma, ``Denial of service attack and defense method on load frequency control system,'' \emph{Journal of the Franklin Institute}, vol. 356, no.~15, pp. 8625--8645, 2019.

\bibitem{DBLP:conf/iasam/HafizAQH19}
\BIBentryALTinterwordspacing
F.~Hafiz, M.~A. Awal, A.~R. de~Queiroz, and I.~Husain, ``Real-time stochastic optimization of energy storage management using rolling horizon forecasts for residential {PV} applications,'' in \emph{2019 {IEEE} Industry Applications Society Annual Meeting, Baltimore, MD, USA, September 29 - Oct. 3, 2019}.\hskip 1em plus 0.5em minus 0.4em\relax {IEEE}, 2019, pp. 1--9. [Online]. Available: \url{https://doi.org/10.1109/IAS.2019.8912315}
\BIBentrySTDinterwordspacing

\bibitem{DBLP:conf/smartgridcomm/MbuwirKD18}
\BIBentryALTinterwordspacing
B.~V. Mbuwir, M.~Kaffash, and G.~Deconinck, ``Battery scheduling in a residential multi-carrier energy system using reinforcement learning,'' in \emph{2018 {IEEE} International Conference on Communications, Control, and Computing Technologies for Smart Grids, SmartGridComm 2018, Aalborg, Denmark, October 29-31, 2018}.\hskip 1em plus 0.5em minus 0.4em\relax {IEEE}, 2018, pp. 1--6. [Online]. Available: \url{https://doi.org/10.1109/SmartGridComm.2018.8587412}
\BIBentrySTDinterwordspacing

\bibitem{DBLP:journals/tsg/QiuNC16}
\BIBentryALTinterwordspacing
X.~Qiu, T.~A. Nguyen, and M.~L. Crow, ``Heterogeneous energy storage optimization for microgrids,'' \emph{{IEEE} Trans. Smart Grid}, vol.~7, no.~3, pp. 1453--1461, 2016. [Online]. Available: \url{https://doi.org/10.1109/TSG.2015.2461134}
\BIBentrySTDinterwordspacing

\bibitem{TUBALLA2016710}
\BIBentryALTinterwordspacing
M.~L. Tuballa and M.~L. Abundo, ``A review of the development of smart grid technologies,'' \emph{Renewable and Sustainable Energy Reviews}, vol.~59, pp. 710--725, 2016. [Online]. Available: \url{https://www.sciencedirect.com/science/article/pii/S1364032116000393}
\BIBentrySTDinterwordspacing

\bibitem{alamaniotis2019elm}
M.~Alamaniotis and G.~Karagiannis, ``Elm-fuzzy method for automated decision-making in price directed electricity markets,'' in \emph{2019 16th International Conference on the European Energy Market (EEM)}.\hskip 1em plus 0.5em minus 0.4em\relax IEEE, 2019, pp. 1--5.

\bibitem{DBLP:journals/tsg/HuL13}
\BIBentryALTinterwordspacing
Q.~Hu and F.~Li, ``Hardware design of smart home energy management system with dynamic price response,'' \emph{{IEEE} Trans. Smart Grid}, vol.~4, no.~4, pp. 1878--1887, 2013. [Online]. Available: \url{https://doi.org/10.1109/TSG.2013.2258181}
\BIBentrySTDinterwordspacing

\bibitem{DBLP:journals/tsg/LiWH20a}
\BIBentryALTinterwordspacing
H.~Li, Z.~Wan, and H.~He, ``Real-time residential demand response,'' \emph{{IEEE} Trans. Smart Grid}, vol.~11, no.~5, pp. 4144--4154, 2020. [Online]. Available: \url{https://doi.org/10.1109/TSG.2020.2978061}
\BIBentrySTDinterwordspacing

\bibitem{li2011predicting}
B.~Li, S.~Gangadhar, S.~Cheng, and P.~K. Verma, ``Predicting user comfort level using machine learning for smart grid environments,'' in \emph{ISGT 2011}.\hskip 1em plus 0.5em minus 0.4em\relax IEEE, 2011, pp. 1--6.

\bibitem{DBLP:journals/tsg/MengistuGCAP19}
\BIBentryALTinterwordspacing
M.~A. Mengistu, A.~A. Girmay, C.~Camarda, A.~Acquaviva, and E.~Patti, ``A cloud-based on-line disaggregation algorithm for home appliance loads,'' \emph{{IEEE} Trans. Smart Grid}, vol.~10, no.~3, pp. 3430--3439, 2019. [Online]. Available: \url{https://doi.org/10.1109/TSG.2018.2826844}
\BIBentrySTDinterwordspacing

\bibitem{DBLP:journals/tsg/CuiLLY19}
\BIBentryALTinterwordspacing
G.~Cui, B.~Liu, W.~Luan, and Y.~Yu, ``Estimation of target appliance electricity consumption using background filtering,'' \emph{{IEEE} Trans. Smart Grid}, vol.~10, no.~6, pp. 5920--5929, 2019. [Online]. Available: \url{https://doi.org/10.1109/TSG.2019.2892841}
\BIBentrySTDinterwordspacing

\bibitem{jiang2017big}
H.~Jiang, Y.~Li, Y.~Zhang, J.~J. Zhang, D.~W. Gao, E.~Muljadi, and Y.~Gu, ``Big data-based approach to detect, locate, and enhance the stability of an unplanned microgrid islanding,'' \emph{Journal of Energy Engineering}, vol. 143, no.~5, p. 04017045, 2017.

\bibitem{manohar2019enhancing}
M.~Manohar, E.~Koley, and S.~Ghosh, ``Enhancing the reliability of protection scheme for pv integrated microgrid by discriminating between array faults and symmetrical line faults using sparse auto encoder,'' \emph{IET Renewable Power Generation}, vol.~13, no.~2, pp. 308--317, 2019.

\bibitem{jurado2015hybrid}
S.~Jurado, {\`A}.~Nebot, F.~Mugica, and N.~Avellana, ``Hybrid methodologies for electricity load forecasting: Entropy-based feature selection with machine learning and soft computing techniques,'' \emph{Energy}, vol.~86, pp. 276--291, 2015.

\bibitem{wen2020load}
L.~Wen, K.~Zhou, and S.~Yang, ``Load demand forecasting of residential buildings using a deep learning model,'' \emph{Electric Power Systems Research}, vol. 179, p. 106073, 2020.

\bibitem{guo2018deep}
Z.~Guo, K.~Zhou, X.~Zhang, and S.~Yang, ``A deep learning model for short-term power load and probability density forecasting,'' \emph{Energy}, vol. 160, pp. 1186--1200, 2018.

\bibitem{tong2018efficient}
C.~Tong, J.~Li, C.~Lang, F.~Kong, J.~Niu, and J.~J. Rodrigues, ``An efficient deep model for day-ahead electricity load forecasting with stacked denoising auto-encoders,'' \emph{Journal of parallel and distributed computing}, vol. 117, pp. 267--273, 2018.

\bibitem{qiu2017empirical}
X.~Qiu, Y.~Ren, P.~N. Suganthan, and G.~A. Amaratunga, ``Empirical mode decomposition based ensemble deep learning for load demand time series forecasting,'' \emph{Applied soft computing}, vol.~54, pp. 246--255, 2017.

\bibitem{DBLP:journals/tsg/KongDJHXZ19}
\BIBentryALTinterwordspacing
W.~Kong, Z.~Y. Dong, Y.~Jia, D.~J. Hill, Y.~Xu, and Y.~Zhang, ``Short-term residential load forecasting based on {LSTM} recurrent neural network,'' \emph{{IEEE} Trans. Smart Grid}, vol.~10, no.~1, pp. 841--851, 2019. [Online]. Available: \url{https://doi.org/10.1109/TSG.2017.2753802}
\BIBentrySTDinterwordspacing

\bibitem{li2020machine}
X.~Li and R.~Yao, ``A machine-learning-based approach to predict residential annual space heating and cooling loads considering occupant behaviour,'' \emph{Energy}, vol. 212, p. 118676, 2020.

\bibitem{boukas2018real}
I.~Boukas, D.~Ernst, and B.~Corn{\'e}lusse, ``Real-time bidding strategies from micro-grids using reinforcement learning,'' 2018.

\bibitem{DBLP:journals/iotj/LuXXDPP19}
\BIBentryALTinterwordspacing
X.~Lu, X.~Xiao, L.~Xiao, C.~Dai, M.~Peng, and H.~V. Poor, ``Reinforcement learning-based microgrid energy trading with a reduced power plant schedule,'' \emph{{IEEE} Internet Things J.}, vol.~6, no.~6, pp. 10\,728--10\,737, 2019. [Online]. Available: \url{https://doi.org/10.1109/JIOT.2019.2941498}
\BIBentrySTDinterwordspacing

\bibitem{DBLP:conf/gamenets/XiaoDLZX17}
\BIBentryALTinterwordspacing
X.~Xiao, C.~Dai, Y.~Li, C.~Zhou, and L.~Xiao, ``Energy trading game for microgrids using reinforcement learning,'' in \emph{Game Theory for Networks - 7th International {EAI} Conference, GameNets 2017 Knoxville, TN, USA, May 9, 2017, Proceedings}, ser. Lecture Notes of the Institute for Computer Sciences, Social Informatics and Telecommunications Engineering, vol. 212.\hskip 1em plus 0.5em minus 0.4em\relax Springer, 2017, pp. 131--140. [Online]. Available: \url{https://doi.org/10.1007/978-3-319-67540-4\_12}
\BIBentrySTDinterwordspacing

\bibitem{DBLP:journals/tii/ZhangMKD19}
\BIBentryALTinterwordspacing
Y.~Zhang, K.~Meng, W.~Kong, and Z.~Y. Dong, ``Collaborative filtering-based electricity plan recommender system,'' \emph{{IEEE} Trans. Ind. Informatics}, vol.~15, no.~3, pp. 1393--1404, 2019. [Online]. Available: \url{https://doi.org/10.1109/TII.2018.2856842}
\BIBentrySTDinterwordspacing

\bibitem{DBLP:journals/tsg/KimZSL16}
\BIBentryALTinterwordspacing
B.~Kim, Y.~Zhang, M.~van~der Schaar, and J.~Lee, ``Dynamic pricing and energy consumption scheduling with reinforcement learning,'' \emph{{IEEE} Trans. Smart Grid}, vol.~7, no.~5, pp. 2187--2198, 2016. [Online]. Available: \url{https://doi.org/10.1109/TSG.2015.2495145}
\BIBentrySTDinterwordspacing

\bibitem{huo2021decision}
Y.~Huo, F.~Bouffard, and G.~Jo{\'o}s, ``Decision tree-based optimization for flexibility management for sustainable energy microgrids,'' \emph{Applied Energy}, vol. 290, p. 116772, 2021.

\bibitem{DBLP:journals/tsg/ChenS19}
\BIBentryALTinterwordspacing
T.~Chen and W.~Su, ``Indirect customer-to-customer energy trading with reinforcement learning,'' \emph{{IEEE} Trans. Smart Grid}, vol.~10, no.~4, pp. 4338--4348, 2019. [Online]. Available: \url{https://doi.org/10.1109/TSG.2018.2857449}
\BIBentrySTDinterwordspacing

\bibitem{zhou2019artificial}
S.~Zhou, Z.~Hu, W.~Gu, M.~Jiang, and X.-P. Zhang, ``Artificial intelligence based smart energy community management: A reinforcement learning approach,'' \emph{CSEE Journal of Power and Energy Systems}, vol.~5, no.~1, pp. 1--10, 2019.

\bibitem{DBLP:conf/isgteurope/ChenB19}
\BIBentryALTinterwordspacing
T.~Chen and S.~Bu, ``Realistic peer-to-peer energy trading model for microgrids using deep reinforcement learning,'' in \emph{2019 {IEEE} {PES} Innovative Smart Grid Technologies Europe, ISGT-Europe 2019, Bucharest, Romania, September 29 - October 2, 2019}.\hskip 1em plus 0.5em minus 0.4em\relax {IEEE}, 2019, pp. 1--5. [Online]. Available: \url{https://doi.org/10.1109/ISGTEurope.2019.8905731}
\BIBentrySTDinterwordspacing

\bibitem{peters2013reinforcement}
M.~Peters, W.~Ketter, M.~Saar-Tsechansky, and J.~Collins, ``A reinforcement learning approach to autonomous decision-making in smart electricity markets,'' \emph{Machine learning}, vol.~92, pp. 5--39, 2013.

\bibitem{DBLP:conf/ssci/SpinolaFFV17}
\BIBentryALTinterwordspacing
J.~Spinola, R.~Faia, P.~Faria, and Z.~A. Vale, ``Clustering optimization of distributed energy resources in support of an aggregator,'' in \emph{2017 {IEEE} Symposium Series on Computational Intelligence, {SSCI} 2017, Honolulu, HI, USA, November 27 - Dec. 1, 2017}.\hskip 1em plus 0.5em minus 0.4em\relax {IEEE}, 2017, pp. 1--6. [Online]. Available: \url{https://doi.org/10.1109/SSCI.2017.8285201}
\BIBentrySTDinterwordspacing

\bibitem{10738107}
\BIBentryALTinterwordspacing
H.~Zhang, S.~Zhang, S.~Maharjan, and Y.~Zhang, ``P4s: Privacy-preserving personalized pricing scheme for smart grid,'' in \emph{2024 IEEE International Conference on Communications, Control, and Computing Technologies for Smart Grids (SmartGridComm)}, 2024, pp. 21--26. [Online]. Available: \url{https://doi.org/10.1109/SmartGridComm60555.2024.10738107}
\BIBentrySTDinterwordspacing

\bibitem{lu2018dynamic}
R.~Lu, S.~H. Hong, and X.~Zhang, ``A dynamic pricing demand response algorithm for smart grid: Reinforcement learning approach,'' \emph{Applied energy}, vol. 220, pp. 220--230, 2018.

\bibitem{oliveira2009mascem}
P.~Oliveira, T.~Pinto, H.~Morais, Z.~A. Vale, and I.~Pra{\c{c}}a, ``Mascem-an electricity market simulator providing coalition support for virtual power players,'' in \emph{2009 15th International Conference on Intelligent System Applications to Power Systems}.\hskip 1em plus 0.5em minus 0.4em\relax IEEE, 2009, pp. 1--6.

\bibitem{salehizadeh2016application}
M.~R. Salehizadeh and S.~Soltaniyan, ``Application of fuzzy q-learning for electricity market modeling by considering renewable power penetration,'' \emph{Renewable and Sustainable Energy Reviews}, vol.~56, pp. 1172--1181, 2016.

\bibitem{chi2021data}
L.~Chi, H.~Su, E.~Zio, M.~Qadrdan, X.~Li, L.~Zhang, L.~Fan, J.~Zhou, Z.~Yang, and J.~Zhang, ``Data-driven reliability assessment method of integrated energy systems based on probabilistic deep learning and gaussian mixture model-hidden markov model,'' \emph{Renewable Energy}, vol. 174, pp. 952--970, 2021.

\bibitem{heuberger2020ev}
C.~F. Heuberger, P.~K. Bains, and N.~Mac~Dowell, ``The ev-olution of the power system: A spatio-temporal optimisation model to investigate the impact of electric vehicle deployment,'' \emph{Applied Energy}, vol. 257, p. 113715, 2020.

\bibitem{10738058}
\BIBentryALTinterwordspacing
S.~Zhang, S.~Maharjan, R.~Shahi, and X.~Ma, ``The impact of integration of renewable energy on imbalance settlement: Resilience analysis,'' in \emph{2024 IEEE International Conference on Communications, Control, and Computing Technologies for Smart Grids (SmartGridComm)}, 2024, pp. 98--104. [Online]. Available: \url{https://doi.org/10.1109/SmartGridComm60555.2024.10738058}
\BIBentrySTDinterwordspacing

\bibitem{DBLP:journals/tcyb/WangHLAC17}
\BIBentryALTinterwordspacing
H.~Wang, T.~Huang, X.~Liao, H.~Abu{-}Rub, and G.~Chen, ``Reinforcement learning for constrained energy trading games with incomplete information,'' \emph{{IEEE} Trans. Cybern.}, vol.~47, no.~10, pp. 3404--3416, 2017. [Online]. Available: \url{https://doi.org/10.1109/TCYB.2016.2539300}
\BIBentrySTDinterwordspacing

\bibitem{DBLP:journals/tsg/XuSNKMH19}
\BIBentryALTinterwordspacing
H.~Xu, H.~Sun, D.~Nikovski, S.~Kitamura, K.~Mori, and H.~Hashimoto, ``Deep reinforcement learning for joint bidding and pricing of load serving entity,'' \emph{{IEEE} Trans. Smart Grid}, vol.~10, no.~6, pp. 6366--6375, 2019. [Online]. Available: \url{https://doi.org/10.1109/TSG.2019.2903756}
\BIBentrySTDinterwordspacing

\bibitem{luo2019two}
S.~Luo and Y.~Weng, ``A two-stage supervised learning approach for electricity price forecasting by leveraging different data sources,'' \emph{Applied energy}, vol. 242, pp. 1497--1512, 2019.

\bibitem{papadimitriou2014forecasting}
T.~Papadimitriou, P.~Gogas, and E.~Stathakis, ``Forecasting energy markets using support vector machines,'' \emph{Energy Economics}, vol.~44, pp. 135--142, 2014.

\bibitem{atef2019comparative}
S.~Atef and A.~B. Eltawil, ``A comparative study using deep learning and support vector regression for electricity price forecasting in smart grids,'' in \emph{2019 IEEE 6th International Conference on Industrial Engineering and Applications (ICIEA)}.\hskip 1em plus 0.5em minus 0.4em\relax IEEE, 2019, pp. 603--607.

\bibitem{DBLP:conf/icmla/DeepaNBPST18}
\BIBentryALTinterwordspacing
S.~N. Deepa, A.~Nagarajan, G.~B, K.~P, J.~S, and A.~A.~V. Tangaradjou, ``Adaptive regularized {ELM} and improved {VMD} method for multi-step ahead electricity price forecasting,'' in \emph{17th {IEEE} International Conference on Machine Learning and Applications, {ICMLA} 2018, Orlando, FL, USA, December 17-20, 2018}, M.~A. Wani, M.~M. Kantardzic, M.~S. Mouchaweh, J.~Gama, and E.~Lughofer, Eds.\hskip 1em plus 0.5em minus 0.4em\relax {IEEE}, 2018, pp. 1255--1260. [Online]. Available: \url{https://doi.org/10.1109/ICMLA.2018.00204}
\BIBentrySTDinterwordspacing

\bibitem{DBLP:journals/tsg/CarriereK19}
\BIBentryALTinterwordspacing
T.~Carriere and G.~Kariniotakis, ``An integrated approach for value-oriented energy forecasting and data-driven decision-making application to renewable energy trading,'' \emph{{IEEE} Trans. Smart Grid}, vol.~10, no.~6, pp. 6933--6944, 2019. [Online]. Available: \url{https://doi.org/10.1109/TSG.2019.2914379}
\BIBentrySTDinterwordspacing

\bibitem{DBLP:conf/iisa/BhagatAF19}
\BIBentryALTinterwordspacing
M.~Bhagat, M.~Alamaniotis, and A.~Fevgas, ``Extreme interval electricity price forecasting of wholesale markets integrating {ELM} and fuzzy inference,'' in \emph{10th International Conference on Information, Intelligence, Systems and Applications, {IISA} 2019, Patras, Greece, July 15-17, 2019}, N.~G. Bourbakis, G.~A. Tsihrintzis, and M.~Virvou, Eds.\hskip 1em plus 0.5em minus 0.4em\relax {IEEE}, 2019, pp. 1--4. [Online]. Available: \url{https://doi.org/10.1109/IISA.2019.8900703}
\BIBentrySTDinterwordspacing

\bibitem{fan2009forecasting}
S.~Fan, J.~R. Liao, R.~Yokoyama, L.~Chen, and W.-J. Lee, ``Forecasting the wind generation using a two-stage network based on meteorological information,'' \emph{IEEE Transactions on Energy Conversion}, vol.~24, no.~2, pp. 474--482, 2009.

\bibitem{DBLP:journals/tsg/LeeB14a}
\BIBentryALTinterwordspacing
D.~Lee and R.~Baldick, ``Short-term wind power ensemble prediction based on gaussian processes and neural networks,'' \emph{{IEEE} Trans. Smart Grid}, vol.~5, no.~1, pp. 501--510, 2014. [Online]. Available: \url{https://doi.org/10.1109/TSG.2013.2280649}
\BIBentrySTDinterwordspacing

\bibitem{salcedo2014feature}
S.~Salcedo-Sanz, A.~Pastor-S{\'a}nchez, L.~Prieto, A.~Blanco-Aguilera, and R.~Garc{\'\i}a-Herrera, ``Feature selection in wind speed prediction systems based on a hybrid coral reefs optimization--extreme learning machine approach,'' \emph{Energy Conversion and Management}, vol.~87, pp. 10--18, 2014.

\bibitem{zhang2016deterministic}
Y.~Zhang, K.~Liu, L.~Qin, and X.~An, ``Deterministic and probabilistic interval prediction for short-term wind power generation based on variational mode decomposition and machine learning methods,'' \emph{Energy Conversion and Management}, vol. 112, pp. 208--219, 2016.

\bibitem{yeh2014forecasting}
W.-C. Yeh, Y.-M. Yeh, P.-C. Chang, Y.-C. Ke, and V.~Chung, ``Forecasting wind power in the mai liao wind farm based on the multi-layer perceptron artificial neural network model with improved simplified swarm optimization,'' \emph{International Journal of Electrical Power \& Energy Systems}, vol.~55, pp. 741--748, 2014.

\bibitem{rahmani2013hybrid}
R.~Rahmani, R.~Yusof, M.~Seyedmahmoudian, and S.~Mekhilef, ``Hybrid technique of ant colony and particle swarm optimization for short term wind energy forecasting,'' \emph{Journal of Wind Engineering and Industrial Aerodynamics}, vol. 123, pp. 163--170, 2013.

\bibitem{chaouachi2010novel}
A.~Chaouachi, R.~M. Kamel, and K.~Nagasaka, ``A novel multi-model neuro-fuzzy-based mppt for three-phase grid-connected photovoltaic system,'' \emph{Solar energy}, vol.~84, no.~12, pp. 2219--2229, 2010.

\bibitem{yang2014weather}
H.-T. Yang, C.-M. Huang, Y.-C. Huang, and Y.-S. Pai, ``A weather-based hybrid method for 1-day ahead hourly forecasting of pv power output,'' \emph{IEEE transactions on sustainable energy}, vol.~5, no.~3, pp. 917--926, 2014.

\bibitem{DBLP:conf/aaai/HeinermannK15}
\BIBentryALTinterwordspacing
J.~Heinermann and O.~Kramer, ``On heterogeneous machine learning ensembles for wind power prediction,'' in \emph{Computational Sustainability, Papers from the 2015 {AAAI} Workshop, Austin, Texas, USA, January 26, 2015}, ser. {AAAI} Technical Report, B.~Dilkina, S.~Ermon, R.~A. Hutchinson, and D.~Sheldon, Eds., vol. {WS-15-06}.\hskip 1em plus 0.5em minus 0.4em\relax {AAAI} Press, 2015. [Online]. Available: \url{http://aaai.org/ocs/index.php/WS/AAAIW15/paper/view/10081}
\BIBentrySTDinterwordspacing

\bibitem{DBLP:journals/tnn/AkFZ16}
\BIBentryALTinterwordspacing
R.~Ak, O.~Fink, and E.~Zio, ``Two machine learning approaches for short-term wind speed time-series prediction,'' \emph{{IEEE} Trans. Neural Networks Learn. Syst.}, vol.~27, no.~8, pp. 1734--1747, 2016. [Online]. Available: \url{https://doi.org/10.1109/TNNLS.2015.2418739}
\BIBentrySTDinterwordspacing

\bibitem{liu2018smart}
H.~Liu, X.~Mi, and Y.~Li, ``Smart deep learning based wind speed prediction model using wavelet packet decomposition, convolutional neural network and convolutional long short term memory network,'' \emph{Energy Conversion and Management}, vol. 166, pp. 120--131, 2018.

\bibitem{zhang2015predictive}
C.-Y. Zhang, C.~P. Chen, M.~Gan, and L.~Chen, ``Predictive deep boltzmann machine for multiperiod wind speed forecasting,'' \emph{IEEE Transactions on Sustainable Energy}, vol.~6, no.~4, pp. 1416--1425, 2015.

\bibitem{wang2016deep}
H.~Wang, G.~Wang, G.~Li, J.~Peng, and Y.~Liu, ``Deep belief network based deterministic and probabilistic wind speed forecasting approach,'' \emph{Applied energy}, vol. 182, pp. 80--93, 2016.

\bibitem{hu2020forecasting}
H.~Hu, L.~Wang, and S.-X. Lv, ``Forecasting energy consumption and wind power generation using deep echo state network,'' \emph{Renewable Energy}, vol. 154, pp. 598--613, 2020.

\bibitem{demolli2019wind}
H.~Demolli, A.~S. Dokuz, A.~Ecemis, and M.~Gokcek, ``Wind power forecasting based on daily wind speed data using machine learning algorithms,'' \emph{Energy Conversion and Management}, vol. 198, p. 111823, 2019.

\bibitem{zhang2020adaptive}
J.~Zhang, Z.~Tan, and Y.~Wei, ``An adaptive hybrid model for day-ahead photovoltaic output power prediction,'' \emph{Journal of Cleaner Production}, vol. 244, p. 118858, 2020.

\bibitem{yan2019uncertainty}
J.~Yan, H.~Zhang, Y.~Liu, S.~Han, and L.~Li, ``Uncertainty estimation for wind energy conversion by probabilistic wind turbine power curve modelling,'' \emph{Applied energy}, vol. 239, pp. 1356--1370, 2019.

\bibitem{shadab2019box}
A.~Shadab, S.~Said, and S.~Ahmad, ``Box--jenkins multiplicative arima modeling for prediction of solar radiation: a case study,'' \emph{International Journal of Energy and Water Resources}, vol.~3, pp. 305--318, 2019.

\bibitem{DBLP:journals/iotj/TangMWN18}
\BIBentryALTinterwordspacing
N.~Tang, S.~Mao, Y.~Wang, and R.~M. Nelms, ``Solar power generation forecasting with a lasso-based approach,'' \emph{{IEEE} Internet Things J.}, vol.~5, no.~2, pp. 1090--1099, 2018. [Online]. Available: \url{https://doi.org/10.1109/JIOT.2018.2812155}
\BIBentrySTDinterwordspacing

\bibitem{marvuglia2012monitoring}
A.~Marvuglia and A.~Messineo, ``Monitoring of wind farms’ power curves using machine learning techniques,'' \emph{Applied Energy}, vol.~98, pp. 574--583, 2012.

\bibitem{lu2020achieving}
Q.~L{\"u}, X.~Liao, H.~Li, and T.~Huang, ``Achieving acceleration for distributed economic dispatch in smart grids over directed networks,'' \emph{IEEE Transactions on Network Science and Engineering}, vol.~7, no.~3, pp. 1988--1999, 2020.

\bibitem{DBLP:journals/corr/ChenWKZ17}
\BIBentryALTinterwordspacing
Y.~Chen, Y.~Wang, D.~S. Kirschen, and B.~Zhang, ``Model-free renewable scenario generation using generative adversarial networks,'' \emph{CoRR}, vol. abs/1707.09676, 2017. [Online]. Available: \url{http://arxiv.org/abs/1707.09676}
\BIBentrySTDinterwordspacing

\bibitem{DBLP:journals/tsg/VandaelCEHD15}
\BIBentryALTinterwordspacing
S.~Vandael, B.~Claessens, D.~Ernst, T.~Holvoet, and G.~Deconinck, ``Reinforcement learning of heuristic {EV} fleet charging in a day-ahead electricity market,'' \emph{{IEEE} Trans. Smart Grid}, vol.~6, no.~4, pp. 1795--1805, 2015. [Online]. Available: \url{https://doi.org/10.1109/TSG.2015.2393059}
\BIBentrySTDinterwordspacing

\bibitem{DBLP:journals/tsg/WanLHP19}
\BIBentryALTinterwordspacing
Z.~Wan, H.~Li, H.~He, and D.~V. Prokhorov, ``Model-free real-time {EV} charging scheduling based on deep reinforcement learning,'' \emph{{IEEE} Trans. Smart Grid}, vol.~10, no.~5, pp. 5246--5257, 2019. [Online]. Available: \url{https://doi.org/10.1109/TSG.2018.2879572}
\BIBentrySTDinterwordspacing

\bibitem{DBLP:journals/tsg/QianSWS20}
\BIBentryALTinterwordspacing
T.~Qian, C.~Shao, X.~Wang, and M.~Shahidehpour, ``Deep reinforcement learning for {EV} charging navigation by coordinating smart grid and intelligent transportation system,'' \emph{{IEEE} Trans. Smart Grid}, vol.~11, no.~2, pp. 1714--1723, 2020. [Online]. Available: \url{https://doi.org/10.1109/TSG.2019.2942593}
\BIBentrySTDinterwordspacing

\bibitem{DBLP:journals/tsg/BedoyaGG19}
\BIBentryALTinterwordspacing
K.~L.~L. Bedoya, C.~Gagn{\'{e}}, and M.~Gardner, ``Demand-side management using deep learning for smart charging of electric vehicles,'' \emph{{IEEE} Trans. Smart Grid}, vol.~10, no.~3, pp. 2683--2691, 2019. [Online]. Available: \url{https://doi.org/10.1109/TSG.2018.2808247}
\BIBentrySTDinterwordspacing

\bibitem{DBLP:journals/tii/LiHCZHCB22}
\BIBentryALTinterwordspacing
S.~Li, W.~Hu, D.~Cao, Z.~Zhang, Q.~Huang, Z.~Chen, and F.~Blaabjerg, ``A multiagent deep reinforcement learning based approach for the optimization of transformer life using coordinated electric vehicles,'' \emph{{IEEE} Trans. Ind. Informatics}, vol.~18, no.~11, pp. 7639--7652, 2022. [Online]. Available: \url{https://doi.org/10.1109/TII.2021.3139650}
\BIBentrySTDinterwordspacing

\bibitem{DBLP:conf/ivs/Zhang21}
\BIBentryALTinterwordspacing
S.~Zhang, ``An energy consumption model for electrical vehicle networks via extended federated-learning,'' in \emph{{IEEE} Intelligent Vehicles Symposium, {IV} 2021, Nagoya, Japan, July 11-17, 2021}.\hskip 1em plus 0.5em minus 0.4em\relax {IEEE}, 2021, pp. 354--361. [Online]. Available: \url{https://doi.org/10.1109/IV48863.2021.9575223}
\BIBentrySTDinterwordspacing

\bibitem{DBLP:journals/tii/WangTWP21}
\BIBentryALTinterwordspacing
Y.~Wang, H.~Tan, Y.~Wu, and J.~Peng, ``Hybrid electric vehicle energy management with computer vision and deep reinforcement learning,'' \emph{{IEEE} Trans. Ind. Informatics}, vol.~17, no.~6, pp. 3857--3868, 2021. [Online]. Available: \url{https://doi.org/10.1109/TII.2020.3015748}
\BIBentrySTDinterwordspacing

\bibitem{qi2019deep}
X.~Qi, Y.~Luo, G.~Wu, K.~Boriboonsomsin, and M.~Barth, ``Deep reinforcement learning enabled self-learning control for energy efficient driving,'' \emph{Transportation Research Part C: Emerging Technologies}, vol.~99, pp. 67--81, 2019.

\bibitem{8333659}
R.~E. Pérez-Guzmán, Y.~Salgueiro-Sicilia, and M.~Rivera, ``Communication systems and security issues in smart microgrids,'' in \emph{2017 IEEE Southern Power Electronics Conference (SPEC)}, 2017, pp. 1--6.

\bibitem{inventions8040084}
\BIBentryALTinterwordspacing
A.~Aghmadi, H.~Hussein, K.~H. Polara, and O.~Mohammed, ``A comprehensive review of architecture, communication, and cybersecurity in networked microgrid systems,'' \emph{Inventions}, vol.~8, no.~4, 2023. [Online]. Available: \url{https://www.mdpi.com/2411-5134/8/4/84}
\BIBentrySTDinterwordspacing

\bibitem{6141833}
Y.~Yan, Y.~Qian, H.~Sharif, and D.~Tipper, ``A survey on cyber security for smart grid communications,'' \emph{IEEE Communications Surveys \& Tutorials}, vol.~14, no.~4, pp. 998--1010, 2012.

\bibitem{SOVACOOL2020109663}
\BIBentryALTinterwordspacing
B.~K. Sovacool and D.~D. {Furszyfer Del Rio}, ``Smart home technologies in europe: A critical review of concepts, benefits, risks and policies,'' \emph{Renewable and Sustainable Energy Reviews}, vol. 120, p. 109663, 2020. [Online]. Available: \url{https://www.sciencedirect.com/science/article/pii/S1364032119308688}
\BIBentrySTDinterwordspacing

\bibitem{CHRISFOREMAN201594}
\BIBentryALTinterwordspacing
J.~{Chris Foreman} and D.~Gurugubelli, ``Identifying the cyber attack surface of the advanced metering infrastructure,'' \emph{The Electricity Journal}, vol.~28, no.~1, pp. 94--103, 2015. [Online]. Available: \url{https://www.sciencedirect.com/science/article/pii/S1040619014002899}
\BIBentrySTDinterwordspacing

\bibitem{PRIYADARSHINI2021107204}
\BIBentryALTinterwordspacing
I.~Priyadarshini, R.~Kumar, R.~Sharma, P.~K. Singh, and S.~C. Satapathy, ``Identifying cyber insecurities in trustworthy space and energy sector for smart grids,'' \emph{Computers \& Electrical Engineering}, vol.~93, p. 107204, 2021. [Online]. Available: \url{https://www.sciencedirect.com/science/article/pii/S0045790621002007}
\BIBentrySTDinterwordspacing

\bibitem{REDA2022112423}
\BIBentryALTinterwordspacing
H.~T. Reda, A.~Anwar, and A.~Mahmood, ``Comprehensive survey and taxonomies of false data injection attacks in smart grids: attack models, targets, and impacts,'' \emph{Renewable and Sustainable Energy Reviews}, vol. 163, p. 112423, 2022. [Online]. Available: \url{https://www.sciencedirect.com/science/article/pii/S1364032122003306}
\BIBentrySTDinterwordspacing

\bibitem{BRETAS201943}
\BIBentryALTinterwordspacing
A.~S. Bretas, N.~G. Bretas, and B.~E. Carvalho, ``Further contributions to smart grids cyber-physical security as a malicious data attack: Proof and properties of the parameter error spreading out to the measurements and a relaxed correction model,'' \emph{International Journal of Electrical Power \& Energy Systems}, vol. 104, pp. 43--51, 2019. [Online]. Available: \url{https://www.sciencedirect.com/science/article/pii/S0142061518303946}
\BIBentrySTDinterwordspacing

\bibitem{RYU2024113851}
\BIBentryALTinterwordspacing
D.-H. Ryu and K.-J. Kim, ``The influence of information privacy concerns and perceived electricity usage habits on the usage intention of advanced metering infrastructure,'' \emph{Renewable and Sustainable Energy Reviews}, vol. 189, p. 113851, 2024. [Online]. Available: \url{https://www.sciencedirect.com/science/article/pii/S1364032123007098}
\BIBentrySTDinterwordspacing

\bibitem{VONLOESSL2023113645}
\BIBentryALTinterwordspacing
V.~{von Loessl}, ``Smart meter-related data privacy concerns and dynamic electricity tariffs: Evidence from a stated choice experiment,'' \emph{Energy Policy}, vol. 180, p. 113645, 2023. [Online]. Available: \url{https://doi.org/10.1016/j.enpol.2023.113645}
\BIBentrySTDinterwordspacing

\bibitem{SCHALLEHN2022112756}
\BIBentryALTinterwordspacing
F.~Schallehn and K.~Valogianni, ``Sustainability awareness and smart meter privacy concerns: The cases of us and germany,'' \emph{Energy Policy}, vol. 161, p. 112756, 2022. [Online]. Available: \url{https://doi.org/10.1016/j.enpol.2021.112756}
\BIBentrySTDinterwordspacing

\bibitem{zhang2020privacy}
\BIBentryALTinterwordspacing
S.~Zhang and E.~M. Schiller, ``Privacy-preservation and the automotives,'' \emph{2020 AutoSPADA workshop, Gothenburg, Sweden}, 2020. [Online]. Available: \url{https://doi.org/10.13140/RG.2.2.20721.86}
\BIBentrySTDinterwordspacing

\bibitem{snow2022solar}
S.~Snow, K.~Chadwick, N.~Horrocks, A.~Chapman, and M.~Glencross, ``Do solar households want demand response and shared electricity data? exploring motivation, ability and opportunity in australia,'' \emph{Energy Research \& Social Science}, vol.~87, p. 102480, 2022.

\bibitem{jaradat2015internet}
M.~Jaradat, M.~Jarrah, A.~Bousselham, Y.~Jararweh, and M.~Al-Ayyoub, ``The internet of energy: smart sensor networks and big data management for smart grid,'' \emph{Procedia Computer Science}, vol.~56, pp. 592--597, 2015.

\bibitem{han2016ip2dm}
W.~Han and Y.~Xiao, ``Ip2dm: integrated privacy-preserving data management architecture for smart grid v2g networks,'' \emph{Wireless Communications and Mobile Computing}, vol.~16, no.~17, pp. 2956--2974, 2016.

\bibitem{DBLP:conf/chi/WilkinsCL20}
\BIBentryALTinterwordspacing
D.~J. Wilkins, R.~Chitchyan, and M.~Levine, ``Peer-to-peer energy markets: Understanding the values of collective and community trading,'' in \emph{{CHI} '20: {CHI} Conference on Human Factors in Computing Systems, Honolulu, HI, USA, April 25-30, 2020}, R.~Bernhaupt, F.~F. Mueller, D.~Verweij, J.~Andres, J.~McGrenere, A.~Cockburn, I.~Avellino, A.~Goguey, P.~Bj{\o}n, S.~Zhao, B.~P. Samson, and R.~Kocielnik, Eds.\hskip 1em plus 0.5em minus 0.4em\relax {ACM}, 2020, pp. 1--14. [Online]. Available: \url{https://doi.org/10.1145/3313831.3376135}
\BIBentrySTDinterwordspacing

\bibitem{photovoltaics2018ieee}
D.~G. Photovoltaics and E.~Storage, ``Ieee standard for interconnection and interoperability of distributed energy resources with associated electric power systems interfaces,'' \emph{IEEE std}, vol. 1547, pp. 1547--2018, 2018.

\bibitem{morstyn2018using}
T.~Morstyn, N.~Farrell, S.~J. Darby, and M.~D. McCulloch, ``Using peer-to-peer energy-trading platforms to incentivize prosumers to form federated power plants,'' \emph{Nature energy}, vol.~3, no.~2, pp. 94--101, 2018.

\bibitem{li2010compressed}
H.~Li, R.~Mao, L.~Lai, and R.~C. Qiu, ``Compressed meter reading for delay-sensitive and secure load report in smart grid,'' in \emph{2010 First IEEE international conference on smart grid communications}.\hskip 1em plus 0.5em minus 0.4em\relax IEEE, 2010, pp. 114--119.

\bibitem{efthymiou2010smart}
C.~Efthymiou and G.~Kalogridis, ``Smart grid privacy via anonymization of smart metering data,'' in \emph{2010 first IEEE international conference on smart grid communications}.\hskip 1em plus 0.5em minus 0.4em\relax IEEE, 2010, pp. 238--243.

\bibitem{antoun2020detailed}
J.~Antoun, M.~E. Kabir, B.~Moussa, R.~Atallah, and C.~Assi, ``A detailed security assessment of the ev charging ecosystem,'' \emph{IEEE Network}, vol.~34, no.~3, pp. 200--207, 2020.

\bibitem{DBLP:journals/tsg/LiDN17}
\BIBentryALTinterwordspacing
H.~Li, G.~D{\'{a}}n, and K.~Nahrstedt, ``Portunes+: Privacy-preserving fast authentication for dynamic electric vehicle charging,'' \emph{{IEEE} Trans. Smart Grid}, vol.~8, no.~5, pp. 2305--2313, 2017. [Online]. Available: \url{https://doi.org/10.1109/TSG.2016.2522379}
\BIBentrySTDinterwordspacing

\bibitem{DBLP:journals/tvt/Pazos-RevillaAG18}
\BIBentryALTinterwordspacing
M.~Pazos{-}Revilla, A.~Alsharif, S.~Gunukula, T.~N. Guo, M.~Mahmoud, and X.~Shen, ``Secure and privacy-preserving physical-layer-assisted scheme for {EV} dynamic charging system,'' \emph{{IEEE} Trans. Veh. Technol.}, vol.~67, no.~4, pp. 3304--3318, 2018. [Online]. Available: \url{https://doi.org/10.1109/TVT.2017.2780179}
\BIBentrySTDinterwordspacing

\bibitem{DBLP:journals/tvt/BansalNCS0G20}
\BIBentryALTinterwordspacing
G.~Bansal, N.~Naren, V.~Chamola, B.~Sikdar, N.~Kumar, and M.~Guizani, ``Lightweight mutual authentication protocol for {V2G} using physical unclonable function,'' \emph{{IEEE} Trans. Veh. Technol.}, vol.~69, no.~7, pp. 7234--7246, 2020. [Online]. Available: \url{https://doi.org/10.1109/TVT.2020.2976960}
\BIBentrySTDinterwordspacing

\bibitem{DBLP:journals/tsg/LiuNZY12}
\BIBentryALTinterwordspacing
H.~Liu, H.~Ning, Y.~Zhang, and L.~T. Yang, ``Aggregated-proofs based privacy-preserving authentication for {V2G} networks in the smart grid,'' \emph{{IEEE} Trans. Smart Grid}, vol.~3, no.~4, pp. 1722--1733, 2012. [Online]. Available: \url{https://doi.org/10.1109/TSG.2012.2212730}
\BIBentrySTDinterwordspacing

\bibitem{DBLP:journals/tdsc/AitzhanS18}
\BIBentryALTinterwordspacing
N.~Z. Aitzhan and D.~Svetinovic, ``Security and privacy in decentralized energy trading through multi-signatures, blockchain and anonymous messaging streams,'' \emph{{IEEE} Trans. Dependable Secur. Comput.}, vol.~15, no.~5, pp. 840--852, 2018. [Online]. Available: \url{https://doi.org/10.1109/TDSC.2016.2616861}
\BIBentrySTDinterwordspacing

\bibitem{DBLP:conf/isgt/ThandiM22}
\BIBentryALTinterwordspacing
R.~Thandi and M.~A. Mustafa, ``Privacy-enhancing settlements protocol in peer-to-peer energy trading markets,'' in \emph{2022 {IEEE} Power {\&} Energy Society Innovative Smart Grid Technologies Conference, {ISGT} 2022, New Orleans, LA, USA, April 24-28, 2022}.\hskip 1em plus 0.5em minus 0.4em\relax {IEEE}, 2022, pp. 1--5. [Online]. Available: \url{https://doi.org/10.1109/ISGT50606.2022.9817551}
\BIBentrySTDinterwordspacing

\bibitem{kong2019privacy}
Q.~Kong, R.~Lu, M.~Ma, and H.~Bao, ``A privacy-preserving sensory data sharing scheme in internet of vehicles,'' \emph{Future Generation Computer Systems}, vol.~92, pp. 644--655, 2019.

\bibitem{DBLP:journals/tsg/LiuZZC15}
\BIBentryALTinterwordspacing
X.~Liu, P.~Zhu, Y.~Zhang, and K.~Chen, ``A collaborative intrusion detection mechanism against false data injection attack in advanced metering infrastructure,'' \emph{{IEEE} Trans. Smart Grid}, vol.~6, no.~5, pp. 2435--2443, 2015. [Online]. Available: \url{https://doi.org/10.1109/TSG.2015.2418280}
\BIBentrySTDinterwordspacing

\bibitem{DBLP:journals/tsg/MuslehCD20}
\BIBentryALTinterwordspacing
A.~S. Musleh, G.~Chen, and Z.~Y. Dong, ``A survey on the detection algorithms for false data injection attacks in smart grids,'' \emph{{IEEE} Trans. Smart Grid}, vol.~11, no.~3, pp. 2218--2234, 2020. [Online]. Available: \url{https://doi.org/10.1109/TSG.2019.2949998}
\BIBentrySTDinterwordspacing

\bibitem{DBLP:journals/tiv/PonnuruRPK22}
\BIBentryALTinterwordspacing
R.~B. Ponnuru, A.~G. Reddy, B.~Palaniswamy, and S.~K. Kommuri, ``Ev-auth: Lightweight authentication protocol suite for dynamic charging system of electric vehicles with seamless handover,'' \emph{{IEEE} Trans. Intell. Veh.}, vol.~7, no.~3, pp. 734--747, 2022. [Online]. Available: \url{https://doi.org/10.1109/TIV.2022.3153658}
\BIBentrySTDinterwordspacing

\bibitem{DBLP:journals/tvt/PonnuruARDSP21}
\BIBentryALTinterwordspacing
R.~B. Ponnuru, R.~Amin, A.~G. Reddy, A.~K. Das, W.~Susilo, and Y.~Park, ``Robust authentication protocol for dynamic charging system of electric vehicles,'' \emph{{IEEE} Trans. Veh. Technol.}, vol.~70, no.~11, pp. 11\,338--11\,351, 2021. [Online]. Available: \url{https://doi.org/10.1109/TVT.2021.3116279}
\BIBentrySTDinterwordspacing

\bibitem{basnet2020deep}
M.~Basnet and M.~H. Ali, ``Deep learning-based intrusion detection system for electric vehicle charging station,'' in \emph{2020 2nd International Conference on Smart Power \& Internet Energy Systems (SPIES)}.\hskip 1em plus 0.5em minus 0.4em\relax IEEE, 2020, pp. 408--413.

\bibitem{gawas2021integrative}
M.~Gawas, H.~Patil, and S.~S. Govekar, ``An integrative approach for secure data sharing in vehicular edge computing using blockchain,'' \emph{Peer-to-Peer Networking and Applications}, vol.~14, no.~5, pp. 2840--2857, 2021.

\bibitem{DBLP:journals/tits/0004OY0022}
\BIBentryALTinterwordspacing
J.~Cui, F.~Ouyang, Z.~Ying, L.~Wei, and H.~Zhong, ``Secure and efficient data sharing among vehicles based on consortium blockchain,'' \emph{{IEEE} Trans. Intell. Transp. Syst.}, vol.~23, no.~7, pp. 8857--8867, 2022. [Online]. Available: \url{https://doi.org/10.1109/TITS.2021.3086976}
\BIBentrySTDinterwordspacing

\bibitem{DBLP:journals/tac/LiSC18}
\BIBentryALTinterwordspacing
Y.~Li, D.~Shi, and T.~Chen, ``False data injection attacks on networked control systems: {A} stackelberg game analysis,'' \emph{{IEEE} Trans. Autom. Control.}, vol.~63, no.~10, pp. 3503--3509, 2018. [Online]. Available: \url{https://doi.org/10.1109/TAC.2018.2798817}
\BIBentrySTDinterwordspacing

\bibitem{islam2018impact}
S.~N. Islam, M.~A. Mahmud, and A.~M.~T. Oo, ``Impact of optimal false data injection attacks on local energy trading in a residential microgrid,'' \emph{Ict Express}, vol.~4, no.~1, pp. 30--34, 2018.

\bibitem{kavousi2021effective}
A.~Kavousi-Fard, A.~Almutairi, A.~Al-Sumaiti, A.~Farughian, and S.~Alyami, ``An effective secured peer-to-peer energy market based on blockchain architecture for the interconnected microgrid and smart grid,'' \emph{International Journal of Electrical Power \& Energy Systems}, vol. 132, p. 107171, 2021.

\bibitem{forbush2011regulating}
\BIBentryALTinterwordspacing
J.~R. Forbush, ``Regulating the use and sharing of energy consumption data: assessing california's sb 1476 smart meter privacy statute,'' \emph{Alb. L. Rev.}, vol.~75, p. 341, 2011. [Online]. Available: \url{https://heinonline.org/HOL/Page?handle=hein.journals/albany75&id=343&collection=journals&index=}
\BIBentrySTDinterwordspacing

\bibitem{doi:10.1177/1783591719895390}
\BIBentryALTinterwordspacing
C.~Ducuing, ``Data as infrastructure? a study of data sharing legal regimes,'' \emph{Competition and Regulation in Network Industries}, vol.~21, no.~2, pp. 124--142, 2020. [Online]. Available: \url{https://doi.org/10.1177/1783591719895390}
\BIBentrySTDinterwordspacing

\bibitem{betti2020share}
\BIBentryALTinterwordspacing
F.~Betti, F.~Bezamat, M.~Fendri, B.~Fernandez, D.~K{\"u}pper, and A.~Okur, ``Share to gain: Unlocking data value in manufacturing,'' \emph{World Economic Forum}, 2020. [Online]. Available: \url{https://www3.weforum.org/docs/WEF_Share_to_Gain_Report.pdf}
\BIBentrySTDinterwordspacing

\bibitem{doi:10.1161/HYPERTENSIONAHA.120.16340}
\BIBentryALTinterwordspacing
A.~Vlahou, D.~Hallinan, R.~Apweiler, A.~Argiles, J.~Beige, A.~Benigni, R.~Bischoff, P.~C. Black, F.~Boehm, J.~Céraline, G.~P. Chrousos, C.~Delles, P.~Evenepoel, I.~Fridolin, G.~Glorieux, A.~J. van Gool, I.~Heidegger, J.~P. Ioannidis, J.~Jankowski, V.~Jankowski, C.~Jeronimo, A.~M. Kamat, R.~Masereeuw, G.~Mayer, H.~Mischak, A.~Ortiz, G.~Remuzzi, P.~Rossing, J.~P. Schanstra, B.~J. Schmitz-Dräger, G.~Spasovski, J.~A. Staessen, D.~Stamatialis, P.~Stenvinkel, C.~Wanner, S.~B. Williams, F.~Zannad, C.~Zoccali, and R.~Vanholder, ``Data sharing under the general data protection regulation,'' \emph{Hypertension}, vol.~77, no.~4, pp. 1029--1035, 2021. [Online]. Available: \url{https://www.ahajournals.org/doi/abs/10.1161/HYPERTENSIONAHA.120.16340}
\BIBentrySTDinterwordspacing

\bibitem{FERNANDES2022113240}
\BIBentryALTinterwordspacing
D.~V. Fernandes and C.~S. Silva, ``Open energy data — a regulatory framework proposal under the portuguese electric system context,'' \emph{Energy Policy}, vol. 170, p. 113240, 2022. [Online]. Available: \url{https://www.sciencedirect.com/science/article/pii/S0301421522004591}
\BIBentrySTDinterwordspacing

\bibitem{10.1093/idpl/ipr004}
\BIBentryALTinterwordspacing
R.~Knyrim and G.~Trieb, ``{Smart metering under EU data protection law},'' \emph{International Data Privacy Law}, vol.~1, no.~2, pp. 121--128, 03 2011. [Online]. Available: \url{https://doi.org/10.1093/idpl/ipr004}
\BIBentrySTDinterwordspacing

\bibitem{havlikova2011smart}
\BIBentryALTinterwordspacing
D.~Havlikova, ``Smart grids in the european data protection legal framework: Smart metering implications for the eu data protection,'' Master's thesis, 2011. [Online]. Available: \url{https://www.duo.uio.no/handle/10852/22927}
\BIBentrySTDinterwordspacing

\bibitem{Cuijpers2013}
\BIBentryALTinterwordspacing
C.~Cuijpers and B.-J. Koops, \emph{Smart Metering and Privacy in Europe: Lessons from the Dutch Case}.\hskip 1em plus 0.5em minus 0.4em\relax Dordrecht: Springer Netherlands, 2013, pp. 269--293. [Online]. Available: \url{https://doi.org/10.1007/978-94-007-5170-5_12}
\BIBentrySTDinterwordspacing

\bibitem{DBLP:conf/isgt/Pallas12}
\BIBentryALTinterwordspacing
F.~Pallas, ``Data protection and smart grid communication - the european perspective,'' in \emph{{IEEE} {PES} Innovative Smart Grid Technologies Conference, {ISGT} 2012, Washington, DC, USA, January 16-20, 2012}.\hskip 1em plus 0.5em minus 0.4em\relax {IEEE}, 2012, pp. 1--8. [Online]. Available: \url{https://doi.org/10.1109/ISGT.2012.6175695}
\BIBentrySTDinterwordspacing

\bibitem{Pallas2013}
\BIBentryALTinterwordspacing
------, \emph{Beyond Gut Level-Some Critical Remarks on the German Privacy Approach to Smart Metering}.\hskip 1em plus 0.5em minus 0.4em\relax Dordrecht: Springer Netherlands, 2013, pp. 313--345. [Online]. Available: \url{https://doi.org/10.1007/978-94-007-5170-5_14}
\BIBentrySTDinterwordspacing

\bibitem{DBLP:conf/ccs/BiselliFC13}
\BIBentryALTinterwordspacing
A.~Biselli, E.~Franz, and M.~P. Coutinho, ``Protection of consumer data in the smart grid compliant with the german smart metering guideline,'' in \emph{SEGS'13, Proceedings of the 2013 {ACM} Workshop on Smart Energy Grid Security, Co-located with {CCS} 2013, November 8, 2013, Berlin, Germany}, B.~Defend and K.~Kursawe, Eds.\hskip 1em plus 0.5em minus 0.4em\relax {ACM}, 2013, pp. 41--52. [Online]. Available: \url{https://doi.org/10.1145/2516930.2516933}
\BIBentrySTDinterwordspacing

\bibitem{Milaj2016}
\BIBentryALTinterwordspacing
J.~Milaj and J.~P. Mifsud~Bonnici, \emph{Privacy Issues in the Use of Smart Meters-Law Enforcement Use of Smart Meter Data}.\hskip 1em plus 0.5em minus 0.4em\relax Cham: Springer International Publishing, 2016, pp. 179--196. [Online]. Available: \url{https://doi.org/10.1007/978-3-319-28077-6_12}
\BIBentrySTDinterwordspacing

\bibitem{doi:10.1080/13600869.2017.1371576}
\BIBentryALTinterwordspacing
A.~Brown and R.~Kennedy, ``Regulating intersectional activity: privacy and energy efficiency, laws and technology,'' \emph{International Review of Law, Computers \& Technology}, vol.~31, no.~3, pp. 340--369, 2017. [Online]. Available: \url{https://doi.org/10.1080/13600869.2017.1371576}
\BIBentrySTDinterwordspacing

\bibitem{10.1093/jwelb/jwx001}
\BIBentryALTinterwordspacing
M.~Mylrea, ``{Smart energy-internet-of-things opportunities require smart treatment of legal, privacy and cybersecurity challenges},'' \emph{The Journal of World Energy Law \& Business}, vol.~10, no.~2, pp. 147--158, 04 2017. [Online]. Available: \url{https://doi.org/10.1093/jwelb/jwx001}
\BIBentrySTDinterwordspacing

\bibitem{DBLP:conf/isdevel/MartinezRPAM19}
\BIBentryALTinterwordspacing
J.~Martinez, A.~Ruiz, J.~Puelles, I.~Arechalde, and Y.~Miadzvetskaya, ``Smart grid challenges through the lens of the european general data protection regulation,'' in \emph{Information Systems Development: Information Systems Beyond 2020, {ISD} 2019 Proceedings, Toulon, France, August 28-30, 2019}.\hskip 1em plus 0.5em minus 0.4em\relax {ISEN} Yncr{\'{e}}a M{\'{e}}diterran{\'{e}}e / Association for Information Systems, 2019. [Online]. Available: \url{https://aisel.aisnet.org/isd2014/proceedings2019/Society/4}
\BIBentrySTDinterwordspacing

\bibitem{doi:10.1080/02646811.2019.1622244}
\BIBentryALTinterwordspacing
K.~Huhta, ``Smartening up while keeping safe? advances in smart metering and data protection under eu law,'' \emph{Journal of Energy \& Natural Resources Law}, vol.~38, no.~1, pp. 5--22, 2020. [Online]. Available: \url{https://doi.org/10.1080/02646811.2019.1622244}
\BIBentrySTDinterwordspacing

\bibitem{vojkovic2020iot}
\BIBentryALTinterwordspacing
G.~Vojkovi{\'c}, M.~Milenkovi{\'c}, and T.~Katuli{\'c}, ``Iot and smart home data breach risks from the perspective of data protection and information security law,'' \emph{Business Systems Research: International journal of the Society for Advancing Innovation and Research in Economy}, vol.~11, no.~3, pp. 167--185, 2020. [Online]. Available: \url{https://hrcak.srce.hr/ojs/index.php/bsr/article/view/12916/6453}
\BIBentrySTDinterwordspacing

\bibitem{9519329}
\BIBentryALTinterwordspacing
V.~Mbonye, P.~R. Subramaniam, and I.~Padayachee, ``{POPIA} compliant regulatory framework for smart grids to secure gaps in existing privacy laws,'' in \emph{International Conference on Artificial Intelligence, Big Data, Computing and Data Communication Systems}, 2021, pp. 1--8. [Online]. Available: \url{https://doi.org/10.1109/icABCD51485.2021.9519329}
\BIBentrySTDinterwordspacing

\bibitem{orlando2021smart}
\BIBentryALTinterwordspacing
D.~Orlando and W.~Vandevelde, ``Smart meters’ roll out, solutions in favour of a trust enhancing law in the eu,'' \emph{Journal of Law, Technology and Trust}, vol.~2, no.~1, 2021. [Online]. Available: \url{https://doi.org/10.19164/jltt.v2i1.1071}
\BIBentrySTDinterwordspacing

\bibitem{Mah2014}
\BIBentryALTinterwordspacing
D.~Mah, K.~P.-y. Leung, and P.~Hills, \emph{Smart Grids: The Regulatory Challenges}.\hskip 1em plus 0.5em minus 0.4em\relax London: Springer London, 2014, pp. 115--140. [Online]. Available: \url{https://doi.org/10.1007/978-1-4471-6281-0_7}
\BIBentrySTDinterwordspacing

\bibitem{DBLP:journals/ei/HolzleitnerR17}
\BIBentryALTinterwordspacing
M.~Holzleitner and J.~Reichl, ``European provisions for cyber security in the smart grid - an overview of the nis-directive,'' \emph{Elektrotech. Informationstechnik}, vol. 134, no.~1, pp. 14--18, 2017. [Online]. Available: \url{https://doi.org/10.1007/s00502-017-0473-7}
\BIBentrySTDinterwordspacing

\bibitem{draffin2017cybersecurity}
\BIBentryALTinterwordspacing
C.~Draffin, ``Cybersecurity white paper,'' 2017. [Online]. Available: \url{https://dspace.mit.edu/bitstream/handle/1721.1/130631/Cybersecurity-White-Paper.pdf?sequence=1&isAllowed=y}
\BIBentrySTDinterwordspacing

\bibitem{LESZCZYNA2018262}
\BIBentryALTinterwordspacing
R.~Leszczyna, ``A review of standards with cybersecurity requirements for smart grid,'' \emph{Computers \& Security}, vol.~77, pp. 262--276, 2018. [Online]. Available: \url{https://www.sciencedirect.com/science/article/pii/S0167404818302803}
\BIBentrySTDinterwordspacing

\bibitem{anderson2018cybersecurity}
\BIBentryALTinterwordspacing
C.~Anderson, ``How cybersecurity regulation for the smart grid could upset the current balance of federal and state jurisdiction in electricity regulation,'' \emph{Nat'l Sec. L. Brief}, vol.~8, p.~43, 2018. [Online]. Available: \url{https://heinonline.org/HOL/Page?handle=hein.journals/natislaw8&id=51&collection=journals&index=}
\BIBentrySTDinterwordspacing

\bibitem{zukowska2020legal}
\BIBentryALTinterwordspacing
A.~{\.Z}ukowska, ``Legal conditions for cybersecurity of the energy sector,'' \emph{Publishing House of Rzeszow University of Technology}, p.~75, 2020. [Online]. Available: \url{https://jbc.bj.uj.edu.pl/Content/855080/NDIGOC079792_2020_003.pdf#page=75}
\BIBentrySTDinterwordspacing

\bibitem{en14237836}
\BIBentryALTinterwordspacing
M.~Krzykowski, ``Legal aspects of cybersecurity in the energy sector: Current state and latest proposals of legislative changes by the eu,'' \emph{Energies}, vol.~14, no.~23, 2021. [Online]. Available: \url{https://www.mdpi.com/1996-1073/14/23/7836}
\BIBentrySTDinterwordspacing

\bibitem{10.1007/978-3-031-15559-8_38}
\BIBentryALTinterwordspacing
T.~Schober and G.~Griessnig, ``Cybersecurity regulations and standards in the automotive domain,'' in \emph{Systems, Software and Services Process Improvement}, M.~Yilmaz, P.~Clarke, R.~Messnarz, and B.~W{\"o}ran, Eds.\hskip 1em plus 0.5em minus 0.4em\relax Cham: Springer International Publishing, 2022, pp. 530--539. [Online]. Available: \url{https://link.springer.com/chapter/10.1007/978-3-031-15559-8_38}
\BIBentrySTDinterwordspacing

\bibitem{DBLP:journals/corr/abs-2210-13119}
\BIBentryALTinterwordspacing
J.~Meyer and G.~Apruzzese, ``Cybersecurity in the smart grid: Practitioners' perspective,'' \emph{CoRR}, vol. abs/2210.13119, 2022. [Online]. Available: \url{https://doi.org/10.48550/arXiv.2210.13119}
\BIBentrySTDinterwordspacing

\bibitem{doi:10.1080/13600869.2022.2094609}
\BIBentryALTinterwordspacing
D.~Markopoulou, ``Tackling cybersecurity challenges in the energy and water sectors in the context of the cybersecurity and sectoral regulatory frameworks: the case of smart metering systems in the new digitalised environment,'' \emph{International Review of Law, Computers \& Technology}, vol.~37, no.~1, pp. 52--77, 2023. [Online]. Available: \url{https://doi.org/10.1080/13600869.2022.2094609}
\BIBentrySTDinterwordspacing

\bibitem{eckardt2024property}
\BIBentryALTinterwordspacing
M.~Eckardt and W.~Kerber, ``Property rights theory, bundles of rights on {IoT} data, and the {EU Data Act},'' \emph{European Journal of Law and Economics}, vol.~57, no.~1, pp. 113--143, 2024. [Online]. Available: \url{https://doi.org/10.1007/s10657-023-09791-8}
\BIBentrySTDinterwordspacing

\bibitem{mokobombang2020open}
\BIBentryALTinterwordspacing
N.~N. Mokobombang, J.~Gutierrez, and K.~Petrova, ``Open government data initiatives: Open by default or publishing with purpose,'' 2020. [Online]. Available: \url{https://aisel.aisnet.org/confirm2020/8}
\BIBentrySTDinterwordspacing

\bibitem{suh2024non}
\BIBentryALTinterwordspacing
J.~Suh, J.~Lee, and J.~Roh, ``On the non-discrimination principles in digital trade,'' \emph{World Trade Review}, vol.~23, no.~1, pp. 72--92, 2024. [Online]. Available: \url{https://doi.org/10.1017/S147474562300037X}
\BIBentrySTDinterwordspacing

\bibitem{hacker2024regulating}
\BIBentryALTinterwordspacing
P.~Hacker, J.~Cordes, and J.~Rochon, ``Regulating gatekeeper artificial intelligence and data: Transparency, access and fairness under the digital markets act, the general data protection regulation and beyond,'' \emph{European Journal of Risk Regulation}, vol.~15, no.~1, pp. 49--86, 2024. [Online]. Available: \url{https://doi.org/10.1017/err.2023.81}
\BIBentrySTDinterwordspacing

\bibitem{kellerbauer2024eu}
M.~Kellerbauer, M.~Klamert, and J.~Tomkin, \emph{The EU Treaties and Charter of Fundamental Rights: a Commentary}.\hskip 1em plus 0.5em minus 0.4em\relax Oxford University Press, 2024.

\bibitem{DBLP:conf/re/RoulandGJ23}
\BIBentryALTinterwordspacing
Q.~Rouland, S.~Gjorcheski, and J.~Jaskolka, ``Eliciting a security architecture requirements baseline from standards and regulations,'' in \emph{31st {IEEE} International Requirements Engineering Conference, {RE} 2023 - Workshops, Hannover, Germany, September 4-5, 2023}, K.~Schneider, F.~Dalpiaz, and J.~Horkoff, Eds.\hskip 1em plus 0.5em minus 0.4em\relax {IEEE}, 2023, pp. 224--229. [Online]. Available: \url{https://doi.org/10.1109/REW57809.2023.00045}
\BIBentrySTDinterwordspacing

\bibitem{hughes2023data}
\BIBentryALTinterwordspacing
N.~Hughes and D.~Kalra, ``Data standards and platform interoperability,'' in \emph{Real-World Evidence in Medical Product Development}.\hskip 1em plus 0.5em minus 0.4em\relax Springer, 2023, pp. 79--107. [Online]. Available: \url{https://doi.org/10.1007/978-3-031-26328-6_6}
\BIBentrySTDinterwordspacing

\bibitem{von2024data}
\BIBentryALTinterwordspacing
F.~von Scherenberg, M.~Hellmeier, and B.~Otto, ``Data sovereignty in information systems,'' \emph{Electronic Markets}, vol.~34, no.~1, p.~15, 2024. [Online]. Available: \url{https://doi.org/10.1007/s12525-024-00693-4}
\BIBentrySTDinterwordspacing

\bibitem{sharma2019data}
\BIBentryALTinterwordspacing
S.~Sharma, \emph{Data privacy and GDPR handbook}.\hskip 1em plus 0.5em minus 0.4em\relax John Wiley \& Sons, 2019. [Online]. Available: \url{https://doi.org/10.1002/9781119594307}
\BIBentrySTDinterwordspacing

\bibitem{Irti2022}
\BIBentryALTinterwordspacing
C.~Irti, \emph{Personal Data, Non-personal Data, Anonymised Data, Pseudonymised Data, De-identified Data}.\hskip 1em plus 0.5em minus 0.4em\relax Singapore: Springer Singapore, 2022, pp. 49--57. [Online]. Available: \url{https://doi.org/10.1007/978-981-16-3049-1_5}
\BIBentrySTDinterwordspacing

\bibitem{doi:10.1080/09640568.2022.2027233}
\BIBentryALTinterwordspacing
J.~K.~K. Ida Dokk~Smith and K.~Szulecki, ``A functional approach to decentralization in the electricity sector: learning from community choice aggregation in california,'' \emph{Journal of Environmental Planning and Management}, vol.~66, no.~6, pp. 1305--1335, 2023. [Online]. Available: \url{https://doi.org/10.1080/09640568.2022.2027233}
\BIBentrySTDinterwordspacing

\bibitem{berntzen2021blockchain}
\BIBentryALTinterwordspacing
L.~Berntzen, Q.~Meng, B.~Vesin, M.~R. Johannessen, T.~Brekke, and I.~Laur, ``Blockchain for smart grid flexibility-handling settlements between the aggregator and prosumers,'' in \emph{International Conference on Digital Society and eGovernments (ICDS)}, 2021. [Online]. Available: \url{https://hdl.handle.net/11250/2787624}
\BIBentrySTDinterwordspacing

\bibitem{bray2019barriers}
\BIBentryALTinterwordspacing
R.~Bray and B.~Woodman, ``Barriers to independent aggregators in europe,'' \emph{ORE Open Research Exeter}, 2019. [Online]. Available: \url{https://ore.exeter.ac.uk/repository/handle/10871/40134}
\BIBentrySTDinterwordspacing

\bibitem{SCHITTEKATTE2021106971}
\BIBentryALTinterwordspacing
T.~Schittekatte, V.~Deschamps, and L.~Meeus, ``The regulatory framework for independent aggregators,'' \emph{The Electricity Journal}, vol.~34, no.~6, p. 106971, 2021. [Online]. Available: \url{https://www.sciencedirect.com/science/article/pii/S1040619021000622}
\BIBentrySTDinterwordspacing

\bibitem{bygrave2021data}
\BIBentryALTinterwordspacing
L.~A. Bygrave, ``Data protection by design and by default,'' \emph{Oxford Encyclopedia of European Union Law (Oxford University Press 2022), Forthcoming, University of Oslo Faculty of Law Research Paper}, no. 2021-19, 2021. [Online]. Available: \url{https://ssrn.com/abstract=3944535}
\BIBentrySTDinterwordspacing

\bibitem{DBLP:journals/frai/NietEV21}
\BIBentryALTinterwordspacing
I.~Niet, R.~van Est, and F.~Veraart, ``Governing {AI} in electricity systems: Reflections on the {EU} artificial intelligence bill,'' \emph{Frontiers Artif. Intell.}, vol.~4, p. 690237, 2021. [Online]. Available: \url{https://doi.org/10.3389/frai.2021.690237}
\BIBentrySTDinterwordspacing

\bibitem{HEYMANN20234538}
\BIBentryALTinterwordspacing
F.~Heymann, K.~Parginos, R.~Bessa, and M.~Galus, ``Operating ai systems in the electricity sector under european’s ai act – insights on compliance costs, profitability frontiers and extraterritorial effects,'' \emph{Energy Reports}, vol.~10, pp. 4538--4555, 2023. [Online]. Available: \url{https://www.sciencedirect.com/science/article/pii/S2352484723015494}
\BIBentrySTDinterwordspacing

\bibitem{https://doi.org/10.1111/reel.12574}
\BIBentryALTinterwordspacing
B.~Espinosa~Apráez and M.~Noorman, ``Regulating ai in the ‘twin transitions’: Significance and shortcomings of the ai act in the digitalised electricity sector,'' \emph{Review of European, Comparative \& International Environmental Law}, vol.~33, no.~3, pp. 367--382, 2024. [Online]. Available: \url{https://onlinelibrary.wiley.com/doi/abs/10.1111/reel.12574}
\BIBentrySTDinterwordspacing

\end{thebibliography}

\end{document}